\documentclass[12pt,amsmath]{iopart}

\usepackage{amsmath}
\usepackage{amsfonts}
\usepackage{slashed}
\usepackage{amssymb}
\usepackage{iopams}  

\usepackage[utf8]{inputenc}
\usepackage{color}
\usepackage[unicode=true,pdfusetitle,bookmarks=true,bookmarksnumbered=false,bookmarksopen=false,
 breaklinks=false,backref=false,colorlinks=false]
 {hyperref}

\usepackage[symbol]{footmisc}
\usepackage{tabularx}
\usepackage{graphicx}
\usepackage{dcolumn}
\usepackage{bm}
\usepackage{hyperref}

\usepackage{graphicx}
\usepackage{bbm}
\usepackage{xcolor}
\usepackage{footnote}
\usepackage[normalem]{ulem}

\newcommand{\tcr}{\textcolor{red}}

\newcommand{\LF}{\left(}
\newcommand{\RF}{\right)}
\newcommand{\LT}{\left[}
\newcommand{\RT}{\right]}
\newcommand{\pd}{\partial}
\newcommand{\Pc}{\mathcal{P}}
\newcommand{\Rc}{\mathcal{R}}
\newcommand{\Tc}{\mathcal{T}}
\newcommand{\Hc}{\mathcal{H}}
\newcommand{\Fc}{\mathcal{F}}
\newcommand{\Oc}{\mathcal{O}}
\newcommand{\Lc}{\mathcal{L}}
\newcommand{\Uc}{\mathcal{U}}
\newcommand{\Vc}{\mathcal{V}}

\newcommand{\Ec}{\mathcal{E}}
\newcommand{\Nc}{\mathcal{N}}
\newcommand{\CPT}{\mathbb{C}\mathbb{P}\mathbb{T}}
\newcommand{\Cb}{\mathbb{C}}
\newcommand{\Pb}{\mathbb{P}}
\newcommand{\Tb}{\mathbb{T}}
\def\Rc{\mathcal{R}}
\def\Ts{\text{T}}

\def\polarizationtensor{\rm e}

\begin{document}

\title[A new understanding of Einstein-Rosen bridges]{A new understanding of Einstein-Rosen bridges}

\author{Enrique Gazta\~naga$^{1,2,3}$, K. Sravan Kumar$^{{\ast} 1}$ and Jo\~ao Marto$^4$}

\address{$^1$ Institute of Cosmology \& Gravitation,
	University of Portsmouth,
	Dennis Sciama Building, Burnaby Road,
	Portsmouth, PO1 3FX, United Kingdom}
\address{$^2$ Institute of Space Sciences (ICE, CSIC), 08193 Barcelona, Spain}
\address{$^3$ Institut d´ Estudis Espacials de Catalunya (IEEC), 08034 Barcelona, Spain}	
\address{$^4$ Departamento de F\'isica, Centro de Matem\'atica e Aplicações (CMA-UBI), Universidade da Beira Interior, Rua Marquês D'Ávila e Bolama, 6201-001 Covilhã, Portugal}	
\ead{enrique.gaztanaga@port.ac.uk, sravan.kumar@port.ac.uk ($^\ast$ corresponding author), jmarto@ubi.pt}
\vspace{10pt}
\begin{indented}
\item[]\today
\end{indented}

\begin{abstract}
The formulation of quantum field theory in Minkowski spacetime, which emerges from the unification of special relativity and quantum mechanics, is based on treating time as a parameter, assuming a fixed arrow of time, and requiring that field operators commute for spacelike distances. This procedure is questioned here in the context of quantum field theory in curved spacetime (QFTCS).
In 1935, Einstein and Rosen (ER), in their seminal paper 
(Einstein and Rosen 1935 Phys. Rev. 48 73–77) proposed that "a particle in the physical Universe has to be described by mathematical bridges connecting two sheets of spacetime" which involved two arrows of time.
Recently proposed direct-sum quantum theory reconciles this ER's vision by introducing geometric superselection sectors associated with the regions of spacetime related by discrete transformations.
We further establish that the quantum effects at gravitational horizons involve the physics of quantum inverted harmonic oscillators that have phase space horizons. This new understanding of the ER bridges is not related to classical wormholes, it addresses the original ER puzzle and promises a unitary description of QFTCS, along with observer complementarity. 
Furthermore, we present compelling evidence for our new understanding of ER bridges in the form of large-scale parity asymmetric features in the cosmic microwave background, which is statistically 650 times stronger than the standard scale-invariant power spectrum from the typical understanding of inflationary quantum fluctuations when compared with the posterior probabilities associated with the model given the data. We finally discuss the implications of this new understanding in combining gravity and quantum mechanics.



\end{abstract}

\maketitle
%
%
%
\renewcommand\contentsname{}
\tableofcontents
\newpage
%
%


\section{Introduction}

{General Relativity (GR) and Quantum Mechanics (QM) formulate the rules of macroscopic and microscopic physical worlds. QM has been successfully merged with the principles of special relativity (SR) and has evolved into quantum field theory (QFT), which has formed the basis for the emergence of what we know now as the standard model (SM) of particle physics. The success of QFT formulation is widely celebrated with quantum electrodynamics, showing an excellent agreement with precise observations of atomic spectra, and 2012's discovery of the Higgs boson at the Large Hadron Collider (LHC). The main pillars of QFT are locality, unitarity, and renormalizability. The unification of quantum mechanics and special relativity involves quantizing classical fields in Minkowski spacetime, a process known as second quantization, which builds upon a deeper understanding of quantum principles. According to QM, the structure of the Sch\"{o}dinger equation reflects the fact that time is a parameter (not an operator) in contrast to the spatial position attached to the position operator. 
Wigner, in 1932 \cite{Wigner1993}, showed that time-reversal symmetry must be implemented by an anti-unitary operator, which further shaped our understanding of time in quantum theory.  Another notable aspect of quantum theory lies in the way we define a positive energy state with respect to an arrow of time. QFT formalism imbibes both of these concepts, along with the causality condition of operators corresponding to spacelike distances that must commute, which ensures compatibility with SR. Thus, even in QFT, time is treated as a parameter in the same way as in QM. The second quantization mimics the position and momentum uncertainty relations of QM with the field and its conjugate momenta. The rest of the developments are further building blocks, such as Feynman propagators, interactions, scattering matrices, etc., which comprise essential elements of perturbative QFT. }

Following the historical footsteps, it is logical to expect that beyond the SM of particle physics and the GR, the immediate new physics one can think of is quantum field theory in curved spacetime (QFTCS). While the QFT in Minkowski spacetime and development of SM was under progress, seminal work of Einstein \& Rosen (ER) in 1935 \cite{Einstein:1935tc}, Schr\"{o}dinger in 1956 \cite{Schrodinger1956}, have shed light on the troubling aspects of QFTCS if one strictly follow the guidelines of quantum theory such as fixing the arrow of time. The seminal works of Hawking on QFTCS in the Schwarzschild black hole (BH) \cite{Hawking:1975vcx,Hawking:1976ra}  inspired by the investigations of Zeldovich and Starobinsky \cite{Zeldovich:1971mw,Starobinsky:1973aij} have led to the issues of unitarity and information-loss paradoxes. Unitarity loss that occurs in the standard formulation of QFT in curved spacetimes (with event or apparent horizons) is associated with pure states evolving to mixed states in the observer's causally accessible physical world \cite{Unruh:2017uaw,Parikh:2002py,Giddings:2022jda}. To be more precise, the standard framework of QFTCS implies an entanglement between "point particle states" across the spacetime horizon, leaving observers on either side accessing only mixed states, even though globally together form a pure state. Due to the fact that an observer causally cut off from a region of spacetime, it is often accepted that unitarity is lost in curved spacetime, i.e., the fate of any observer's accessibility is only to mixed states, not pure states. The challenge here is whether we can formulate a new understanding of QFTCS where pure states evolve into pure states in a way that every observer's description of the causally accessible physical world is complete. This question was first posed by Schrödinger in 1956 \cite{Schrodinger1956} in the context of an expanding Universe, and it has been the formidable barrier to constructing a successful theory of quantum gravity \cite{Witten:2001kn}.   
Even after decades of research and numerous explorations of Planck-scale quantum gravity\footnote{{Throughout this paper, our reference to Planck scale quantum gravity aligns with the conventional expectation of a renormalizable, ultraviolet (UV) complete quantum theory of gravity, applicable up to and beyond Planck length scales, where the graviton is typically treated as a fluctuation around Minkowski spacetime. However, if one aims to develop a Planck-scale quantum gravity framework within a curved spacetime context (i.e., treating graviton fluctuations around a curved background such as de Sitter space), it becomes crucial to address the foundational issues of quantum fields in curved spacetime, which is the central focus of this paper.} }, the fundamental questions about QFTCS, such as the loss of unitarity and black hole information paradox \cite{Raju:2020smc,Raju:2021lwh}, still loom around. 
{The emergence of these issues challenges our current approach to quantizing fields in curved spacetime. Traditionally, frameworks aiming to unify gravity and quantum mechanics, such as string theory, loop quantum gravity, and others, have suggested that a not-yet-discovered theory of quantum gravity will restore unitarity and ultimately resolve the black hole information paradox at Planck scales \cite{Almheiri:2019hni,Kim:2022pfp,Giddings:2022jda,Ashtekar:2025ptw}. In contrast, recent work by Gerard ’t Hooft emphasizes that the path to quantum gravity may lie in a deeper understanding and reformulation of  QFTCS \cite{tHooft:2022umh}. Similar perspectives are emerging in the investigations of Witten \cite{Witten:2021jzq}, Giddings \cite{Giddings:2022jda}, and others, who have also underscored the need to revisit the foundational structures of quantum theory in the presence of gravity. Decades of effort have not yielded a fully unitary formulation of QFTCS, suggesting that gravity may compel us to reconsider the very foundations of quantum mechanics. We propose a perspective that redefines our understanding of quantum mechanics and quantum field theory by introducing a new structure of Hilbert and Fock spaces, each associated with discrete regions of spacetime of a given manifold.}

In this paper, we highlight the crucial observations of Einstein-Rosen (ER) in an attempt to combine GR and QM  \cite{Einstein:1935tc}. The basic essence of ER investigation is the incompatibility between gravity and quantum theory due to the possibility of two arrows of time describing one physical world. ER demanded that there should only be one physical world, but they were not in favor of choosing an arrow of time by hand. Since the QM and standard QFT requires fixing the arrow of time (or the arrow of causality) \cite{Donoghue:2019ecz,tHooft:2018jeq}, to solve the particle problem in GR, ER conjectured that a description of the particle (quantum field) in one physical world has to be described by mathematical bridges between two sheets of spacetime. A similar conclusion was obtained independently by Schr\"{o}dinger in 1956 \cite{Schrodinger1956} and 't Hooft \cite{tHooft:2016qoo} in 2016 in the context of cosmological (de Sitter) and black hole (BH) spacetimes. The occurrence of two arrows of time in an attempt to describe one physical world is not only limited to the (quantum) physics at the gravitational horizons but also bound to occur in the context of phase space horizons of an inverted harmonic oscillator (IHO). The seminal work of Berry and Keating (BK) in 1999 \cite{Berry1999} uncovered the intricacies in the quantum physics of IHO. As a way out, they proposed the identification of phase space regions. There is an intriguing similarity between BK's proposal in the context of quantum IHO phase space and Schr\"{o}dinger and 't Hooft's proposals in the context of quantum physics at gravitational horizons. In other words, the first quantisation of IHO and the second quantisation in curved spacetime are fundamentally related. The purpose of the paper is to juxtapose all these foundational developments that independently emerged across decades and identify the universal features connecting them. 

We discuss the relations between ER bridges and analogous proposals in different contexts with the recently developed framework of direct-sum quantum theory and its applications to early Universe cosmology, and BH physics \cite{Kumar:2023ctp,Kumar:2023hbj,Kumar:2024oxf,Gaztanaga:2024vtr,Kumar:2022zff,Gaztanaga:2024whs,Kumar:2024ahu}. Direct-sum quantum theory is based on the discrete spacetime (such as parity ($\Pc$) and time reversal ($\Tc$)) (a)symmetries of the physical system to formulate a description of the quantum state as a direct-sum of components corresponding to (geometric) superselection sectors (SSS) of Hilbert space. {Geometric SSS are Hilbert spaces that describe quantum states corresponding to regions of physical space related by discrete spacetime transformations. If a Hilbert space is a direct-sum of geometric SSS, a state vector in that Hilbert space becomes a direct-sum of components, corresponding to geometric SSS. The same applies to operators in the Hilbert space. This is called the geometric superselection rule.  }   
We show that the "direct-sum" is the mathematical bridge that matches the expectations of ER bridges in describing one physical world with two arrows of time. The two arrows of time here operate at the parity conjugate regions of physical space embedded with the geometric construction of SSS. This framework restores unitarity in curved spacetime, and it is tested against the latest observations of the cosmic microwave background from the Planck satellite data. 

The paper is organized as follows. {In Sec .~\ref {sec:QFTmin} we provide an overview of the concept of time as we understand it through QFT in Minkowski spacetime and how the whole subject is developed with (quantum) harmonic oscillator physics and highlight the undercurrent role of the inverted harmonic oscillator (IHO) in the standard model (SM) of particle physics.} 
In Sec.~\ref{sec:IHO}, we discuss the quantum physics of Berry and Keating's IHO and connection to the non-trivial zeros of the Riemann zeta function. We present in particular the conceptual conundrums associated with the quantization of IHO. 
In Sec.~\ref{sec:history}, we analyze the discrete spacetime symmetries of curved spacetimes and their essential role in understanding quantum fields in curved spacetimes. We show in particular how quantum fields in curved spacetime require understanding the quantum physics of IHO. 
We also discuss the origins of ER's proposal of mathematical bridges, which has links with later discoveries by Schr\"{o}dinger (1956) and 't Hooft (2016).  In Sec.~\ref{sec:disumQT}, we present the basic elements of direct-sum quantum field theory (QFT) and demonstrate the new understanding of spacetime with geometric SSS. In Sec.~\ref{sec:BKDQ}, we study the implications of direct-sum quantum theory for understanding IHO and show how the construction resonates well with BK's quantization proposal. In Sec.~\ref{sec:ERDQFT}, we uncover the relation between ER bridges and the direct-sum QFTCS in the contexts of Rindler, de Sitter, and Schwarzschild spacetimes.  
 In Sec.~\ref{Sec:DSI}, we provide observational support for our new understanding of ER bridges with early Universe cosmology that leads to temperature fluctuations in the cosmic microwave background (CMB). In Sec.~\ref{Sec:conc}, we summarize by highlighting important aspects of our studies and outline future directions. Furthermore, we discuss non-trivial implications for the open challenges we have in all the theories of Planck scale quantum gravity\cite{deBoer:2022zka,Loll:2022ibq}.

 Throughout the paper, we follow the units $\hbar=c=1$ and metric signature $(-+++)$. Throughout this paper, we use a dot over a variable to indicate differentiation with respect to the time parameter $t$ and a prime over a variable to denote differentiation with respect to conformal time $\tau$. 

 \section{Quantum fields in Minkowski spacetime and examples of (inverted) harmonic oscillator physics} 
 \label{sec:QFTmin}

 Before we step into the concerning issues of QFTCS in the next sections, it is useful to recall some specific foundations of QFT in Minkowski spacetime. The elements of foundations we discuss in this section are not only the textbook material \cite{Coleman:2018mew}, but also the core concepts of current investigations \cite{Donoghue:2019ecz,Donoghue:2020mdd}. QFT in Minkowski spacetime (merge of QM and special relativity) is about formulating quantum fields on the  manifold defined by
 \begin{equation}
     ds^2 = -dt_p^2+d\textbf{x}^2\,,
     \label{minmet}
 \end{equation}
 where $t_p$ is the parametric time coordinate and $\textbf{x}$ denotes 3 dimensional space. 
 The spacetime \eqref{minmet} is invariant under discrete transformations such as parity $\Pc$ and time reversal $\Tc$ 
 \begin{equation}
     \Tc: t_p\to -t_p,\quad \Pc: \textbf{x}\to -\textbf{x}
     \label{dissymmin}
 \end{equation}
 However, at its foundational stage relies on the following steps: 
\begin{itemize}
    \item Assume an arrow of time $t_p: -\infty \to \infty$ 
    \item Define a positive energy ($\Ec>0$) state with respect to the arrow of time: $\vert \Psi\rangle_t = e^{-i\Ec t_p}\vert\Psi\rangle_0$ 
    \item Field operators for space-like distances must commute. For example, let us take a Klein-Gordon (KG) field; we have 
    \begin{equation}
        \left[ \hat \phi\LF x \RF,\,\hat\phi\LF y \RF  \right] = 0,\quad \LF x-y \RF^2>0\,. 
        \label{commspace}
    \end{equation}
    \item The canonical commutation relations between the field and its conjugate momenta are 
    \begin{equation}
        \LT \hat{\phi}\LF t_p,\,\textbf{x} \RF,\, \hat\Pi_\phi\LF t_p,\,\textbf{y} \RF  \RT = i\delta^{3}\LF \textbf{x}-\textbf{y} \RF
    \end{equation}
\end{itemize}
It is already noted in \cite{Donoghue:2019ecz,Donoghue:2020mdd} that the arrow of time in QFT in Minkowski spacetime is strongly associated with the conventional structure of Schr\"{o}dinger equation 
\begin{equation}\label{scheq}
    i\frac{\pd\vert \Psi\rangle}{\pd t_p} = \hat H \Psi
\end{equation}
where $\hat H$ is the Hamiltonian operator. We could have "$-i$" instead of "$+i$" on the left hand side of Schr\"{o}dinger equation \eqref{scheq}. If we change  "$+i$" with "$-i$" everywhere in quantum theory, we essentially change our notion of the arrow of time, which is typically $t_p: -\infty\to \infty$, but it gets changed to $t_p: \infty\to -\infty$ with the $-i$ convention. So practically no results of observables such as scattering amplitudes and decay rates would change with this \cite{Donoghue:2019ecz}.  

The Schr\"{o}dinger equation, though it is widely taught as a non-relativistic QM, we borrow the concept of fixing the arrow of time and the definition of positive energy state from it when we construct the QFT. The essential step that merges QM with special relativity is the mathematical operation \eqref{commspace}, which quantum mechanically preserves the basic lessons we learn from special relativity, i.e., there cannot be any communication between space-like distances. All these steps guide us in expressing the field operator in the following well-known structure 
\begin{equation}
   \hat\phi\LF x \RF \equiv \hat \phi\LF t_p,\, \textbf{x}  \RF = \int \frac{d^3k}{\LF 2\pi \RF^{3/2}\sqrt{2\vert k_0\vert}} \Bigg[a_\textbf{k} e^{-ik_0t_p+i\textbf{k}\cdot \textbf{x}}+a_\textbf{k}^\dagger e^{+ik_0t_p-i\textbf{k}\cdot \textbf{x}}\Bigg]
\end{equation}
where the first term with the annihilation operator $a_\textbf{k}$ is a positive energy term, whereas the second term with the creation operator $a_\textbf{k}^\dagger$ is a negative energy term with respect to the assumption of the arrow of time $t_p: -\infty \to \infty$. 
The relation between QM and QFT also lies in the elegant extension of the single quantum harmonic oscillator to an infinite collection of quantum harmonic oscillators. Indeed, let us consider an example of a KG field in 1+1 dimensions 
\begin{equation}
\begin{aligned}
S^{1+1}_{KG} & = \frac{1}{2}\int dt_p \Bigg\{\int dx\Bigg[ \LF \frac{\pd\phi}{\pd t_p} \RF^2- \LF \frac{\pd\phi}{\pd x} \RF^2 -m^2\phi^2 \Bigg]\Bigg\}  \\ 
 & = \frac{1}{2}\int dt_p \Bigg\{\int dx\, \phi\Bigg[ \LF -\frac{\pd^2}{\pd t_p^2} \RF +\LF \frac{\pd^2}{\pd x^2} \RF -m^2 \Bigg]\phi \Bigg\}
\end{aligned}
\label{11KGaction}
\end{equation}
This is a continuous approximation of $n$ harmonic oscillators separated by an infinitesimal distance $s$, which is known as a lattice model 
\begin{equation}
S^{1+1}_{KG} = \frac{1}{2}\int dt_p \Bigg\{\sum_n s\Bigg[ \LF \frac{\pd\phi_n}{\pd t_p} \RF^2-\frac{1}{s^2}\LF \phi_{n+1}-\phi_n \RF^2 -  m^2\phi_n^2 \Bigg]\Bigg\}
\label{lattice-action}
\end{equation}
In the limit $n\to \infty$ and $s\to 0$ the action \eqref{lattice-action} becomes \eqref{11KGaction}. 
The important lesson in this simple step is the understanding of the gradient term in the KG action, which represents the coupling between infinitely many harmonic oscillators. The mass term and the kinetic term are analogous to the single harmonic oscillator case given by the Lagrangian 
\begin{equation}
    S_{HO} = \frac{1}{2} \int dt \LT \dot{q}^2-\omega^2 q^2 \RT
    \label{HOaction}
\end{equation}
Here we can relate the mass $m$ with the frequency $\omega$ by equating the relativistic energy with unit quantum energy $E=mc^2=\hbar \omega$. 
We can easily now extend the KG field to 1+3 dimensions as 
\begin{equation}
S^{1+3}_{KG} = \frac{1}{2}\int dt_p \Bigg\{\int d^3x\Bigg[ \LF \frac{\pd\phi}{\pd t_p} \RF^2- \LF \nabla\phi \RF^2 -m^2\phi^2 \Bigg]\Bigg\}
\end{equation}
which is again an infinite collection of harmonic oscillators in 3 spatial dimensions. What if we consider the case $\omega\to i\omega?$ i.e., $\tilde\omega^2 = -\omega^2>0$ in \eqref{HOaction} and correspondingly $m\to im?$ (i.e., $\mu^2 = -m^2>0$ in \eqref{11KGaction}. This would turn the harmonic oscillator into an IHO and the corresponding field into a tachyonic field (i.e., an infinite collection of coupled IHOs)
\begin{equation}
    S_{IHO} = \frac{1}{2} \int dt \LT \dot{q}^2-(-\tilde\omega^2) q^2 \RT,\quad S^{1+1}_{KG}  = \frac{1}{2}\int dt_p \Bigg\{\int dx\Bigg[ \LF \frac{\pd\phi}{\pd t_p} \RF^2- \LF \frac{\pd\phi}{\pd x} \RF^2 -(-\mu^2)\phi^2 \Bigg]\Bigg\}
    \label{IHOac}
\end{equation}
and the KG field into a tachyonic field
\begin{equation}
S^{1+3}_{T} = \frac{1}{2}\int dt_p \Bigg\{\int d^3x\Bigg[ \LF \frac{\pd\phi}{\pd t_p} \RF^2- \LF \nabla\phi \RF^2 -(-\mu^2)\phi^2 \Bigg]\Bigg\},\quad \mu^2>0\,.
\label{tachyonac}
\end{equation}
The IHO and the tachyonic field configurations play a fundamental role in the standard model of particle physics. The famous Higgs potential 
\begin{equation}
    V_H = -\frac{\mu_H^2}{2} H_{\rm sm}^\dagger H_{\rm sm}+ \frac{\lambda_H}{4}\LF H_{\rm sm}^\dagger H_{\rm sm} \RF^2
    \label{Higgs-pot}
\end{equation}
where $H_{\rm sm} = \frac{1}{\sqrt{2}} h e^{i\theta_h}$ here is the Higgs field which is a complex scalar field (${\rm SU}(2)$ doublet) with 4 components, $\mu_H^2>0$ which is not the physical mass of the Higgs boson and $\lambda_H>0$ (dimensionless) indicates the self-interaction of the Higgs field. The potential has two degenerate minima $v= \pm \sqrt{\frac{\mu_H^2}{\lambda_H}}$. This is due to the $Z_2$ symmetry $H_{\rm sm}\to -H_{\rm sm}$ of the potential \eqref{Higgs-pot}. Substituting $\mu_H^2 = \lambda_H v^2$ and adding a suitable constant ($-\frac{\lambda_H}{4} v^4$ since it does not affect the dynamics)  to the potential in \eqref{Higgs-pot}, we obtain the famous Mexican-hat form of the potential, which still carries the tachyonic mass term 
\begin{equation}
   V_H =  
   \frac{\lambda_H}{4} \LF H_{\rm sm}^\dagger H_{\rm sm}-v^2 \RF^2\,. 
\end{equation}
The physical mass of the Higgs boson in the SM arises from perturbative expansion around the minima of the potential in the unitary gauge as (after the so-called spontaneous symmetry breaking and the subsequent absorption of 3 Goldstone bosons by the $W^\pm$ and $Z$ bosons, resulting in the electroweak symmetry breaking) 
\begin{equation}
    H_{\rm sm} = \frac{1}{\sqrt{2}}\begin{pmatrix}
        0 \\ 
        v+\phi_h
    \end{pmatrix},\quad V_{\phi_h}= \frac{1}{2}m_h^2\phi_h^2+\lambda_H  vh^3+\frac{\lambda_H}{4}\phi_h^4,\quad m_h^2 = 2\lambda_H v^2.
    \label{exphiggs}
\end{equation}
where $m_h^2$ is the physical mass of the Higgs boson. Examining \eqref{Higgs-pot} with \eqref{tachyonac}, we can clearly see that the Higgs field before the spontaneous symmetry breaking should be viewed as an infinite collection of self-interacting IHOs. After the $Z_2$ symmetry breaking (subsequently the electroweak symmetry breaking), what we see is the massive (physical) Higgs field $\phi_h$, which can be seen as an infinite collection of harmonic oscillators in Minkowski spacetime. The tachyonic instability (IHO physics) plays a significant role in SM. It is usually studied as a non-perturbative transition from false vacuum to the true vacuum using an instanton solution obtained in Euclidean spacetime ($t_p\to it_p$) \cite{Coleman:1977py}. 
 In the SM, for all the practical calculations, understanding the IHO physics is not needed, even though its relevance is absolutely clear in its construction. This is because the majority of the standard model calculations involve expanding the Higgs field perturbatively around the minimum \eqref{exphiggs}. However, it is worth noting here is that the quantum physics of IHO is an absolutely non-trivial subject in QM that surfaced from the seminal works of Berry and Keating \cite{Berry1999}, also it is still an active subject of investigation across the spectrum of problems in theoretical physics \cite{Subramanyan:2020fmx}. 

The appearance of IHO physics is not just limited to SM of particle physics, but it does appear in QFTCS and plays a significant role in shaping our understanding of unitarity.  
In the later sections, we will discuss in detail the physics of IHO and its relation to the fundamental understanding of nature. 

\subsection{Time translations and Killing vector of Minkowski spacetime:} 

Understanding the symmetries of spacetime is fundamental in GR. Minkowski spacetime, as a maximally symmetric solution, possesses the highest possible number of continuous symmetries, each associated with a Killing vector field. If a vector $\xi^\mu$ is a Killing vector of a given manifold endowed with a metric $\LF \mathcal{M},\,g_{\mu\nu} \RF$ then the Lie derivative of the metric tensor along $\xi^\mu$ vanishes as
\begin{equation}
\begin{aligned}
(\mathcal{L}_\xi g)_{\mu\nu} & = \xi^\lambda \partial_\lambda g_{\mu\nu} + g_{\lambda\nu} \partial_\mu \xi^\lambda + g_{\mu\lambda} \partial_\nu \xi^\lambda = 0 \\ 
& = \nabla_\mu\xi_\nu+\nabla_\nu\xi_\mu =0\,.
\end{aligned}
\label{Lieder}
\end{equation}
In Miknowski spacetime there are 10 generators associated with spacetime translations, rotations and Lorentz boosts. We particularly focus on the generator of time translations 
\begin{equation}
    \xi^\mu = \LF \frac{\pd}{\pd t_p} \RF^\mu = (1,\,0,\,0,\,0)
    \label{killingMin}
\end{equation}
which is a Killing vector of Minkowski manifold. For every continuous symmetry there is a conserved quantity by Noether theorem. The time translation symmetry gives the following conserved energy ($\mathbb E$) 
\begin{equation}
   \mathbb E = \int d^3x J^0 = \int d^3x\, T^{00}\,, 
\end{equation}
where $J^0$ is the zeroth component of conserved current $J^\mu$ and $T^{00}$ is the $00$th component of energy-momentum tensor $T^{\mu\nu}$. 
The crucial point we must notice here is that the Killing vector \eqref{killingMin} retains its property \eqref{Lieder} under the time reversal transformation $t_p\to -t_p$. Thus, the conserved energy $\mathbb E$ remains the same under $t_p\to -t_p$. This is not surprising as the time reversal operation is (discrete) symmetry \eqref{dissymmin} of the Minkowski manifold \eqref{minmet}.

\section{Review of inverted harmonic oscillator (IHO), quantization and Riemann Hypothesis}

\label{sec:IHO}

As we have seen in the previous section, the Higgs Mexican hat potential \eqref{Higgs-pot} indicates the first appearance of IHO in SM physics. As we will further demonstrate in the later sections, in the context of early Universe cosmology, the inflationary quantum fluctuations, which manifestly appear in terms of canonical quantum variables (known as Mukhanov-Sasaki variables) \cite{Albrecht:1992kf}, can be understood through QFT in terms of inverted harmonic oscillators (IHOs). Even in the context of Black Hole physics, IHOs are found to be the fundamental building blocks to describe Hawking radiation \cite{Betzios:2020wcv,Ullinger:2022xmv,Subramanyan:2020fmx}. The role of IHOs even extends to the Rindler spacetime and also in the context of the quantum Hall effect, molecular physics, and even in biophysics (See \cite{Subramanyan:2020fmx,Sundaram:2024ici} and references therein). {Moreover, the quantum aspects of the IHO manifest in relation to the nontrivial zeros of the Riemann zeta function} \cite{Schumayer:2011yp}. {This section provides a comprehensive brief review of IHO physics, quantization, and its non-trivial relation to the Riemann zeta function. In particular, we emphasize how foundational questions regarding the quantization of the IHO are intimately connected to the challenge of establishing its energy spectrum as corresponding to the nontrivial zeros of the Riemann zeta function. The elements discussed in this section, especially the wave functions and phase space regions of the IHO, are closely tied to the physics of quantum fields in curved spacetime, which we will develop in the sections that follow.}

\begin{figure}
    \centering
\includegraphics[width=0.5\linewidth]{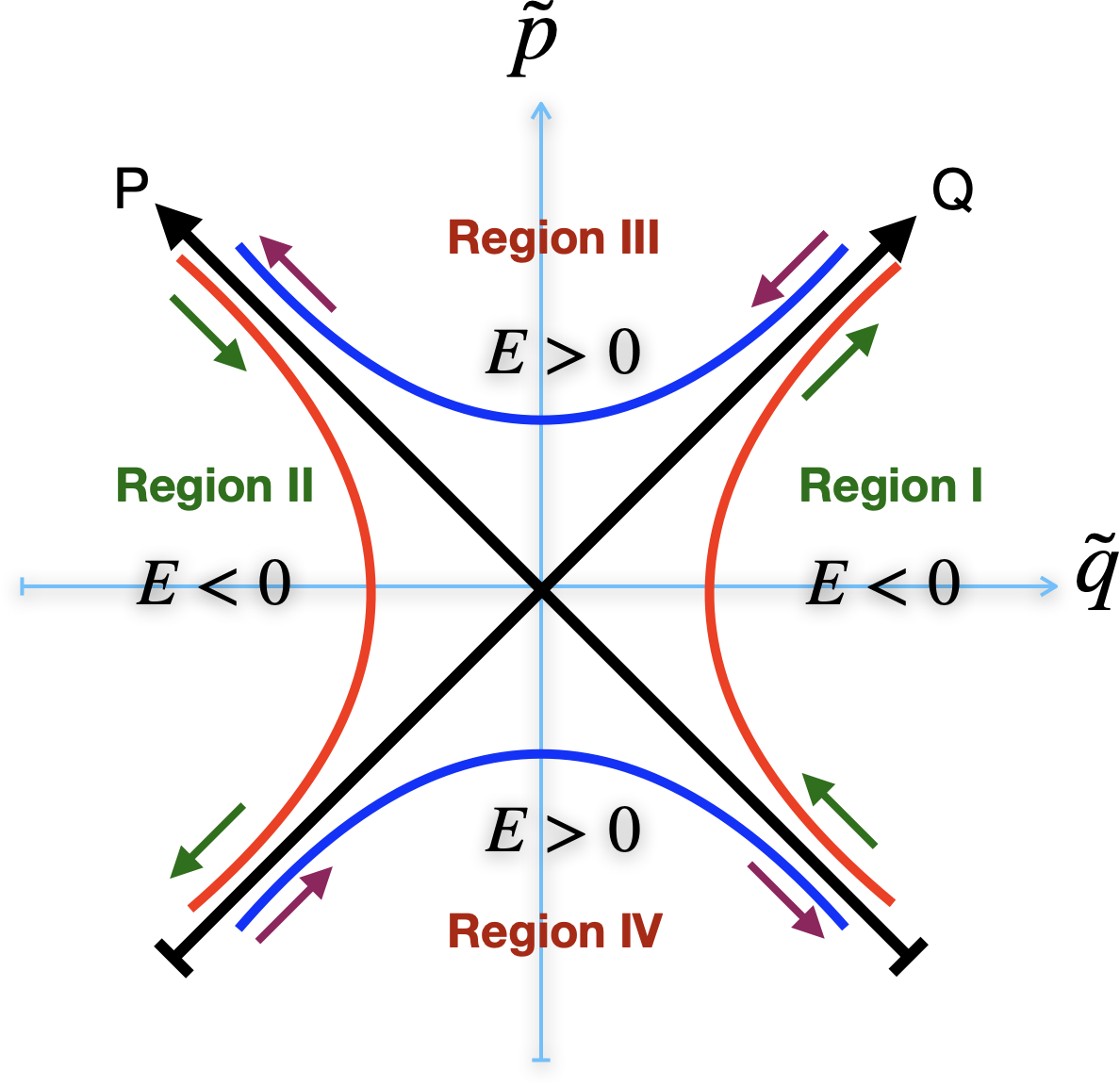}
    \caption{Phase space of inverted harmonic oscillator representing doubly degenerate positive and negative energy solutions in \eqref{BKsol1} and \eqref{sol2BK}. The negative energy trajectories are given by $Q>0,\, P<0$ and $Q<0,\, P>0$ whereas the positive energy trajectories are $Q>0,\,P>0$ and $Q<0,\,P<0$. These double degenerate trajectories are related by \eqref{discreteTPQ} whereas the positive and negative energy regions are related by \eqref{BKPE}.}
    \label{fig:IHOPS}
\end{figure}

In 1999, M. V. Berry and J. Keating (BK) found a remarkable relation between the energy spectrum of the IHO and the zeros of the Riemann zeta function \cite{Berry1999} along $Re[s]=1/2$. This is in line with Hilbert-P\'olya's conjecture \cite{Sierra:2016rgn}.  The following classical Hamiltonian describes the IHO
\begin{equation}
	H_{iho} =
\frac{ \omega}{2} \LF \Tilde{p}^2-\Tilde{q}^2 \RF, \quad \Tilde{p} = \frac{p}{\sqrt{ m\omega}},\quad \Tilde{q}= \sqrt{m\omega} q
	\label{IHOhamil}
\end{equation}
The Hamiltonian equations of motion are
\begin{equation}
    \dot{\tilde{q}} = \frac{\pd H_{iho}}{\pd \tilde{p}},\quad \dot{\tilde{p}}= -\frac{\pd H_{iho}}{\pd \tilde{q}}\,.
    \label{ihoem}
\end{equation}
Like the Harmonic oscillator case, the IHO is symmetric under $\Pc\Tc$, i.e., $t\to -t,\,\tilde{q}\to -\tilde{q}$ with a crucial difference that energy is not bounded from below, {which lead us to interpret it as an instability. The classical solutions of \eqref{ihoem} can be written as 
\begin{equation}
    H_{iso} =  \frac{ \omega}{2} E \implies \begin{cases}
        \Tilde{p}= \pm \sqrt{\vert E\vert }\cosh{\omega t},\quad \Tilde{q}= \pm \sqrt{\vert E\vert }\sinh{\omega t},\quad E>0 \\  \Tilde{p}= \pm \sqrt{\vert E\vert }\sinh{\omega t},\quad \Tilde{q}= \pm \sqrt{\vert E\vert }\cosh{\omega t},\quad E<0
    \end{cases}
    \label{BKsol1}
\end{equation}
where $\omega>0$ and $E\omega$ characterize the energy of the physical system, which can be both positive ($E>0$) and negative ($E<0$).\footnote{{In the case of a harmonic oscillator, where position and momentum are harmonic functions of time, the system energy is positive definite  
\begin{equation}
    H_{ho} = \frac{\omega}{2}\LF \tilde{p}^2+\tilde{q}^2 \RF,\quad \tilde{p}= \sqrt{\vert E\vert }\cos\LF \omega t \RF,\quad \tilde{q}= \sqrt{\vert E\vert }\sin\LF \omega t \RF
\end{equation}}} Since the physical system can span a range of infinitely positive and negative energies, the Hamiltonian of the IHO is said to be unbounded from below. This is a clear contrast with usual quantum mechanics, where we always deal with physical systems whose energies are bounded from below (even when we have situations where the potential is negative). Since the potential of IHO is unbounded from below (can take infinite negative values), there are conceptual limitations to treating IHO as a scattering problem \cite{BALAZS1990123}. The phase space trajectories in Fig.~\ref{fig:IHOPS} define four regions separated by phase space horizons or separatrices \cite{BALAZS1990123,Ullinger:2022xmv} $\tilde p= \pm \vert \tilde{q}\vert $
\begin{equation}
    \begin{aligned}
        \tilde p & = \sqrt{\vert E\vert }\sinh\LF \omega t \RF,\,\tilde q = \sqrt{\vert E\vert }\cosh\LF \omega t \RF,\,t: -\infty\to \infty\,(\rm Region\,I,\,E<0)  \\
          \tilde p & = \sqrt{\vert E\vert }\sinh\LF \omega t \RF,\,\tilde q = -\sqrt{\vert E\vert }\cosh\LF \omega t \RF,\,t: \infty\to -\infty\,(\rm Region\,II,\,E<0) \\ 
        \tilde p & = \sqrt{\vert E\vert }\cosh\LF \omega t \RF,\,\tilde q = -\sqrt{\vert E\vert }\sinh\LF \omega t \RF,\,t: -\infty\to \infty\,(\rm Region\,III,\,E>0)\\
          \tilde p & = -\sqrt{\vert E\vert }\cosh\LF \omega t \RF,\,\tilde q = -\sqrt{\vert E\vert }\sinh\LF \omega t \RF,\,t: \infty\to -\infty\,(\rm Region\,IV,\,E>0) \,,
    \end{aligned}
    \label{phihoeq}
\end{equation}
where we can notice the behavior of position and momentum swap when changing from a region of negative energy to positive energy and vice versa. The arrows of time in \eqref{phihoeq} define the arrows of phase space trajectories in Fig.~\ref{fig:IHOPS}.
}

One can rewrite \eqref{IHOhamil} in terms of the so-called canonically rotated coordinates 
\begin{equation}
	H_{iho} = \frac{ \omega}{2}\LF Q\cdot P+P\cdot Q  \RF,\quad Q= \frac{\Tilde{p}+\Tilde{q}}{\sqrt{2}},\, \quad P= \frac{\Tilde{p}-\Tilde{q}}{\sqrt{2}}
 \label{BKHamilt}
\end{equation}
which is known as the Berry-Keating Hamiltonian \cite{Berry1999} whose equations of motion give the following solutions 
\begin{equation}
	Q= Q_0 e^{\omega t},\quad P= P_0 e^{-\omega t},\quad H_{iho} = Q_0P_0 = \frac{\omega}{2}E\,.
 \label{sol2BK}
\end{equation}
From the phase space trajectories of IHO Fig.~\ref{fig:IHOPS}, we may conclude that the Hamiltonian is unbounded and the system is highly unstable. Depending on the initial conditions, the phase space exhibits doubly degenerate time evolutions with both positive and negative energies, separated by phase space horizons or separatrices \cite{BALAZS1990123,Ullinger:2022xmv} $\Tilde{p}= \pm\Tilde{q}$. Furthermore, we can also notice that the doubly degenerate trajectories are associated with opposite arrows of time ($t: -\infty \to \infty$ and $t: \infty \to -\infty $) together with the following discrete transformations  
\begin{equation}
    Q\to -Q,\quad P\to -P,\quad t\to -t.  
    \label{discreteTPQ}
\end{equation}
which leaves the Hamiltonian \eqref{BKHamilt} invariant. It is worth noting that \eqref{discreteTPQ} with $t\to -t$ in \eqref{sol2BK} becomes the $\Pc\Tc$ transformation in our notation. {Furthermore, the positive and negative energy regions in Fig.~\ref{fig:IHOPS} are related by 
\begin{equation}
    Q\to \mp P,\quad P\to \pm Q\implies \tilde{p}\to \mp \tilde{q},\quad \tilde{q}\to \pm \tilde p\,.
    \label{BKPE}
\end{equation}}
{An interesting observation in IHO phase space in Fig.~\ref{fig:IHOPS} is that a parity transformation $\Pc: \tilde q\to -\tilde q$ alone does not take us from Region I to Region II; we must apply simultaneously the time reversal $\Tc: t\to -t$. }

As a consequence of Heisenberg's uncertainty principle, we have (restoring here the factors of $\hbar$)
\begin{equation}
    [\hat{\Tilde{q}},\,\hat{\Tilde{p}}] = i\hbar,\quad [\hat Q,\,\hat P]=i\hbar,\quad \hat{H}_{iho}= -i\hbar\omega \LF Q\pd_Q+\frac{1}{2} \RF\,.
    \label{IHOhamilQ}
\end{equation}

Notice in Fig.~\ref{fig:IHOPS} that the parity conjugate regions of physical space, with opposite arrows of time, are separated by the lines of phase space horizons or separatrices $\tilde{p} = \pm \vert \tilde{q}\vert $. Quantum mechanically, IHO has been understood in two ways \cite{Sierra:2007du}: (i) With the BK's quantization: by applying the identification for doubly degenerate points in phase space $\LF Q,\,  P \RF$ and $\LF - Q,\, - P \RF$ along with boundary conditions based on the dilatation symmetries. Interestingly, these lead to matching the spectrum of IHO with the non-trivial zeros of the Riemann zeta function along the line $Re[s]=1/2$. (ii) Considering the IHO as a scattering problem with incoming and outgoing states. This allows the quantum states to reflect and tunnel from one region to another region. However, this consideration faces fundamental challenges at the foundational level due to the presence of phase space horizons \cite{BALAZS1990123}. Furthermore, the connection between IHO and the Riemann zeta function is unclear in the scattering approach.  Notice that the analysis of IHO's Wigner function\footnote{Wigner function is a function of position and the momentum, it is a quasi-probability that describes quantum states in a phase space from which we can derive position space and momentum space wave functions of the Hamiltonian. In the context of IHO \eqref{BKHamilt} the Wigner function is defined as 
\[
W_E(Q, P) = \frac{1}{2\pi \hbar} \int_{-\infty}^{\infty} dy \; e^{-i P y / \hbar} \left( Q + \frac{y}{2} \right)^{-i\lambda - \frac{1}{2}} \left( Q - \frac{y}{2} \right)^{i\lambda - \frac{1}{2}}
\]} and the corresponding conditions for scattering, it was found in \cite{BALAZS1990123} that the tunneling from the left ($E<0$) to the right region ($E<0$) of phase space depicted in Fig.~\ref{fig:IHOPS} is not possible unless one invokes an evolution of quantum states from negative to positive energy. QM does not allow this because the $E<0$ and $E>0$ regions of phase space involve distinct time evolutions. 
On the other hand, BK's quantum description of IHO suffers from issues related to quantum chaos \cite{Berrychaos} because Hamiltonian is unbounded with regions of phase space containing different arrows of time. Furthermore, BK's identification and boundary conditions lack physical and (phase space) geometrical understanding.  

The position $Q$ and momentum $P$ wavefunctions of the IHO Hamiltonian operator (for the region $Q>0$ and $E<0$) are (in the units of setting $\omega=1$) \cite{Sierra:2016rgn,Berry1999,Ullinger:2022xmv}
\begin{equation}
    \Psi \LF Q \RF = \frac{C}{\sqrt{2\pi\hbar}}\vert Q\vert^{-\frac{1}{2}+\frac{i\vert E\vert }{\hbar }},\quad  \Psi \LF P \RF = \frac{1}{\sqrt{2\pi\hbar}}  \vert P\vert^{-\frac{1}{2}-\frac{i\vert E\vert }{\hbar}} \LF 2\hbar \RF^{\frac{i\vert E\vert }{\hbar}}\frac{\Gamma\LF \frac{1}{4}+\frac{i\vert E\vert}{2\hbar} \RF}{\Gamma\LF \frac{1}{4}-\frac{i\vert E\vert}{2\hbar} \RF}
\end{equation}
which satisfy the orthogonal and completeness properties \cite{Ullinger:2022xmv,Sundaram:2024ici}. The wavefunctions as a function of $ \Tilde{p}$ and $\Tilde{q}$ can be found explicitly, along with a detailed discussion of probability densities without any singularities at the phase space horizons, can be found in \cite{BARTON1986322,Sundaram:2024ici,Ullinger:2022xmv}. The wave function of IHO becomes delocalized with time evolution. Thus, one cannot have the usual interpretation of a particle. This resonates with the situation in {describing quantum fields in} curved spacetime, where we cannot have usual particle interpretations {as we will discuss in the later sections}. 

Notable features of quantum IHO are 
\begin{itemize}
    \item With the quantum mechanical limitation $\vert Q\vert \geq \ell_Q$ and $\vert P\vert \geq \ell_P$ such that $\ell_Q\ell_P = 2\pi\hbar $, the energy spectrum of IHO becomes discrete. Counting the number of states between $0$ and $\vert E\vert >0$ one gets 
    \begin{equation}
        N\LF E \RF =  \frac{\vert E\vert  }{2\pi\hbar} \LF \ln\frac{\vert E\vert }{2\pi\hbar}-1 \RF+\frac{7}{8}
    \end{equation}
    which matches with the average number of non-trivial zeros of the Riemann zeta function $\zeta\LF \frac{1}{2}\pm i\bar T \RF$ for $\bar T\gg 1$ with the identification $\bar T \to \frac{\vert E\vert }{\hbar}$. 
    \item The relation between IHO energy eigenstates and Riemann zeros was shown to be more than a coincidence with the analysis of scale transformations and the discrete symmetries of the IHO's phase space, which form the dihedral group\cite{Aneva:1999fy} $D_4$ of order 8. These symmetries render a boundary condition (for either $E>0$ or $E<0$)
    \begin{equation}
        Q^{1/2}\zeta\LF \frac{1}{2}-\frac{iE}{\hbar} \RF\Psi(Q) + P^{1/2}\zeta\LF \frac{1}{2}+\frac{iE}{\hbar} \RF \Psi(P)=0\,.
        \label{boundarycondiBK}
    \end{equation}
      The condition \eqref{boundarycondiBK} implies the position and momentum wave function are time reversals of each other \cite{Berry1999}. However, the geometric and physical interpretation of this condition in association with the entire region of phase space was stated as an open problem by Berry and Keating \cite{Berry1999}.
    \item The wavefunction $\Psi\LF Q \RF$ is also an Eigen function of the Weyl reflected Laplace-Beltrami operator (expressed here in the units of setting $\omega=1$) 
    \begin{equation}
        L_R = -Q^2\pd_Q^2-2Q\pd_Q = \LF \frac{1}{2}-\frac{i\hat{H}_{iho}}{\hbar} \RF \LF \frac{1}{2}+\frac{i\hat{H}_{iho}}{\hbar} \RF
        \label{LBOp}
    \end{equation}
    with positive definite Eigenvalues $\LF \frac{1}{4}+\frac{E^2}{4\hbar^2} \RF$. 
    \item BK proposes identifying the discrete set of points in phase space, which are $\LF Q,\, P\RF$, and $\LF -Q,\,-P \RF$. This is very much similar to the antipodal identification in de Sitter spacetime proposed by Schr\"{o}dinger and the one of 't Hooft in the context of Schwarzschild spacetime \cite{Schrodinger1956,tHooft:2016qoo}. As we will discuss in the later sections, the antipodal identification is similar to the ER's mathematical bridge. Thus, what BK proposes is another "mathematical bridge" to join the IHO's phase space regions with opposite arrows of time. 
    \item {The phase space of IHO has several symmetries, the most important symmetry is the dilatation defined by the following scaling transformation
    \begin{equation}
        Q \to e^\lambda Q\,,\quad P \to e^{-\lambda} P 
    \end{equation}
    which preserve the Hamiltonian \eqref{BKHamilt}. This is called the hyperbolic scaling symmetry. 
    The generator of these scaling transformations is known as the Hamiltonian vector field defined by
    \begin{equation}
        X_H = Q\frac{\pd}{\pd Q}- P\frac{\pd}{\pd P}
        \label{dilatonicop}
    \end{equation}
    This dilatation operator's (also called Liouville operator) \eqref{dilatonicop} eigenfunction ($\Psi_D$) has a universal structure 
    \begin{equation}
        \Psi_D(Q,\,P) = \LF \frac{Q}{P} \RF^{\lambda} g\LF QP \RF 
    \label{dilEigen}
    \end{equation}
    where $g(QP)$ is an arbitrary function of the scale invariant quantity $QP$. The Wigner function of the BK Hamiltonian \eqref{BKHamilt} has the following generic structure as well 
 \begin{equation}
        W_E(Q,\,P) = \Nc_E \LF \frac{Q}{P} \RF^{iE/\hbar} g_W\LF QP \RF\,,
    \label{dilEigen}
    \end{equation}
    where $g_W(QP)$ is an arbitrary function modulated along constant $QP$ curves in the phase space and $\Nc_E$ is a normalization constant. Thus, if we have a physical system that deals with eigenfunctions of dilatation \eqref{dilEigen}, we can anticipate the role of IHO physics.  We witness this in the context of QFT in black hole, de Sitter, and Rindler spacetimes that we shall discuss in the next section.  }
\end{itemize}

\section{QFT in curved spacetime: discrete symmetries, unitarity loss, and new insights}

\label{sec:history} 

{In flat spacetime, QFT is elegantly formulated by decomposing the field into Fourier modes, where each mode behaves like a harmonic oscillator with a well-defined frequency. This framework rests on the presence of a global timelike Killing vector field along with a presumption on the arrow of time, which guarantees a definition of energy and particle states. The vacuum and excited states of the field are then naturally described in terms of the quantized energy levels of these oscillators.
However, in curved spacetime, particularly in the presence of event or dynamical horizons, such as those associated with black holes, de Sitter or accelerated observers (as in the Unruh effect), the situation changes dramatically. The lack of a global timelike Killing vector makes the notion of a unique vacuum ambiguous, and the field modes evolve with time-dependent or even imaginary frequencies near the horizon. These features suggest, as the central idea of this paper, a shift from ordinary harmonic oscillators to IHO. The IHO has a hyperbolic potential, unlike the parabolic potential of the standard harmonic oscillator. Its solutions exhibit exponential instability rather than bounded oscillations, mirroring the behavior of quantum field modes near horizons. In fact, studies of quantum fields in Rindler, de Sitter, or Schwarzschild spacetimes often reduce to analyzing effective IHO dynamics. This inverted potential captures crucial aspects of horizon physics, such as mode amplification and thermal particle creation. Furthermore, the IHO structure underpins the Bogoliubov transformations that relate different field quantizations (e.g., between inertial and accelerated observers), leading to phenomena like Hawking radiation and the Unruh effect. The logarithmic phase singularities and time asymmetry of the IHO reflect the causal disconnection and thermality introduced by the horizon.
In this light, the IHO is not just a mathematical curiosity but a fundamental component in the description of quantum fields in curved spacetimes. It represents a shift from conservative, oscillatory systems to unstable, scattering-like systems that encode the irreversible and horizon-induced nature of quantum phenomena in gravity. Thus, while harmonic oscillators are central to flat spacetime QFT, IHOs become indispensable in curved spacetimes where horizons, their deep thermodynamic and quantum implications, come into play. }

{Quantizing fields in curved spacetime builds directly upon the foundational techniques of quantum field theory in flat Minkowski spacetime, as outlined in the previous section. In this section, we focus on a conceptual overview of QFTCS, emphasizing key developments in the contexts of de Sitter space, Schwarzschild black hole (SBH), and Rindler spacetime. We also highlight the role of quantum fluctuations in inflationary cosmology, which offer important observational consequences. Beyond reviewing the foundational literature, we draw attention to several enduring conceptual challenges, particularly those related to discrete spacetime symmetries, that have complicated the formulation of a consistent and unitary QFTCS. Crucially, we demonstrate how a common mathematical structure emerges across these spacetimes, tied to the hyperbolic nature of their coordinate transformations and the associated inverted harmonic oscillator dynamics governing field modes near horizons. These (a)symmetry-related structural insights serve as a conceptual bridge to the new framework we propose in the following sections. }

\subsection{QFTCS in BH spacetime}

The ER paper \cite{Einstein:1935tc} of 1935 is the first work in history that looked for quantum effects in curved spacetime. 
{ER attempted to formulate quantum theory in SBH spacetime described by}
\begin{equation}
    ds^2 = -\LF 1-\frac{2GM}{r} \RF dt^2 + {\LF 1-\frac{2GM}{r} \RF}^{-1} dr^2+ r^2 d\Omega^2
    \label{eq:SBH}
\end{equation}
where $d\Omega^2 = d\theta^2+\sin^2\theta d\varphi^2$ describes two dimensional sphere, $\LF t,\,r \RF$ are time and radial cordinates. There is a coordinate singularity at the Horizon $r=r_S=2GM$, and the physical singularity is at $r=0$. 
{ER noted that one cannot write quantum theory in a spacetime with a coordinate singularity. Thus, they found a new coordinate system to remove coordinate singularity by a redefinition of the radial coordinate as
\begin{equation}
    r= u^2+2GM \implies u\in \LT -\infty, \infty \RT
\end{equation}
which renders the SBH metric \eqref{eq:SBH} in to the following form  
\begin{equation}
    ds^2 = -\LF \frac{u^2}{u^2+2GM} \RF dt^2+4\LF u^2+2GM \RF du^2+ \LF u^2+2GM \RF^2 d\Omega^2\,.
    \label{ERmetr}
\end{equation}
These new coordinates $\LF t,\,u,\,,\theta,\,\varphi \RF$ which we call the Einstein-Rosen coordinates represent the spacetime $r>2GM$ since $u= \pm \sqrt{r-2GM}$ and $u\in \LT -\infty,\,\infty \RT$.
In the Schwarzschild coordinates $\LF t,\,r,\,\theta,\,\varphi \RF$, the metric \eqref{eq:SBH} is invariant under discrete transformations such as parity and time reversal defined by
\begin{equation}
    \Tc: t\to -t,\quad \Pc: \LF \theta,\,\varphi \RF\to \LF \pi-\theta,\,\pi+\varphi \RF,\quad r\to r\,.
    \label{SBHsym}
\end{equation}
The discrete symmetries in the Einstein-Rosen coordinates are  
\begin{equation}
    \Tc: t\to -t,\quad \Pc: \LF \theta,\,\varphi \RF\to \LF \pi-\theta,\,\pi+\varphi \RF, \quad r\to r,\quad \Pc^{\rm new}: u\to -u\,.
\end{equation}
where $\Pc^{\rm new}$ is a new "parity" transformation that requires a new interpretation, which is what the ER paper was about. As ER demands both $u>0$ and $u<0$ could represent two sheets of spacetime, but in reality, there is only one exterior $r>2GM$. So they proposed an aesthetic solution 
\begin{center}
    A particle in the physical world should be represented by the mathematical bridge between two sheets of spacetime.
\end{center}
However, the ER metric has a peculiarity in comparison with the Schwarzschild metric \eqref{eq:SBH} in the following two aspects 
\begin{itemize}
    \item $\sqrt{-g}\Big\vert_{r\to 2GM} = \sqrt{r_S^4\sin^2\theta}$ (for \eqref{eq:SBH}) while for the Einstein-Rosen metric $\sqrt{-g}\Big\vert_{r\to 2GM\implies u\to 0} = \sqrt{4u^2 \LF u^2+2GM \RF^4\sin^2\theta}\Big\vert_{u\to 0}\to 0$. Since $\int\sqrt {-g} d^4x$ is a measure of the volume integral, if it vanishes for a finite radius, we cannot define consistently any action for matter fields in Schwarzschild spacetime. 
    \item There is an additional discrete symmetry $u\to -u$ which is absent in the \eqref{eq:SBH} metric (See \eqref{SBHsym}).  
\end{itemize}
We can reduce this additional discrete symmetry by the following identification 
\begin{equation}
    t\to -t \Longleftrightarrow u\to -u
\end{equation}
But still, we are left with the first problem about $\sqrt{-g}$, which cannot be solved with the ER metric form. Kruskal-Sz\'ekeres solves the above two problems and gives us the same number of discrete symmetries as the original Schwarzschild metric, as can be seen below. } 

The Kruskal-Sz\'ekers (KS) coordinates $\LF U,\, V\RF$ 
\begin{equation}
       U = \pm \,4GM \sqrt{\Big\vert 1- \frac{r}{2GM}\Big\vert } \exp\LF -\frac{t-r}{4GM} \RF,\, V = \pm \,4GM \sqrt{\Big\vert 1-\frac{r}{2GM}\Big\vert } \exp\LF \frac{t+r}{4GM} \RF
       \label{Kruskalcoord}
   \end{equation}
   which obey 
   \begin{equation}
   	\begin{aligned}
   		UV& = 16G^2M^2\LF 1-\frac{r}{2GM}\RF \exp\LF {\frac{r}{2GM}}\RF,\quad \frac{U}{V} =  \pm\, \exp\LF{-\frac{t}{2GM}}\RF 
   	\end{aligned}
   	\label{eq:KS}
   \end{equation}
remove the $r=2GM$.
ER worried about the appearance of two identical sheets of spacetime when one aims to describe a quantum field in the exterior of the Schwarzschild BH (SBH). 
  With the redefinition
\begin{equation}
T = \frac{U+V}{2\sqrt{e}},\,X=\frac{V-U}{2\sqrt{e}}
\label{TXdef}
\end{equation}
 The SBH becomes 
\begin{equation}
    ds^2 = \frac{2GM}{r}e^{1-\frac{r}{2GM}}\LF -dT^2+dX^2\RF + r^2d\Omega^2 
    \label{KSmetr}
\end{equation}
 From \eqref{eq:KS} we can notice that
\begin{equation}
    r>2GM \implies \begin{cases}
        U<0,\, V>0 \\ 
        U>0,\,V<0
    \end{cases},\quad r<2GM \implies \begin{cases}
        U>0,\, V>0 \\ 
        U<0,\,V<0
    \end{cases}
    \label{eq:extint}
\end{equation}
where we can notice the following discrete symmetry in both regions $r<2GM$ and $r>2GM$ 
\begin{equation}
   T\to -T \implies U\to -U,\, V\to -V \implies X\to -X
   \label{TtomT}
\end{equation}
{Thus, the $\Pc\Tc$ operation in KS coordinates become
\begin{equation}
    \Pc: \LF \theta,\,\varphi \RF\to \LF \pi-\theta,\,\pi+\varphi \RF,\quad \Tc: T\to -T,\,X\to -X
    \label{PTKS}
\end{equation}}
This implies there are two arrows of time $T: \mp\infty \to \pm\infty$,  to describe the exterior (interior) of the SBH related by discrete transformations. 
If we consider one arrow of time, say $T: -\infty \to \infty$, to do quantum physics with positive energy states, then one ends up with another physical spacetime with the opposite arrow of time and negative energy states. {Though Einstein-Rosen uses a different coordinate system \eqref{ERmetr} to write the Schwarzschild metric non-singular at $r=2GM$, the sheets of spacetime describing $r>2GM$ are similar to what we describe here in terms of KS coordinates. Most importantly, the two sheets representing the same physical world (outside the SBH) are related by a discrete coordinate transformation ($U\to -U,\,V\to -V$).
Since in the near-horizon approximation \eqref{KSmetr} looks very similar to some "flat" spacetime metric as
\begin{equation}
    ds^2 \Bigg\vert_{r\approx 2GM} = -dT^2+dX^2+r_S^2d\Omega^2
    \label{localflat}
\end{equation}
With its resemblance to "Minkowski", we can perceive the form of the metric \eqref{localflat} to realize quantum fields in this spacetime. For this, as we learned at the beginning of Sec.~\ref{sec:QFTmin}, we must define a positive energy state and an arrow of time for $"T"$ here. Thus, 
we can interpret the ER concerns as the appearance of a (quantum mechanically) negative energy state that comes by reversing the arrow of time ($T\to -T$ i.e, $T: \infty \to -\infty$ in the (naive writing of) Schr\"{o}dinger equation (in the $r\approx 2GM$ approximation) $i\frac{\pd\vert \Psi\rangle }{\pd T}=E\vert\Psi\rangle $. Thus, we have both positive and negative energies possible due to the discrete symmetry \eqref{eq:extint}. 
The conceptual conundrums here are 
\begin{itemize}
    \item If we choose $T: -\infty\to \infty$ we first break by hand the symmetry of the metric \eqref{KSmetr}. Then we are bound to interpret the parallel identical regions \eqref{eq:extint} by the transformation \eqref{TtomT}, either a nonphysical or a parallel Universe. This is the interpretation that the majority of developments have adopted ever since the seminal works of Hawking \cite{Hawking:1975vcx,Hawking:1976ra,Almheiri:2019hni,Almheiri:2020cfm,Maldacena:2013xja}. 
    \item ER paper emphasized the importance of defining only one physical region, but without breaking any of the discrete symmetries of the manifold. ER, conjectured:
\begin{center}
\it A quantum field in physical space has to be described by mathematical bridges between two sheets of spacetime
\end{center}
\end{itemize}
In the later sections, we shall return to the new conception of time reversal that can describe the same positive energy state with the opposite arrow of time. }

{It is worth noting that the Einstein-Rosen paper is literally about the quantum mechanical understanding of the Schwarzschild horizon. After nearly 20 years of ER paper, some elements of discussion in the ER paper motivated Misner, Morris, Thorne, and Yurtsever to find "wormhole" solutions with GR modifications or introducing exotic matter on the right-hand side of the Einstein equations \cite{Misner:1957mt,Morris:1988tu}. The result of these findings led to the interpretation of the ER paper's "mathematical bridges between two sheets of spacetime" as "non-traversable wormholes" connecting various regions of spacetime, and those new wormhole solutions in modified gravity and/or with exotic matter, that violate energy conditions, became known as "traversable wormholes". 
See the books on Lorentzian wormholes by Visser \cite{Visser:1995cc} and the one on Wormholes and Warpdrives (in modified gravity) by Lobo in \cite{Lobo:2017cay}. In recent years, using the NFW dark matter profiles, the existence of wormhole geometries in the galactic halos has been proposed \cite{Rahaman:2013xoa,Rahaman:2015wpx}. Though all of these developments are interesting in their respective themes of investigation, in this paper, we stick to uncovering the original motivations of the ER paper, which is not only about Einstein field equations, but it is also about GR and QM. Thus, we focus on deriving new (quantum) realizations of the ER bridges by formulating a unitary QFTCS (See Sec.~\ref{sec:ERDQFT} and Sec.~\ref{sec:QBH} in particular and also Fig.~\ref{fig:ERB}).} 

\subsubsection{{Occurrence of IHO in BH physics:}}

Here, we would like to highlight how the physics of IHO is most relevant for the subject of QFTCS in the BH spacetime. We illustrate this by the massless scalar field example, which is also an important tool used in the original work on Hawking radiation \cite{Hawking:1975vcx}. Due to the spherical symmetry of Schwarzschild spacetime, one can expand the massless scalar field in spherical harmonics as 
\begin{equation}
    \phi\LF U, V,\,\theta,\, \varphi \RF = \sum \frac{\Phi_{\ell m}\LF U,\,V \RF}{r}Y_{\ell m}\LF \theta,\, \varphi \RF
\end{equation}
Thus, the massless Klein-Gordon field action in near horizon approximation along with neglecting the effective mass term for a sufficiently large BH becomes 
\begin{equation}
    S^{2D}_{KG} = \int dU dV \Phi_{\ell m}\LF -\pd_U\pd_V \RF \Phi_{\ell m}
\end{equation}
The solutions for the field $\Phi_{\ell m}$ for the case $U<0,\,V>0$ are 
\begin{equation}
\Phi_{\ell m}(U, V) = A_\omega \, (-U)^{i\omega/\kappa} + B_\omega \, V^{-i\omega/\kappa} 
\label{solsUV}
\end{equation}
where $\kappa = \frac{1}{4GM}$. 
It is easy to check that \eqref{solsUV} are the eigenfunctions of the Killing vector of the Schwarzschild manifold 
\begin{equation}
    K = \pd_t= \frac{1}{4GM} \LF V\frac{\pd}{\pd V}-U\frac{\pd}{\pd U} \RF
    \label{boostUV}
\end{equation}
that generates time translations. 
We can immediately notice the similarity between the structure of the boost operator in Schwarzschild spacetime \eqref{boostUV} and the dilatation operator of BK IHO \eqref{dilatonicop}. Furthermore, in Black hole physics, the appearance of IHO  can be intuitively seen through the behavior of Kruskal coordinates $U,\, V$ \eqref{Kruskalcoord} as a function of $t$ (with $U$ decaying exponentially and $V$ growing exponentially with $t$), which scale similarly to $P,\, Q$ of IHO as a function of $t$ \eqref{sol2BK}. {It means any solution for $\Phi_{\ell m}\LF U,\,V \RF$ would have a similar scaling behavior as the phase space wavefunction (or Wigner function) of IHO. In other words, the KS manifold \eqref{KSmetr}, whose discrete symmetries can be explicitly seen through \eqref{eq:extint} and \eqref{TtomT}, is analogous to the phase space of IHO, which carries a similar set of discrete symmetries \eqref{discreteTPQ} and \eqref{BKPE}.} 
't Hooft has explicitly derived the gravitational backreaction effects between {\it in} going state (at position $V_{in}$ around $r\lesssim 2GM$) and {\it out} going state (at position $U_{out}$ around $r\gtrsim 2GM$)
 near the horizon of SBH and applied first quantization, which yielded \cite{tHooft:2015pce,tHooft:2016rrl} (See Section 6.1 in \cite{Kumar:2023hbj} for a relevant derivation that is more apt to the context of physics we discuss in this paper) 
 \begin{equation}
 	\LT \hat V^{in}_{\ell m},\, \hat U^{out}_{\ell m} \RT = i\hbar \frac{8\pi G}{r_S^2 \LF \ell^2+\ell+1 \RF}
 	\label{uvcom}
 \end{equation}
where $r_S=2GM$ is called the Schwarzschild radius. The above result is obtained from the GR equations of motion with the partial wave expansion 
 \begin{equation}
 	U_{out} = 4GM \sum U^{out}_{\ell m} Y^\ell_m\LF \theta,\, \varphi \RF,\quad  V_{in} = 4GM \sum V^{in}_{\ell m} Y^\ell_m \LF \theta,\, \varphi \RF
 \end{equation}
 where $Y^\ell_m$'s are the spherical harmonics. From \eqref{uvcom} and \eqref{Kruskalcoord}, one can deduce that the following Hamiltonian, which is analogous to IHO, describes the quantum effects in the Black hole horizon (See \cite{Betzios:2020wcv,Ullinger:2022xmv} for more details)
 \begin{equation}
 	\hat H_{BH} = \frac{\hbar \omega_{BH} }{2}\LF \hat U^{out}_{\ell m}\hat V^{in}_{\ell m} + \hat V^{in}_{\ell m}\hat U^{out}_{\ell m} \RF
 	\label{BHIho}
 \end{equation}
 The Eq.~\eqref{BHIho} establish the connection between IHO and quantum effects in gravity and the need for understanding ER bridges.

\subsection{QFTCS in de Sitter spacetime}

After 20 years, Schr\"{o}dinger in 1956 encountered a similar conundrum in the context of "Expanding Universes" in de Sitter (dS) spacetime \cite{Schrodinger1956}. We can understand this by the following dS metric in the flat Friedmann-Lema\^itre-Robertson-Walker (FLRW) coordinates 
\begin{equation}
    ds^2 = -dt^2+a^2(t)d\textbf{x}^2=\frac{1}{H^2\tau^2}\LF -d\tau^2+d\textbf{x}^2 \RF,\quad a(t)=e^{Ht}=-\frac{1}{H\tau},\quad R_{dS}=12H^2\,. 
    \label{FLRWdS}
\end{equation}
where $R_{dS}$ is the curvature scalar of de Sitter space and $\tau= \int \frac{dt}{a}$ being the conformal time and $H=\frac{\dot a}{a}$ is the Hubble parameter. One thumb rule in physics is to make use of symmetries. The origin of physical conundrums often occurs, throwing away any symmetries by hand. Similar to SBH, dS spacetime, too \eqref{FLRWdS}, describes the physical world with two possible arrows of time given by 
\begin{equation}
	{\rm Expansion\,\, of \,\,Universe}  \implies \begin{cases}
	 H>0,\quad t: -\infty \to \infty \implies \tau<0 \\ 
	 H<0,\quad t: \infty \to -\infty \implies \tau>0
	\end{cases}
 \label{tdSsym}
\end{equation}
In the same manner, we can have the contracting Universe (decreasing scale factor), which can also be formulated with two arrows of time 
\begin{equation}
	{\rm Contracting\,\, \,\,Universe}  \implies \begin{cases}
	 H<0,\quad t: -\infty \to \infty \implies \tau>0 \\ 
	 H>0,\quad t: \infty \to -\infty \implies \tau<0
	\end{cases}
 \label{contrtdSsym}
\end{equation}
The dS spacetime \eqref{FLRWdS} is invariant under the discrete symmetry $\Pc\Tc$ 
\begin{equation}
   \Tc: \tau\to -\tau \implies t\to -t,\,H\to -H,\quad \Pc: \textbf{x}\to -\textbf{x}
   \label{disdS}
\end{equation}
Since de Sitter is a maximally symmetric spacetime, one can use it to build various topologies. de Sitter spacetime in closed FLRW coordinates is expressed with scale factor $a(t)=\cosh\LF Ht \RF$ and the positive spatial curvature $k = \vert H\vert >0$ as
\begin{equation}
\begin{aligned}
	ds^2 & = -dt^2 + \cosh^2\LF Ht \RF\LT \frac{dr^2}{1-H^2r^2}+ r^2d\Omega^2 \RT  \\ 
    & = -dt^2+ \frac{1}{H^2}\cosh^2\LF Ht \RF \LT d\chi^2+ \sin^2\chi d\Omega^2 \RT
    \end{aligned}
	\label{closeddS}
\end{equation}
where $r = \frac{1}{H}\sin\chi\in \LT 0,\, \frac{1}{H}\RF$. 
The Ricci scalar for the above metric is $R=12H^2$ where $H$ here is the asymptotic (constant) value of the Hubble parameter as $t\to \pm \infty$ suggested by
\begin{equation}
	H_{\rm closed} = \frac{\dot{a}}{a} = H\tanh\LF H t  \RF\,,\quad R= 12H_{\rm closed}^2+6\dot{H}_{\rm closed}+\frac{6H^2}{a^2}= 12H^2\,. 
	\end{equation}
With the definition of conformal time $\tau$
\begin{equation}
 t(\tau) = \frac{1}{H} \cosh^{-1} \left( \frac{1}{\cos(H \tau)} \right)   
\end{equation}
The metric \eqref{closeddS} becomes
\begin{equation}
    ds^2 = \frac{1}{\cos^2(H \tau)} \left(
  -d\tau^2 + d\chi^2 + \sin^2 \chi d\Omega^2
\right)
\end{equation}
In this picture, the Universe evolves from contraction to expansion (bounce), which is compatible with two arrows of time. Such a bounce can arise from a closed FLRW cloud evolving from contraction to expansion through a degenerate quantum ground state of constant energy density (see \cite{Gaztanaga:2025cun}).
\begin{equation}
	{\rm Contraction\,to\,Expansion} \implies \begin{cases}
		t: -\infty \to +\infty, \quad H>0,\quad \LF \tau: -\frac{\pi}{2\vert H\vert } \to \frac{\pi}{2\vert H\vert } \RF \\ 
		t: +\infty \to -\infty, \quad H<0,\quad \LF \tau: \frac{\pi}{2\vert H\vert } \to -\frac{\pi}{2\vert H\vert } \RF
	\end{cases}
	\label{clldS}
\end{equation}
{It is worth noting that the two arrows of time \eqref{clldS} of a bouncing Universe \cite{Gaztanaga:2025cun} fit nicely with James Hartle's proposal \cite{Hartle:2013tm} of the Universe emerging from a time-symmetric quantum phase, which is a concept well-rooted in the no boundary proposal in quantum cosmology \cite{Hartle:1983ai,Hartle:2007gi}. }

The dS metric 
\eqref{FLRWdS} in the static coordinates $\LF t_s,\,r \RF$ can be expressed as  \cite{Griffiths:2009dfa}
 \begin{equation}
\begin{aligned}
    ds^2 & = -\LF 1-H^2r^2 \RF dt_s^2 + \frac{1}{\LF 1-H^2r^2 \RF}dr^2 + r^2d\Omega^2 \\ 
    &  = \frac{1}{H^2\LF 1-\Tilde{\Uc}\Tilde{\Vc} \RF^2} \LF -4d\Tilde{\Uc}d\Tilde{\Vc}+ \LF 1+\Tilde{\Uc}\Tilde{\Vc} \RF^2 d\Omega^2 \RF
    \end{aligned}
    \label{statdS}
\end{equation}
where 
\begin{equation}
    \tilde{\Uc}(r, t_s) = \pm e^{-H t_s} \sqrt{ \frac{1 - H r}{1 + H r} },\quad 
\tilde{\Vc}(r, t_s) = \pm e^{H t_s} \sqrt{ \frac{1 - H r}{1 + H r} }
\label{UVdS}
\end{equation}
All of this indicates that we can identify the scale factor as the classical or thermodynamic clock, which can be compatible with two arrows of coordinate time. We will learn later that coordinate time acts as the quantum mechanical time parameter.   

{In the standard description of QFTCS in de Sitter spacetime, one often chooses $\tau<0$ (or Poincaré patch \cite{Hartman:2017}) (for \eqref{FLRWdS}), which breaks the discrete symmetry \eqref{disdS} by a strict assumption on the arrow of time.  In literature, \cite{Bousso:2002fq,Hartman:2020khs,Shaghoulian:2021cef,Shaghoulian:2022fop,Balasubramanian:2001rb,Balasubramanian:2021wgd,Colas:2024xjy,Brandenberger:2021pzy,Brandenberger:2022pqo}, $\tau<0$ and $\tau>0$ choices are usually considered as 
causally disconnected entangled Universes where unitarity (pure states evolving into pure states) is lost for an observer living in either of the Universes. This approach has led to the construction of so-called thermofield double states in the dual Minkowski and dual FLRW spacetimes \cite{Hartman:2020khs}, which share close similarities with Maldacena-Susskind's proposal of ER=EPR \cite{Maldacena:2013xja} in the context of AdS BHs that we shall discuss later. }

\subsubsection{Occurrence of IHO physics in de Sitter spacetime:}

The action for the massless KG field
\begin{equation}
    \phi = \frac{1}{\cos(H\tau)}\Phi\LF \tau,\,r,\,\theta,\,\varphi \RF
\end{equation}
in closed de Sitter can be expressed as
\begin{equation}
    S_{KG}^{\rm dS} = \frac{1}{2} \int d\tau \, d^3x \, \sqrt{\gamma} \left[
(\partial_\tau \Phi)^2 - \gamma^{ij} \partial_i \Phi \, \partial_j \Phi -(-\mu_{eff}^2) \, \Phi^2
\right]
\label{csdSiho}
\end{equation}
where $\gamma_{ij}$ is metric correspinding to $d\Sigma_3^2 = d\chi^2+ \sin^2\chi d\Omega^2$. 
Comparing \eqref{csdSiho} with the structure of \eqref{tachyonac}, we can deduce easily that massless scalar fields in de Sitter (closed FLRW) are tachyons with an effective time-dependent mass 
\begin{equation}
 \mu_{eff}^2= \frac{a''}{a}
=H^2\!\left(2\sec^2(H\tau)-1\right)
=H^2\!\left(1+2\tan^2(H\tau)\right)
>0 
\end{equation}
Extending this to a massive KG field, we will get  
\begin{equation}
\begin{aligned}
\mu_{\mathrm{eff}}^2(\tau)
=\frac{a''}{a}-m^2 a^2
=H^2\!\left(1+2\tan^2(H\tau)\right)-m^2 a^2
\end{aligned}
\end{equation}
which is a tachyonic in behavior for large $H$ (high curvature) and $a\ll 1$ (in the units of current value of the scale factor set as $a_0=1$), which is the scenario for the early Universe. 
We can do a similar exercise for the massive KG field $\phi = -\frac{1}{H\tau}\Phi$ for de Sitter in flat FLRW coordinates \eqref{FLRWdS}, which gives us 
\begin{equation}
    S_{KG}^{\rm fdS} = \frac{1}{2} \int d\tau \, d^3x \left[ (\partial_\tau \Phi)^2 - (\nabla \Phi)^2 -(- \tilde{\mu}_{\text{eff}}^2(\tau))\, \Phi^2 \right]
    \label{flatdSKG}
\end{equation}
with 
\begin{equation}
   \tilde{\mu}_{\text{eff}}^2(\tau)  
= \frac{1}{\tau^2} \left( 2-\frac{m^2}{H^2}  \right)
\end{equation}
From which we can deduce 
\begin{equation}
    m^2\ll H^2 \implies  \tilde{\mu}_{\text{eff}}^2 \thickapprox \frac{2}{\tau^2} >0\,.
\end{equation}
Again, we can compare \eqref{flatdSKG} with \eqref{tachyonac} and deduce that the massless fields and sub-Hubble $m^2\ll H^2$ massive fields have the tachyonic instability with the mass square term, which is negative compared to the positive mass square term of the standard KG field. This is the same as IHO instability discussed in Sec.~\ref{sec:IHO}. 

Also in the context of static de Sitter spacetime, the null coordinates \eqref{UVdS} do have a scaling behavior with time $t_s$ similar to the Schwarzschild case \eqref{Kruskalcoord}. Thus, a quantum field in static coordinates of de Sitter also involves the IHO physics. Furthermore, similar to the Schwarzschild BH case, the static de Sitter too has a killing vector 
\begin{equation}
    K = \pd_{t_s} = H\LF \tilde \Vc\frac{\pd}{\pd \tilde\Vc}-\tilde\Uc\frac{\pd}{\pd \tilde\Uc} \RF
\end{equation}
which does indicate a lack of a unique global arrow of time. 

\subsection{QFTCS versus Rindler spacetime} 

Rindler spacetime elegantly mimic the spacetime horizons in curved spacetime. This is why QFT in Rindler spacetime that leads to Unruh radiation is analoguous to the Hawking radiation in SBH \cite{Mukhanov:2007zz}. The essence of Rindler spacetime can be understood by starting with 1+1 dimensional Minkowski spacetime
\begin{equation}
    ds^2 = -dt^2+ dz^2
    \label{min2}
\end{equation}
The Rindler spacetime can be realized as
\begin{equation}
\begin{aligned}
 z^2-t^2 & = \frac{1}{a^2} e^{2a\xi} \implies \begin{cases}
     z  = \frac{1}{a} e^{a\xi} \cosh{a t_R},\quad  t= \frac{1}{a} e^{a\xi} \sinh{a t_R} \quad \LF \textrm{Right Rindler} \RF \\   z  = -\frac{1}{a} e^{a\xi} \cosh{a t_R},\quad  t= \frac{1}{a} e^{a\xi}\sinh{at_R} \quad \LF \textrm{Left Rindler} \RF
 \end{cases} \\ &\implies \boxed{ds^2 = e^{2a\xi}\LF -dt_R^2+d\xi^2\RF} \\   
  t^2-z^2 & = \frac{1}{a^2} e^{2a t_R} \implies \begin{cases}
     t  = \frac{1}{a} e^{a t_R} \cosh{a\xi},\quad  z= \frac{1}{a} e^{a t_R} \sinh{a\xi}\quad \LF \textrm{Future Kasner} \RF  \\  t  = -\frac{1}{a} e^{a t_R} \cosh{a\xi},\quad  z= \frac{1}{a} e^{a t_R} \sinh{a\xi}\quad \LF \textrm{Past Kasner} \RF 
 \end{cases} \\ & \implies \boxed{ds^2 = e^{2a t_R}\LF -dt_R^2+d\xi^2\RF}
    \end{aligned}
    \label{Robserco}
\end{equation}
We can express the whole Rindler spacetime $ds^2 = -dU_RdV_R$
in a coordinate system defined by 
\begin{equation}
    \begin{aligned}
        U_R &= -\frac{1}{a}e^{-au}<0,\quad &&V_R= \frac{1}{a}e^{av}>0\quad &&(\rm Right\, Rindler)\\
        U_R&= \frac{1}{a}e^{-au}>0,\quad &&V_R= -\frac{1}{a}e^{av}<0\quad &&(\rm Left\, Rindler) \\
         U_R&= \frac{1}{a}e^{-au}>0,\quad &&V_R= \frac{1}{a}e^{av}>0\quad &&(\rm Future\, Kasner) \\
          U_R&= -\frac{1}{a}e^{-au}<0,\quad &&V_R= -\frac{1}{a}e^{av}<0\quad &&(\rm Past\, Kasner)
    \end{aligned}
    \label{UVcoord}
\end{equation}
where 
\begin{equation}
\begin{aligned}
    u&= t_R-\xi,\quad v= t_R+\xi \\ 
    U_R&= t-z,\quad V_R=t+z
    \end{aligned}
\end{equation}
We can visually see the structure of Rindler spacetime in Fig.~\ref{fig:Rindler}.

\subsubsection{IHO analogy with Rindler spacetime}

\begin{figure}
    \centering
    \includegraphics[width=0.5\linewidth]{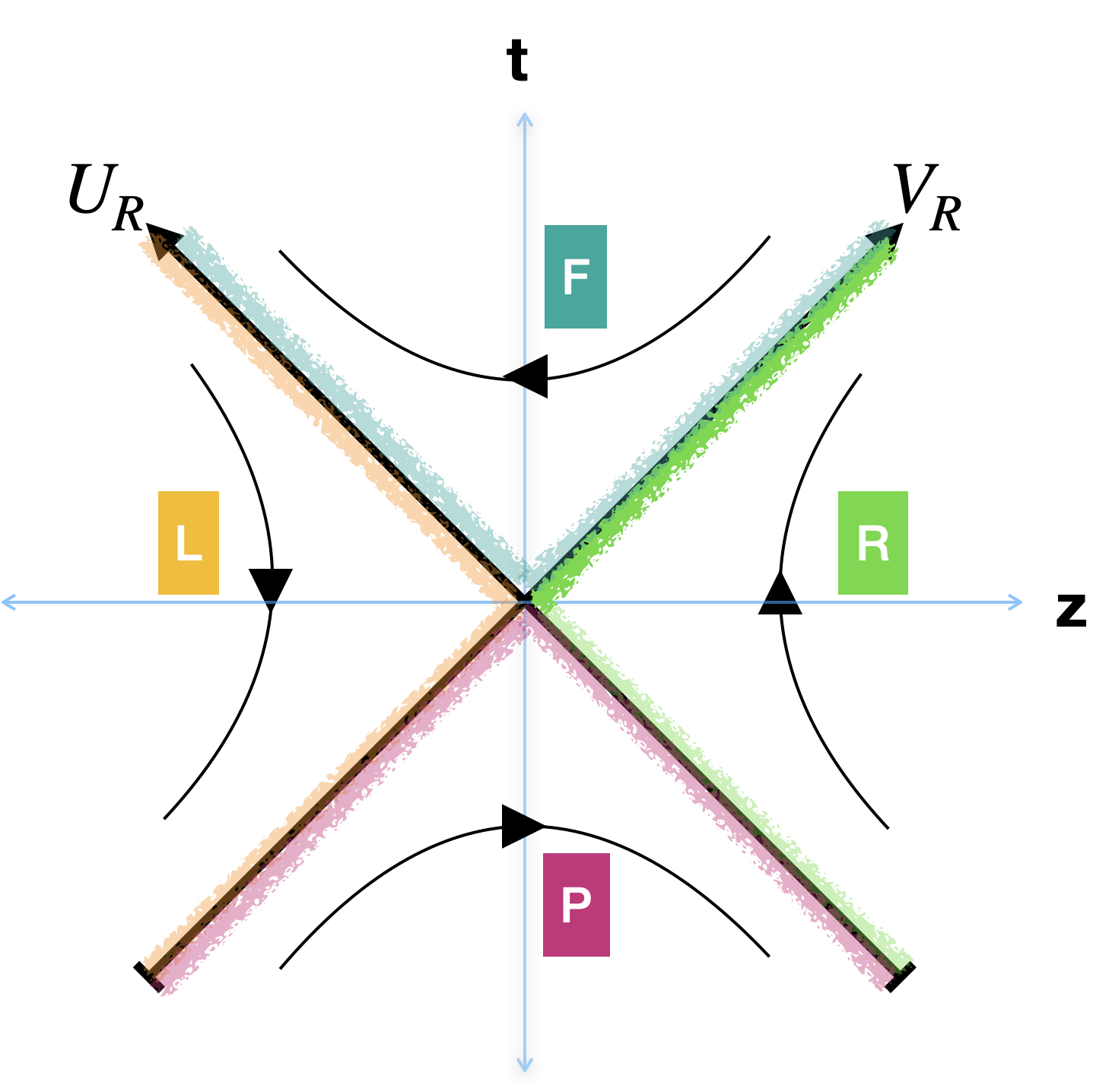}
    \caption{The figure represents the Left, Right ($z^2\gtrsim t^2$) and Future, Past ($t^2\gtrsim z^2$) regions of Rindler spacetime. The curved lines in the Left and Right regions are constant acceleration $ae^{-a\xi}$ curves with arrows of time $t_R: \infty \to -\infty$ (Left) and $t_R: -\infty \to \infty$ (Right). Future and Past Rindler with arrows indicate changing $z: \pm\infty \to \mp\infty$, $t_R: \mp\infty \to \pm\infty$. The Fuzzy colored lines indicate the Rindler Horizons for Left (Yellow), Right (Green), Future (Cyan), and Past (Pink).}
    \label{fig:Rindler}
\end{figure}

IHO physics in Rindler spacetime is very appealing. As we have seen with the above steps, coordinate redefinitions of 1+1D Miknowski metric create 4 regions of flat spacetime (See Fig.~\ref{fig:Rindler}) by the discrete transformations on the null coordinates
\begin{equation}
    U_R = \pm \frac{1}{a}e^{a\xi-at_R},\quad V_R= \pm \frac{1}{a}e^{a\xi+at_R}
    \label{URVR}
\end{equation}
Juxtaposing \eqref{URVR} with \eqref{eq:KS}, \eqref{UVdS}, we can observe the similarities associated with exponential scaling of null coordinates with parametric time. Thus, a quantum field in Rindler spacetime is very closely related to the phase space wave function of IHO, which we shall discuss in more detail in the later sections. 

\subsection{Inflationary (quasi-de Sitter) quantum fluctuations}

Inflationary spacetime is, by definition, a quasi-dS in character \cite{Starobinsky:1980te}. Thus, the spacetime during inflation would depart slightly from the metric \eqref{FLRWdS} by the non-zero slow-roll parameters 
\begin{equation}
    \epsilon = \frac{d}{d t}\LF \frac{1}{H} \RF = -\frac{\dot H}{H^2}\,,\quad \eta = \frac{\dot\epsilon}{H\epsilon}
    \label{epsiloneta}
\end{equation}
that gauge the adiabatically evolving Hubble parameter $H=\frac{\dot a}{a}$. Inflationary cosmology requires an additional new (scalar) degree of freedom that must be added to GR, which can be done either by a scalar field or modification of gravity (Starobinsky's $R+R^2$ inflation, for example). 
The metric fluctuations ($g_{\mu\nu}=\bar{g}_{\mu\nu}+\delta g_{\mu\nu}$) and the scalar field fluctuations ($\phi = \bar{\phi}+\delta\phi$) around the background dynamics in flat FLRW spacetime ($\bar g_{\mu\nu},\,\bar\phi$) lead to primordial seeds of temperature fluctuations and polarization features in the CMB \cite{Baumann:2018muz}.
 These fluctuations can be realized by a metric of the form 
\begin{equation}\label{ADMmetric}
		ds^2 =  a^2\LF \tau \RF \Big(-\LF 1+2\Phi \RF d\tau^2+  \LT \LF 1-2\Psi \RF \delta_{ij}+ h_{ij}\RT dx^idx^j\Big)\,.
	\end{equation}
	where $ \Phi$ and $\Psi$ coincide with the Bardeen potentials in Newtonian or Longitudinal gauge. $h_{ij}$ is the transverse and traceless spin-2 fluctuation. Using the perturbed linearized Einstein equations ($ij$ and $0i$ components), we get the following constraints 
    \begin{equation}
		\begin{aligned}
			\Phi=\Psi,\quad \dot{\Psi} + H\Phi = \frac{1}{2} \, \dot{\bar{\phi}} \, \delta\phi
		\end{aligned}
	\end{equation}
using which we get the second-order perturbed action for metric scalar fluctuations during inflation as 
\begin{equation}\label{scalar}
		\delta^{(2)}S_{s} = \frac{1}{2}\int d\tau d^3x a^2\frac{\dot{\bar\phi}^{2}}{H^2} \Bigg[ \zeta^{\prime 2} -\LF \pd\zeta \RF^2 \Bigg]
		\,\text{,}
	\end{equation}
    where $\zeta = \Psi+\frac{\dot{\bar\phi}}{H}\delta\phi$ is the curvature perturbation \cite{Mukhanov:2005sc}. 
We write the above action in the canonical form as 
\begin{equation}
	\delta^{(2)}S_{s} = \frac{1}{2}\int d\tau d^3x\,  V_{MS} \LF -\pd_\tau^2+\pd_i^2 -(-\mu_{eff}^2) \RF V_{MS},\quad \mu_{eff}^2 =  \frac{1}{\tau^2}\LF 2+2\epsilon+\eta \RF
	\label{ihoinf}
\end{equation}
where $V_{MS} = a\frac{\dot{\bar\phi}}{H}\zeta$ is called the Mukhanov-Sasaki variable, which effectively represents a scalar field with time-dependent negative mass square term since during inflation $\epsilon\ll 1, \, \eta\ll 1$. This time dependent mass of $V_{MS}$ is a manifestation of gravity causing quasi-dS expansion.
Thus, inflationary quantum (scalar) fluctuation in terms of the Mukhanov-Sasaki variable ($V_{MS}$) is related to IHO physics which was first discussed in \cite{Albrecht:1992kf}. 
  Similarly, graviton fluctuations during inflation are also described by inverted harmonic oscillators as we can deduce from the following second order action 
\begin{equation}
	\delta^{(2)}S_{h} = \frac{1}{2}\int d\tau d^3x\,  u_{h} \LF -\pd_\tau^2+\pd_i^2 -(-\mu_{eff}^2) \RF u_{h},\quad \mu_{eff}^2 =  \frac{1}{\tau^2}\LF 2+2\epsilon \RF
	\label{ihoinfhij}
\end{equation}
where the tensor fluctuation $h_{ij}$ expanded in terms of the polarization tensor $\polarizationtensor^{s}_{ij}$ as
\begin{equation}
	h_{ij} = \sum_{s= \times,\,+} \polarizationtensor^{s}_{ij} u_{h}
    \label{hijexp}
\end{equation}

\subsection{Unitarity of standard QFTCS: echoing ideas from Einstein-Rosen, Schr\"{o}dinger and 't Hooft}

In all the spacetime manifolds we discussed in the previous sections, which are BH, Rindler, de Sitter, and inflationary spacetimes, we witnessed identical degenerate regions of spacetimes just related by discrete transformations. Historically, QFT in curved spacetime \cite{Birrell:1982ix} has always been carried out by choosing one of the regions as physical to the real world description. {For example, in the context of BH spacetime, Hawking's assumption \cite{Hawking:1975vcx} is only the regions $U<0,\,V>0$ and $U>0,\,V>0$ are considered to be physical. In the context of dS and inflationary spacetimes, the common practice is to choose $\tau<0$ for an expanding Universe \cite{Mukhanov:2007zz}. This procedure not only breaks the symmetries of the manifolds but also the unitarity and observer complementarity.    }

Schrödinger demanded that there cannot be two expanding Universes; there should only be one Universe, which is similar to ER, who demanded one physical world. There cannot be two exteriors to SBH. Analogous to ER bridge Schr\"{o}dinger proposed the so-called antipodal identification (i.e., to identify $\LF \tau,\,\textbf{x} \RF$ and $\LF -\tau,\,-\textbf{x} \RF$ to represent a single physical event), often called the Elliptic interpretation of de Sitter space. {To be more precise, Schrödinger insists the two realizations of expanding Universe in dS spacetime \eqref{tdSsym} or \eqref{clldS} should be seen as one Universe, not two. According to this, the past light cone of one observer has to be identified with the future light cone of another observer at the antipodal point.} 
Schr\"{o}dinger's conjecture is:

\begin{center}
\it	Every event in dS has to be described by thin, rigid rods connecting the antipodal ($\Pc\Tc$ conjugate) points in spacetime 
\end{center}
We notice the one-to-one correspondence between ER bridges and Schr\"{o}dinger's rods. After 60 years, following the seminal works of Norma G. Sanch\'ez and Whiting \cite{Sanchez:1986qn}, Gerard 't Hooft, too, arrived at a similar idea in the context of SBH \cite{tHooft:2016qoo} i.e., to identify $\LF U,\,V \RF$ and $\LF -U, -V\RF$ together with parity conjugate points $\LF \theta,\varphi\RF$ and $\LF \pi-\theta,\,\pi+\varphi \RF$. {In the recent formulations, 't Hooft called the parallel region outside ($r>2GM$), which is obtained by $U\to -U,\, V\to -V$, as "hidden" region of SBH and interpreted that these regions contain quantum "clones" \cite{tHooft:2024auh}. This is analogous to Israel's thermofield double states \cite{Israel:1976ur} formalism, which augments the physical Fock space states with "fictitious" dual Fock space states related by the same discrete transformation. }

All these developments spanned over 90 years, and have a common goal of achieving unitary quantum physics in curved spacetime defined by:
\begin{center}
	\it An imaginary observer bounded by a gravitational horizon has to witness pure states evolving into pure states. 
\end{center}
Another concept called observer complementarity is tied to the unitarity definition above, which requires different observers in curved spacetime to share complementary information in the form of pure states. This leads to information reconstruction beyond the spacetime horizons that the observer cannot causally access. Both unitarity and observer complementarity are the essential requirements for QFTCS and quantum gravity. 

{Unitarity is broken in standard QFTCS because an entangled state across the spacetime horizon (both inside and outside regions) is realized through the vacuum structure of the curved spacetime 
\begin{equation}
    \vert 0\rangle_{H} = \sum_{n=0}^\infty e^{-\frac{\beta \omega n}{2}} \, |n\rangle_{\text{inside}} \otimes |n\rangle_{\text{outside}},\quad |n\rangle_{\text{inside}} \otimes |n\rangle_{\text{out}} \in \Hc_{\rm global}= \mathcal{H}_{\text{inside}} \otimes \mathcal{H}_{\text{outside}}
    \label{vacSTQFT}
\end{equation}
This leads to a local observer accessing only part of the entanglement (mixed state) and it is violation of unitarity \cite{Almheiri:2019hni,Brandenberger:2022pqo}. The emergence of \eqref{vacSTQFT} is deeply connected to the treatment of spacetime regions bounded by gravitational horizons as open quantum systems. This is known as the central dogma in cosmology and black hole physics \cite{Almheiri:2020cfm,Shaghoulian:2021cef}. 
Furthermore, unitarity in standard QFTCS is broken also because of time reversal symmetries in dS and BH spacetimes allow us to describe physical world with two arrows of time for both inside and outside the horizon. As we discussed earlier, this has led to the conundrum of entangled Universes separated by spacelike distances. 
}
All the investigations over the decades have admitted the inevitability of unitarity loss in curved spacetime unless a new physics is built from an unknown theory of quantum gravity at the Planck scales. Even then, the particle description problem initiated by ER remained unsolved. Thus, QFTCS remains one of the deepest problems in theoretical physics that impedes the progress in achieving quantum gravity \cite{Giddings:2022jda}. 

With all the significant developments in cosmology and astrophysics, both in theoretical and observational aspects, the importance of discovering the true nature of quantum fields in curved spacetime is the need of the hour. Every development of this subject, starting from Zel'dovich and Starobinsky's revelation of particle production in cosmological backgrounds \cite{Zeldovich:1977vgo}, Starobinsky's later formulation of cosmic inflation and the generation of quantum fluctuations \cite{Starobinsky:1980te,Sasaki:1986hm,Mukhanov:1988jd}, Hawking's BH radiation (that was followed from Starobinsky's work on Kerr BHs \cite{Starobinsky:1973aij}) \cite{Hawking:1975vcx} has pushed significantly the field of theoretical and observational physics. 

ER's proposal of the mathematical bridge later evolved into classical possibilities and interpretations of wormholes connecting different universes or space-like distances in a single universe with the need for exotic matter or modifications of gravity \cite{Misner:1957mt,Morris:1988cz,Morris:1988tu,Cramer:1994qj,Visser:1995cc}.  However, the paper of ER is majorly concerned with gravity and quantum mechanics in the sense of QFTCS in the vicinity of gravitational horizons. The exact realization of "a mathematical bridge" (quantum mechanically) to represent a physical Universe has been unclear over these decades. Our recent attempts in this direction show a promising outcome both from theoretical and observational points of view, which forms the crux of this paper.

\section{Direct-sum Quantum Theory and geometric superselection sectors} 

\label{sec:disumQT}

{The preceding discussion highlighted how Schrödinger’s antipodal identification, Einstein–Rosen’s insistence on a single physical universe, and 't Hooft’s reinterpretation of black hole horizons all converge on a deeper, unresolved tension in quantum field theory in curved spacetime: the apparent breakdown of unitarity when gravitational horizons are involved. These ideas suggest that what appear to be causally disconnected or duplicated regions, such as the two asymptotic exteriors of a Schwarzschild black hole or antipodal points in de Sitter space, may, in fact, represent different facets of a single, global quantum event. This observation motivates a radical departure from the traditional Hilbert space framework, where quantum states are localized to a single causal patch. Instead, we propose a direct-sum Quantum Theory, in which physical states are defined across parity and time conjugate regions, encoding complementary information accessible to distinct observers. This structure not only respects the observer complementarity principle but also opens a pathway to restore unitarity without invoking unknown Planck scale physics. In what follows, we develop the mathematical and conceptual foundation for this extended quantum framework.}

\subsection{Direct-sum quantum mechanics}

In the previous section, we discussed how SBH and dS spacetimes can allow a description of one physical world with two arrows of time. Similar realization occurs even with the Schr\"{o}dinger equation, which is an order differential equation in time
\begin{equation}
	i\frac{\partial \vert \Psi\rangle }{\partial t_p} = \hat{H}\vert \Psi \rangle = \Ec\vert\Psi\rangle,\quad t_{p}: -\infty \to \infty
	\label{eq:Sch1}
\end{equation}
where $\hat{H}$ here is assumed to be time-independent parity symmetric Hamiltonian for simplicity.
The Schr\"{o}dinger equation \eqref{eq:Sch1} sets the definition of positive energy state with a presumption on the arrow of time 
\begin{equation}
	\vert \Psi\rangle_{t_p} = e^{-i\Ec t_p}\vert \Psi\rangle_0,\quad \Ec>0,\quad t_p: -\infty \to \infty
\end{equation}
Suppose one assumes an opposite arrow of time; an equivalent definition of a positive energy state becomes
\begin{equation}
	\vert \Psi\rangle_{t_p} = e^{i\Ec t_p}\vert \Psi\rangle_0,\quad \Ec>0,\quad t_p: \infty \to -\infty
\end{equation}
This would emerge from the Schr\"{o}dinger equation with a sign change of the complex number, which is obvious because we reversed the arrow of time. 
\begin{equation}
	-i\frac{\partial \vert \Psi\rangle }{\partial t_p} = \hat{H} |\Psi \rangle = \Ec \vert\Psi\rangle,\quad t_p: \infty \to -\infty
\label{eq:Sch2}
\end{equation}
The entire QFT is built on the definition of a positive energy state. Thus, one must define an arrow of time before specifying the quantum theory. 
Thus, there is an ambiguity in fixing the arrow time, which is associated with whether to have "$+i$" or "$-i$" in the description of the Schr\"{o}dinger equation. Nature does not distinguish between "$+i$" and "$-i$"; Quantum theory (without gravity) is known to be time-symmetric. Thus, it does not matter what convention we use for the arrow of time; we would arrive at the same physics. This is what a recent work by J. Donoghue and G. Menezes shows \cite{Donoghue:2019ecz}, that is, the entire QFT can be reconstructed with $-i$ convention with opposite arrow time by replacing everywhere $+i$ with $-i$. 

This crucial observation is the basis for building a direct-sum quantum theory, which removes the requirement of defining an arrow of time to declare a positive energy state. We formulate here the description of a quantum state by (geometric) superselection rule  \cite{Kumar:2023ctp,Kumar:2023hbj,Kumar:2024oxf,Gaztanaga:2024vtr,Kumar:2024ahu} involving $\Pc\Tc$.

\begin{figure}
    \centering
    \includegraphics[width=0.4\linewidth]{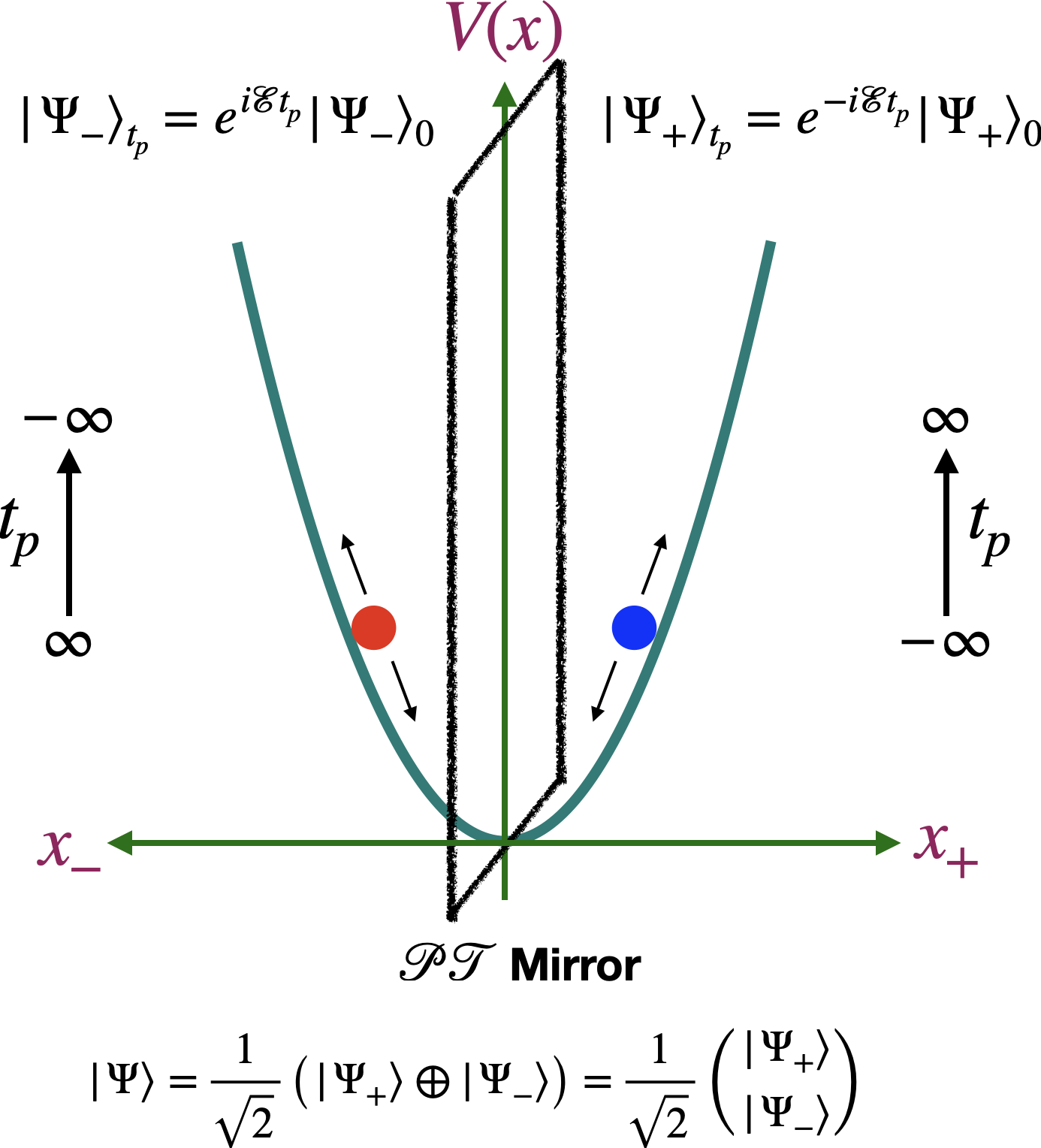}
    \caption{The picture depicts the new understanding of quantum harmonic oscillator in a direct-sum Hilbert space. Time is a parameter in quantum theory. In contrast, the spatial position is an operator. A quantum state here is described by a direct-sum of two components in parity conjugate points in physical space. }
    \label{fig:HO}
\end{figure}
Here, we formulate a quantum state as a direct-sum of two orthogonal components\footnote{Note that direct-sum operation is different from superposition.} $\vert \Psi_\pm\rangle$ 
\begin{equation}
\begin{aligned}
    \vert \Psi\rangle = \frac{1}{\sqrt{2}}\LF \vert \Psi_+\rangle \oplus \vert \Psi_-\rangle \RF = \frac{1}{\sqrt{2}}\begin{pmatrix}
        \vert \Psi_+\rangle \\ \vert \Psi_-\rangle
    \end{pmatrix}
    \end{aligned}
\end{equation}
that are positive energy states with opposite arrows of time at parity conjugate points in physical space governed by the direct-sum Schr\"{o}dinger equation \cite{Kumar:2023hbj}
\begin{equation}
    i\frac{\pd }{\pd t_p}\begin{pmatrix}
        \vert \Psi_+\rangle \\ \vert \Psi_-\rangle
    \end{pmatrix} = \begin{pmatrix}
        \hat{H}_+ & 0 \\ 
        0 & -\hat{H}_-
    \end{pmatrix}\begin{pmatrix}
        \vert \Psi_+\rangle \\ \vert \Psi_-\rangle
    \end{pmatrix}
    \label{schdisum}
\end{equation}
defined in a direct-sum Hilbert space $\Hc = \Hc_+ \oplus \Hc_-$. The Hilbert spaces $\Hc_\pm$ are called geometric superselection sectors (SSS) describing quantum states in the parity conjugate regions. Here $\hat H = \hat H_+\LF \hat x_+,\,\hat p_+ \RF \oplus \hat H_-\LF \hat x_-,\,\hat p_- \RF$ is the time-independent Hamiltonian of the physical system split as a direct sum of two. The position operator here becomes $\hat x = \frac{1}{\sqrt{2}}\LF \hat{x}_+\oplus \hat x_- \RF$ with eigenvalues being $x_+ = x\gtrsim 0$ and $x_- = x\lesssim 0$. Similarly, the momentum operator becomes  $\hat p = \frac{1}{\sqrt{2}}\LF \hat{p}_+\oplus \hat p_- \RF$ with $\hat p_{\pm} = \mp i\frac{\pd}{\pd x_{\pm}}$. 
The \eqref{schdisum} can also be written as 
\begin{equation}
\begin{pmatrix}
        i\frac{\pd}{\pd_{t_p}}\vert \Psi_+\rangle \\ 
       -i \frac{\pd}{\pd_{t_p}} \vert \Psi_-\rangle \end{pmatrix}  = \begin{pmatrix}
	    \hat{H}_+ && 0 \\ 
        0 && \hat{H}_-
	\end{pmatrix} \begin{pmatrix}
	    \vert \Psi_+ \rangle \\ 
        \vert \Psi_-\rangle
	\end{pmatrix}
    \label{schdisum1}
\end{equation}
The $-$ sign in \eqref{schdisum} or \eqref{schdisum1} indicates the state $\vert \Psi_-\rangle$ evolves with the opposite arrow of time $t_p:\infty \to -\infty$.
With direct-sum QM, we describe the wave function and the probabilities as 
\begin{equation}
\Psi(x) = \frac{1}{\sqrt{2}} \begin{pmatrix}
       \langle x_+\vert \quad \langle x_- \vert 
    \end{pmatrix}\begin{pmatrix}
        \,\,\,\,\vert \Psi_+\rangle_0 \,e^{-i\Ec t} \\
        \vert\Psi_-\rangle_0 \,e^{i\Ec t}  
    \end{pmatrix} \implies \begin{cases}
        \frac{1}{\sqrt{2}}\Psi_+\LF x_+ \RF e^{-i\Ec t},\quad x_+ = x\gtrsim 0 \\  \frac{1}{\sqrt{2}}\Psi_-\LF x_- \RF e^{i\Ec t},\quad x_- = x\lesssim 0\,.
    \end{cases}
    \label{eq:wavefunction}
\end{equation}
and
\begin{equation}
  \int_{-\infty}^\infty dx  \langle \Psi\vert \Psi \rangle =  \frac{1}{2} \int^{0}_{-\infty} dx_-  \langle \Psi_-\vert \Psi_- \rangle +\frac{1}{2} \int^{\infty}_0 dx_+  \langle \Psi_+\vert \Psi_+ \rangle   =1\,. 
\end{equation}
The canonical commutation relations are 
\begin{equation}
\LT \hat x_\pm,\, \hat p_{\pm}\RT = \pm i\,,\quad \LT \hat x_+,\, \hat x_- \RT = \LT \hat p_+,\, \hat p_- \RT = \LT \hat x_+,\, \hat p_- \RT = \LT \hat p_+,\, \hat x_- \RT =0
\end{equation}
We note that $\Pc\Tc$ operations remain the same irrespective of any coordinate translations. Thus, one can shift the origin $x=0$, but the direct-sum quantum theory is unaltered. Once we divide the quantum state by the above direct-sum operation into sectoral Hilbert space, we can still perform time reversal and parity operations individually in $\Hc_\pm$. The $\Tc$ operation in each SSS turns the positive energy states to negative ones and changes the direction of momenta, whereas the $\Pc$ operation changes only the direction of momenta. In Fig.~\ref{fig:HO} we illustrate how the quantum harmonic oscillator can be viewed in a new way with our direct-sum quantum theory.

\subsection{Direct-sum QFT in Minkowski spacetime}

\label{sec:Dtheory}

Minkowski spacetime $ds^2 = -dt_p^2+d\textbf{x}^2$  is $\Pc\Tc$ symmetric ($t_p\to -t_p$ and $\textbf{x}\to -\textbf{x}$). Thus, extending the first quantization approach by direct-sum Schr\"{o}dinger equation to the second quantization is straightforward, and we call it direct-sum quantum field theory (DQFT) \cite{Kumar:2023ctp,Kumar:2024oxf}. For example, the Klein-Gordon (KG) field operator now becomes a direct-sum of two components as a function of $\Pc\Tc$ conjugate points 
\begin{equation}
    \hat \phi(x)  = \frac{1}{\sqrt{2}}\begin{pmatrix}
        \hat \phi_+ & 0 \\ 
        0 & \hat \phi_-
    \end{pmatrix}
    \label{eq:qfdisum}
\end{equation}
where 
\begin{equation}
    \begin{aligned}
    \hat \phi_{+}( x) & =     \int \frac{d^3k}{\LF 2\pi \RF^{3/2}}\frac{1}{\sqrt{2\vert k_0\vert }}\LT a_{(+)\textbf{k}}e^{ik\cdot  x}+a_{(+)\textbf{k}}^\dagger e^{-ik\cdot x}   \RT \\ 
    \hat \phi_-(- x) & =    \int \frac{d^3k}{\LF 2\pi \RF^{3/2}}\frac{1}{\sqrt{2\vert k_0\vert }}\LT a_{(-)\textbf{k}}e^{-ik\cdot x}+a_{(-)\textbf{k}}^\dagger e^{ik\cdot  x}\RT 
    \end{aligned}
    \label{eq:fppm}
\end{equation}
where $k\cdot x =-k_0 t_p+\textbf{k}\cdot \textbf{x}$ and
the creation and annihilation operators obey 
\begin{equation}
    \LT a_{(\pm)\textbf{k}},\, a^\dagger_{(\pm)\textbf{k}}\RT = 1,\quad  \LT a_{(\pm)\textbf{k}},\, a^\dagger_{(\mp)\textbf{k}}\RT= \LT a_{(\pm)\textbf{k}},\, a_{(\mp)\textbf{k}}\RT =0\,.
    \label{commre}
\end{equation}
This gives a new causality condition
\begin{equation}
	\LT \hat \phi_+(x),\, \hat \phi_-(-y) \RT =0\,.
\end{equation}
 along with the standard condition, which demands the operators to commute for space-like distances 
 \begin{equation}
	\LT \hat{\phi}_{\pm}(x),\,\hat {\phi}_{\pm}(y) \RT=0,\quad \LF x-y \RF^2>0.  
\end{equation}
Note that the $\hat \phi_{\pm}$ are field operators exclusively defined for parity conjugate points in physical space with positive energy states defined with opposite arrows of time. The direct-sum of these two operators results in the description of the quantum field \eqref{eq:qfdisum} everywhere in Minkowski spacetime. The construction here is based on $\Pc\Tc$ and any Lorentz transformations and translations on \eqref{eq:qfdisum} preserve $\Pc\Tc$ symmetric feature of DQFT Minkowski vacuum\footnote{Positive energy state in vacuum $\vert 0_{M+}\rangle$ is $\vert \phi_k\rangle = e^{-i\Ec t}\vert\phi_k\rangle_0$ whereas in $\vert 0_{M-}\rangle$ it is $\vert \phi_k\rangle = e^{i\Ec t}\vert\phi_k\rangle_0$. Here in this notation $k_0=\Ec$ and $k=\vert \textbf{k}\vert$.} 
\begin{equation}
    \vert 0 \rangle_M = \begin{pmatrix}
        \vert 0_{+}\rangle_M \\ 
        \vert 0_{-}\rangle_M 
    \end{pmatrix},\quad a_{(+)\textbf{k}}\vert 0_{+}\rangle_M =0,\quad a_{(-)\textbf{k}}\vert 0_{-}\rangle_M =0\,. 
    \label{minvac}
\end{equation}
Correspondingly, the Fock space of DQFT is a direct-sum of geometric superselection sectors (SSS) $\Fc = \Fc_+\oplus \Fc_-$ describing quantum states in parity conjugate regions of Minkowski space\footnote{In spatial 3D, parity is a discrete transformation totally different from rotation. In spherical coordinates, parity operation takes a point at a radial distance $r$ to its antipode i.e., $\LF \theta,\,\varphi \RF\to \LF \pi-\theta,\,\pi+\varphi \RF$ which can never be achieved by rotations.}. The two-point function in DQFT is given by
\begin{equation}
\begin{aligned}
    \langle 0\vert \hat{\phi}\LF x \RF  \hat{\phi}\LF x^\prime \RF \vert 0\rangle & = \frac{1}{2} {}_{M}\langle 0_+\vert \hat{\phi}_+\LF x \RF  \hat{\phi}_+\LF x^\prime \RF \vert 0_+\rangle_M + \frac{1}{2} {}_M\langle 0_-\vert \hat{\phi}_-\LF -x \RF  \hat{\phi}_-\LF -x^\prime \RF \vert 0_- \rangle_{M}
    \end{aligned}
    \label{twopointf}
\end{equation}
A similar structure is followed for the propagator, which is a time-ordered product of two field operators. {Thus, the propagator of a quantum field between any two points in Minkowski becomes the sum of two terms, each describing the field propagation in parity conjugate regions of physical space.}  
In DQFT, all the interactions are divided into direct-sum; for example, a cubic interaction would look like
\begin{equation}
  \frac{\lambda}{3} \hat{\phi}^3 = \frac{\lambda}{3} \begin{pmatrix}
      \hat{\phi}_+^3 & 0 \\ 
      0 & \hat{\phi}_-^3
  \end{pmatrix}
\end{equation}
This means we will never have any mixing between $\hat{\phi}_+$ and $\hat{\phi}_-$. As a consequence, all the standard QFT calculations extended to DQFT give the same results, which is obvious because of $\Pc\Tc$ symmetry of Minkowski spacetime (See \cite{Kumar:2023ctp,Kumar:2024oxf} for more details). {Furthermore, it is important to note that any entangled state in DQFT splits according to the direct-sum vacuum structure \eqref{minvac}. Because of the $\Pc\Tc$ based geometric SSS, the Reeh-Schlieder theorem that reveals the entanglement properties of quantum fields in Minkowski  \cite{Witten:2018zxz} can be applied in DQFT separately to each geometric vacuum $\vert 0_\pm\rangle_M$ \cite{Kumar:2024oxf}. This elegant construction would become extremely useful to retain unitarity in Rindler as well as various curved spacetimes, which we show in the later sections.} According to DQFT, the standard model degrees of freedom, such as particles ($\vert SM\rangle$) and antiparticles ($\vert \overline{SM}\rangle$ get represented according to the direct-sum split of the SM vacuum. 
\begin{equation}
    \vert 0_{SM}\rangle = \begin{pmatrix}
        \vert 0_{SM+}\rangle \\ 
        \vert 0_{SM-}\rangle 
    \end{pmatrix} \quad \vert SM\rangle = \frac{1}{\sqrt{2}}\begin{pmatrix}
        \vert SM_+\rangle \\ 
        \vert SM_-\rangle \end{pmatrix} \quad \vert \overline{SM}\rangle = \frac{1}{\sqrt{2}}\begin{pmatrix}
        \vert \overline{SM}_+\rangle \\ 
        \vert \overline{SM}_-\rangle 
    \end{pmatrix}
\end{equation}
{Note that the geometric superselection rule is the same for all Fock spaces of the SM degrees of freedom, i.e., the parity conjugate regions are uniquely defined for all states of the SM. {We provided DQFT quantization of a real scalar field, but construction is very straightforward for the complex scalar, fermion, and gauge fields. Every quantum field is written as direct-sum of two components, which are $\Pc\Tc$ mirror images of each other spanning the entire Minkowski spacetime. Thus, we can easily extend the standard quantization \cite{Buchbinder:2021wzv} to DQFT as follows:
\begin{itemize}
    \item Complex scalar field operator $\hat{\phi}_c$ in DQFT is expanded as 
   \begin{equation}
    \hat{\phi}_c = \frac{1}{\sqrt{2}}\LF \hat{\phi}_{c\,+}\oplus  \hat{\phi}_{c\,-}  \RF,\quad \hat{\phi}_{c\,\pm} = \int  \frac{d^3k}{\LF 2\pi \RF^{3/2}}\frac{1}{\sqrt{2\vert k_0\vert }}\Bigg[ a_{(\pm)\textbf{k}}e^{\pm ik\cdot  x}+b_{(\pm)\textbf{k}}^\dagger e^{\mp ik\cdot x}   \Bigg],
   \end{equation}
   where $\LT \hat{\phi}_{c\,+},\, \hat{\phi}_{c\,-} \RT=0$, 
   $a_{(\pm)\textbf{k}},\, a^\dagger_{(\pm)\textbf{k}}$ and $b_{(\pm)\textbf{k}},\, b^\dagger_{(\pm)\textbf{k}}$ are canonical creation and annihilation operators of the parity conjugate regions (denoted by subscripts $_{(\pm)}$) attached with geometric SSS.
   All the cross commutation relations of $a_{(\pm)},\, a^\dagger_{(\pm)}$ and $b_{(\pm)},\, b^\dagger_{(\pm)}$ vanish.  
   \item Fermionic field operator in DQFT becomes 
   \begin{equation}
   \begin{aligned}
  \hat \psi & = \frac{1}{\sqrt{2}}\LF \hat \psi_+\oplus \hat \psi_- \RF \\
     \hat  \psi_{\pm} & = \sum_{{\tilde s}} \int \frac{d^3k}{\LF 2\pi \RF^{3/2}\sqrt{2\vert k_0\vert}} \Bigg[ c_{{\tilde s}(\pm)\textbf{k}} u_{\tilde s}(\textbf{k}) e^{\pm ik\cdot x} + d_{{\tilde s}(\pm)\textbf{k}}^\dagger v_{\tilde s}(\textbf{k}) e^{\mp ik\cdot x}\Bigg]
     \end{aligned}
   \end{equation}
where ${\tilde s}=1,2$ correspond to the two independent solutions of $\LF \slashed k+m\RF u_{\tilde{s}}=0$ and $\LF -\slashed k+m\RF v_{\tilde{s}}=0$ corresponding to spin-$\pm\frac{1}{2}$. The creation and annihilation operators of geometric SSS of Fock space here satisfy the anti-commutation relations $\Big\{ c_{\tilde{s}(\pm)\textbf{k}},\,c_{\tilde{s}(\pm)\textbf{k}}^\dagger \Big\}=1,\, \Big\{ c_{\tilde{s}(\mp)\textbf{k}},\,c_{\tilde{s}(\pm)\textbf{k}}^\dagger \Big\}=\Big\{ c_{\tilde{s}(\mp)\textbf{k}},\,c_{\tilde{s}(\pm)\textbf{k}} \Big\}=0$ leading to the new causality condition $\Big\{ \hat\psi_+,\,\hat \psi_-\Big\} =0$.
\item The vector field operator in DQFT expressed as 
\begin{equation}
	\hat A_\mu = \frac{1}{\sqrt{2}}\LF \hat{A}_{+\mu}\oplus \hat A_{-\mu} \RF,\quad \hat{A}_{\pm \mu}= \int \frac{d^3k}{\LF 2\pi \RF^{3/2}\sqrt{2\vert k_0\vert }} e^{(\lambda)}_\mu\Bigg[ c_{(\pm \lambda)\textbf{k}} e^{\pm ik\cdot x}+c^\dagger_{(\pm \lambda)\textbf{k}} e^{\mp ik\cdot x}  \Bigg]
	\end{equation} 
	where $e^{(\lambda)}_\mu$ is the polarization vector satisfying the transverse and traceless conditions. The creation and annihilation operators $c_{(\pm \lambda)\textbf{k}},\,c_{(\pm \lambda)\textbf{k}}^\dagger$ satisfy the similar relations as \eqref{commre}.
\end{itemize} }

All the SM calculations remain the same because all the interaction terms are split into direct-sum in the following way. 
\begin{equation}
   \Lc_c \sim\Oc_{SM}^3=\begin{pmatrix}
        \Oc_{SM_+}^3 & 0 \\ 
        0 & \Oc_{SM_-}^3
    \end{pmatrix} \quad \Lc_q \sim \Oc_{SM}^4 = \begin{pmatrix}
        \Oc_{SM_+}^4 & 0 \\ 
        0 & \Oc_{SM_-}^4
    \end{pmatrix}
\end{equation}
Here, $\Oc_{SM}$ is an arbitrary operator involving any SM fields and their derivatives. \footnote {Remember that any derivative operators must be split into components joined by a direct-sum operation.}  }Evidently, the DQFT framework does not alter the QFT calculations in Minkowski due to the spacetime being $\Pc\Tc$ symmetric. 
If we compute any scattering amplitude, say, N particles to M particles, the DQFT gives 
\begin{equation}
    A_{N\to M} = \frac{A^{N\to M}_+ \LF p_a, -p_b\RF  + A^{N\to M}_-\LF -p_a, p_b \RF}{2},\quad A^{N\to M}_+ \LF p_a, -p_b\RF = A^{N\to M}_-\LF -p_a, p_b \RF,
\end{equation}
where $p_a,\,p_b$ with $a=1,\cdots N$ and $b=1,\cdots M$ represent the 4-momenta of all the states involved in the scattering. 
$A_{\pm}$ represents amplitudes as a function of 4-momenta of initial and final states
computed in both vacuums $\vert 0_{SM\pm}\rangle$. Notice that the in (out) states in $\vert 0_{SM\pm}\rangle$ come with the opposite sign, which is due to the arrow of time being opposite in both vacuums. The amplitudes $A_{\pm}$ are equal at any order in perturbation theory due to the $\Pc\Tc$ symmetry of Minkowski spacetime. 
The famous $\CPT$ (charge conjugation $\mathbb{C}$, Parity $\Pb$, and Time reversal $\Tb$) invariance of scattering amplitudes \cite{Coleman:2018mew} also holds in both vacuums, which means
\begin{equation}
    A^{N\to M}_+(p_a, -p_b) = A^{M\to N}_+(-p_a, p_b) , \quad A^{N\to M}_-(-p_a, p_b) = A^{M\to N}_-(p_a, -p_b) \,.
\end{equation}
This is attributed to the fact that the $\CPT$ operation of any scattering process would turn the outgoing anti-particles into in-going particles and vice-versa \cite{Coleman:2018mew}. {In the standard description of QFT, though an arrow of time is assumed (as discussed in Sec.~\ref{sec:QFTmin}), $\CPT$ invariance holds in all particle interactions. The discrete operations here (in the contexts of $\CPT$ invariant scattering processes or $\Cb\Pb$ violating decays in weak interactions) are defined in terms of momentum space (see the discussion in chapter 11, from page 225 of \cite{Coleman:2018mew}). To be precise, the charge conjugation operation $\Cb$ conjugates the charge, the parity $\Pb$ operation changes the 3-momenta $\textbf{k}\to -\textbf{k}$, the $\Tb$ operation interchange the ingoing state as an outgoing state and vice-versa along with $\textbf{k}\to -\textbf{k}$.  }

{In the framework of DQFT, once we express the quantum field operator as direct-sum of components in geometric superselection sectors \eqref{eq:qfdisum} (based on $\Pc\Tc$), we can apply the discrete transformations such as the parity $\Pb$ in each SSS in the following manner 
\begin{equation}
\begin{aligned}
\Pb\hat\phi \Pb^{-1} & = \Pb\hat \phi_+ \Pb^{-1} \oplus \Pb\hat \phi_- \Pb^{-1} \\ 
    \Pb\hat \phi_+ \Pb^{-1} & = \int \frac{d^3k}{\LF 2\pi \RF^{3/2}}\frac{1}{\sqrt{2\vert k_0\vert }}\LT a_{(+)-\textbf{k}}e^{ik\cdot  x}+a_{(+)-\textbf{k}}^\dagger e^{-ik\cdot x}   \RT  \\ 
 \Pb\hat \phi_- \Pb^{-1}    & =  \int \frac{d^3k}{\LF 2\pi \RF^{3/2}}\frac{1}{\sqrt{2\vert k_0\vert }}\LT a_{(-)-\textbf{k}}e^{-ik\cdot x}+a_{(-)-\textbf{k}}^\dagger e^{ik\cdot  x}\RT 
    \end{aligned}
    \label{PQFT}
\end{equation}
where we applied the effect of parity $\Pb$ operation that changes 3-momenta $\textbf{k}\to -\textbf{k}$ and position $\textbf{x}\to -\textbf{x}$. Notice in particular that the creation and annihilation operators in \eqref{PQFT} correspond to $-\textbf{k}$. Similarly, the time reversal operation $T$ in each geometric SSS can be deduced as 
\begin{equation}
\begin{aligned}
\Tb\hat\phi \Tb^{-1} & = \Tb\hat \phi_+ \Tb^{-1} \oplus \Tb\hat \phi_- \Tb^{-1} \\
    \Tb\hat \phi_+ \Tb^{-1} & = \int \frac{d^3k}{\LF 2\pi \RF^{3/2}}\frac{1}{\sqrt{2\vert k_0\vert }}\LT a_{(+)-\textbf{k}}e^{-ik\cdot  x}+a_{(+)-\textbf{k}}^\dagger e^{ik\cdot x}   \RT  \\ 
 \Tb\hat \phi_- \Tb^{-1}    & =  \int \frac{d^3k}{\LF 2\pi \RF^{3/2}}\frac{1}{\sqrt{2\vert k_0\vert }}\LT a_{(-)-\textbf{k}}e^{ik\cdot x}+a_{(-)-\textbf{k}}^\dagger e^{-ik\cdot  x}\RT 
    \end{aligned}
\end{equation}
where we can notice the change $i\to -i$ in the mode functions due to the anti-unitary character of time reflection. Also the 3-momenta changes as $\textbf{k}\to -\textbf{k}$. These are $\Pb\Tb$ operations in QFT, which reflect in changes in Fourier modes and coefficients (the creation-annihilation operators). 
\begin{equation}
\begin{aligned}
(\Pb\Tb)\hat\phi (\Pb\Tb)^{-1} & = (\Pb\Tb)\hat \phi_+ (\Pb\Tb)^{-1} \oplus (\Pb\Tb)\hat \phi_- (\Pb\Tb)^{-1} \\
    (\Pb\Tb)\hat \phi_+ (\Pb\Tb)^{-1} & = \int \frac{d^3k}{\LF 2\pi \RF^{3/2}}\frac{1}{\sqrt{2\vert k_0\vert }}\LT a_{(+)\textbf{k}}e^{-ik\cdot  x}+a_{(+)\textbf{k}}^\dagger e^{ik\cdot x}   \RT  \\ 
 (\Pb\Tb)\hat \phi_- (\Pb\Tb)^{-1}    & =  \int \frac{d^3k}{\LF 2\pi \RF^{3/2}}\frac{1}{\sqrt{2\vert k_0\vert }}\LT a_{(-)\textbf{k}}e^{ik\cdot x}+a_{(-)\textbf{k}}^\dagger e^{-ik\cdot  x}\RT 
    \end{aligned}
\end{equation}
The $\CPT$ invariance, $\Cb\Pb$ violations of SM are associated with implications of discrete spacetime transformations in Fourier space, which differ from the $\Pc\Tc$ operations in DQFT, which are tied to spacetime (geometric) symmetry of the Minkowski manifold and the geometric superselection rules.  To make this distinction apparent, we use different notation for $\Pc,\,\Tc$ (geometric aspect) and the $\Cb,\,\Pb,\,\Tb$ (on scattering states). Furthermore, the meaning of these operations in the form of vacuum structure can be understood through the following steps 
\begin{equation}
\begin{aligned}
 \vert 0_+\rangle & =   \LF \Pc\Tc \RF \vert 0_-\rangle \LF \Pc\Tc \RF^{-1} \implies a_{+,\,\textbf{k}}\to a_{-,\,\textbf{k}} \\ 
  \vert 0_{+}\rangle & = \LF \Pb\Tb \RF \vert 0_{+}\rangle  \LF \Pb\Tb \RF^{-1} \implies  a_{+,\,\textbf{k}}\to a_{+,\,\textbf{k}} \\
  \vert 0_{-}\rangle & = \LF \Pb\Tb \RF \vert 0_{-}\rangle  \LF \Pb\Tb \RF^{-1} \implies  a_{-,\,\textbf{k}}\to a_{-,\,\textbf{k}}
 \end{aligned}
\end{equation}
Therefore, the $\CPT$ invariance and $\Cb\Pb$ violations in weak interactions of the standard model remain the same in the DQFT picture. 
The standard model vacuum in DQFT is $\CPT$ invariant as 
\begin{equation}
    {\LF \CPT\RF }\vert 0_{\rm SM}\rangle {\LF \CPT\RF}^{-1} =  \LF {\CPT}\RF \vert 0_{\rm SM_+}\rangle \LF{\CPT}\RF^{-1} \oplus \LF {\CPT}\RF \vert 0_{\rm SM-}\rangle {\LF \CPT\RF}^{-1} = \vert 0_{\rm SM}\rangle 
\end{equation}.
} 
\begin{figure}[ht!] 
	\centering
	\includegraphics[width=0.7\linewidth]{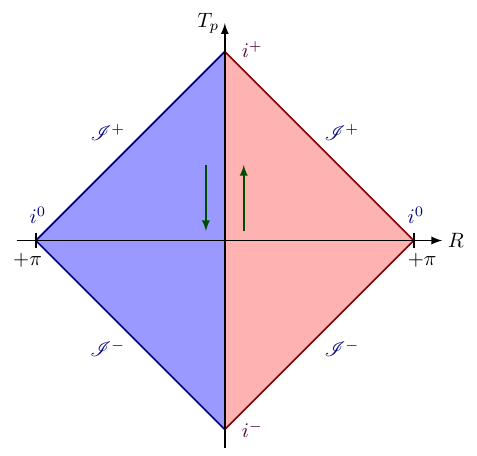}
	\caption{
	The DQFT representation of Minkowski spacetime in terms of compactified coordinates $T_p   = \arctan\LF t_p+r \RF + \arctan\LF t_p-r \RF$, $T_p\in \LF -\pi+R\to \pi+R \RF$ and 
		$R = 	\arctan\LF t_p+r \RF -\arctan\LF t_p-r \RF $, $R\in \LF 0,\,\pi \RF$ where $r$ is being the radial coordinate. The left and right triangles are $\Pc\Tc$ conjugates of each other. A quantum field operator in DQFT is a direct-sum of two components corresponding to parity-conjugate regions of physical space, with positive energy states defined with opposite arrows of time.  }
	\label{fig:minkowski-np}
\end{figure}

In summary, we presented a new understanding of quantum (field) theory with a direct-sum (mathematical bridge) between $\Pc\Tc$ conjugate sheets of spacetime. Fig.~\ref{fig:minkowski-np} depicts the conformal diagram of DQFT in Minkowski spacetime. 
Using the geometric superselection rules formulated by parity conjugate regions of physical space, we have successfully incorporated two arrows of time in a single quantum state description. 
DQFT does not change the practical results in SM particle physics, but it gives a new feature of understanding the role of "time" in quantum theory.  
We will witness in the next sections that this structure will lead us to a novelty of building the connection between gravity and quantum mechanics, and tackle the new challenges associated with problems like IHO. 

\subsubsection{The concept of geometric superselection sectors (SSS):}  

{We employ this concept for quantization of IHO and
in different contexts of QFTCS in the upcoming sections. Here, we clarify further the meaning of geometric SSS in the abstract mathematical form.  }

{Let $\Hc$ be the total Hilbert space of the quantum theory to describe the evolution of a quantum state in a spacetime with disjoint regions related by discrete spacetime transformations. Then, we define a decomposition:}
{
\begin{equation}
    \mathcal{H} = \bigoplus_{i} \Hc_i,
\end{equation}
where each $\Hc_i$ is a geometric superselection sector corresponding to a 
region related to others via a discrete transformation.
A generic state $\vert\Psi\rangle \in \mathcal{H}$ is expressed as:
\begin{equation}
    \vert\Psi\rangle = \bigoplus_{i} \vert\Psi_i\rangle, \qquad \vert\Psi_i\rangle \in \mathcal{H}_i,
\end{equation}
and operators $\mathcal{O}$ act block-diagonally:
\begin{equation}
    \mathcal{O} = \bigoplus_i \mathcal{O}_i, \qquad \mathcal{O}_i : \Hc_i \rightarrow \Hc_i.
\end{equation}}
{This structure imposes a geometric superselection rule: transitions between sectors $\Hc_i$ and $\Hc_j$ ($i \neq j$) are forbidden by symmetry, as no local operator connects them.}
{The direct-sum structure provides a mathematical realization of ER bridges, interpreting entanglement across spacetime regions with opposing arrows of time not via geometric wormholes, but through correlated states in this extended Hilbert space. This notion of geometric SSS is straightforwardly applicable to Fock spaces as well. We discuss this part in detail in the later sections. }

\subsection{Direct-sum quantization of Berry and Keating IHO}

\label{sec:BKDQ}

This section aims to elucidate how we can build a new construction of quantum IHO with direct-sum operation, which echoes consistently with the absolutely crucial observations made by Berry and Keating, B. Aneva \cite{Berry1999,Aneva:1999fy}. As discussed in Sec.~\ref{sec:IHO} and as we can see in Fig.~\ref{fig:IHOPS}, the regions of phase space are related by the following discrete group of transformations, which form the dihedral group\cite{Aneva:1999fy} $D_4$ of order $8$
\begin{equation}
\begin{aligned}
T_1^\pm: Q^\prime & = - \frac{h}{Q},\quad P^\prime = \pm\frac{PQ^2}{h} \\ 
    T_2^\pm: Q^\prime & = - \frac{h}{P},\quad P^\prime = \mp\frac{QP^2}{h} 
    \end{aligned},
\end{equation}
along with $-T_1^\pm$ and $-T_2^\pm$ {that include transformations in \eqref{discreteTPQ} and \eqref{BKPE}}. These transformations include dilatations and preserve the quantization conditions $Q\geq \ell_Q$, and $P\geq \ell_P$ discussed in Sec.~\ref{sec:IHO}.
Due to the presence of regions related by discrete operations, we formulate quantum theory with geometric superselection sectors in the phase space. This observation was also recently made in a work on generalized Born oscillators, which contains BK IHO as a special case \cite{Giordano:2023wgx}. The relation between IHO Hamiltonian \eqref{IHOhamil} and the Weyl reflected Laplace-Beltrami operator \eqref{LBOp} motivates us to write positive ($E>0\equiv +E$) and negative energy ($E<0\equiv -E$) quantum states as direct-sum of two components corresponding to direct-sum Hilbert space ($\Hc_{iho}$)
\begin{equation}
    \vert \Psi_{iho}\rangle  = \LF\big\vert \Psi^{(+E)}_{iho}\big\rangle \oplus  \big\vert \Psi^{(-E)}_{iho} \big\rangle \RF = \begin{pmatrix}
         \big\vert \Psi^{(+E)}_{iho}\big\rangle \\  
         \big\vert \Psi^{(-E)}_{iho}\big\rangle
    \end{pmatrix},\quad {\Hc}_{iho}= {\Hc}^{(+E)}_{iho}\oplus {\Hc}^{(-E)}_{iho}
    \label{PNEIHO}
\end{equation}
The doubly degenerate trajectories in phase space imply a further direct-sum split of the above states into the respective components as 
\begin{equation}
\begin{aligned}
 \big\vert \Psi^{(-E)}_{iho}\big\rangle &= \frac{1}{\sqrt{2}}\LF \big\vert \Psi^{(-E)}_{I}\big\rangle \oplus  \big\vert \Psi^{(-E)}_{II} \big\rangle\RF,\quad {\Hc}^{(-E)} = {\Hc}^{(-E)}_I\oplus {\Hc}^{(-E)}_{II}\\ 
    \big\vert \Psi^{(+E)}_{iho}\big\rangle &= \frac{1}{\sqrt{2}}\LF \big\vert \Psi^{(+E)}_{III}\big\rangle \oplus  \big\vert \Psi^{(+E)}_{IV} \big\rangle\RF,\quad {\Hc}^{(+E)} = {\Hc}^{(+E)}_{III}\oplus {\Hc}^{(+E)}_{IV}
    \end{aligned}
    \label{ERBIHO}
\end{equation}
The rules of direct-sum quantum theory rely on dividing the physical space by parity operation, which we do separately for all the regions of the phase space. In particular, the regions III and IV of the phase space individually contain parity conjugate regions ($\pm\tilde{q}$) (See Fig.~\ref{fig:IHOPS}). But in contrast, the regions $I$ and $II$ together cover $\pm\tilde{q}$. 

 We split the position and momentum operators of the entire phase space as 
\begin{equation}
\begin{aligned}
    \hat{Q} & = \frac{1}{\sqrt{2}} \LF\hat{Q}_{(+E)}\oplus \hat{Q}_{(-E)}\RF,\quad \hat{Q}_{(+E)} & = \frac{1}{\sqrt{2}} \LF\hat{Q}_{III}\oplus \hat{Q}_{IV}\RF,\quad \hat{Q}_{(-E)} & = \frac{1}{\sqrt{2}} \LF\hat{Q}_{I}\oplus \hat{Q}_{II}\RF  \\
  \hat{P} & = \frac{1}{\sqrt{2}} \LF\hat{P}_{(+E)}\oplus \hat{P}_{(-E)}\RF,\quad \hat{P}_{(+E)} & = \frac{1}{\sqrt{2}} \LF\hat{P}_{III}\oplus \hat{P}_{IV}\RF,\quad \hat{P}_{(-E)} & = \frac{1}{\sqrt{2}} \LF\hat{P}_{I}\oplus \hat{P}_{II}\RF 
    \end{aligned}
    \label{posmomiho}
\end{equation}
with the only non-zero commutation relations being
\begin{equation}
\begin{aligned}
  \LT \hat Q_{III},\,\hat P_{III} \RT &= i\hbar,\quad    \LT \hat Q_{IV},\,\hat P_{IV} \RT = -i\hbar \\
    \LT \hat Q_{I},\,\hat P_{I} \RT &= i\hbar,\quad    \LT \hat Q_{II},\,\hat P_{II} \RT = -i\hbar
    \end{aligned}
\end{equation}
which can be realized with four commuting sets of creation and annihilation operators. With \eqref{posmomiho} the Hamiltonian of IHO \eqref{IHOhamil} split into direct-sum of four components describing the four regions of the phase space in Fig.~\ref{fig:IHOPS}.  Finally, the quantization of IHO is governed by the direct-sum Schr\"{o}dinger equation of the following
\begin{equation}
    i\hbar \frac{\pd}{\pd t}\begin{pmatrix}
         \big\vert \Psi^{(+E)}_{iho}\big\rangle \\  
         \big\vert \Psi^{(-E)}_{iho}\big\rangle
    \end{pmatrix} = \begin{pmatrix}
        \hat H^{(+E)}_{iho} && 0 \\
        0 && \hat H^{(-E)}_{iho} 
    \end{pmatrix}\begin{pmatrix}
         \big\vert \Psi^{(+E)}_{iho}\big\rangle \\  
         \big\vert \Psi^{(-E)}_{iho}\big\rangle
    \end{pmatrix}
    \label{di-sumiho}
\end{equation}
which describes the evolution of positive and negative energy quantum states. Due to the doubly degenerate regions $I$ ($III$) and $II$ ($IV$) related by \eqref{discreteTPQ} in the separatrix phase space (Fig.~\ref{fig:IHOPS}), the states $\vert \Psi_{iho}^{\pm E}\rangle $ would then be governed by
\begin{equation}
    \begin{aligned}
       i\hbar \frac{\pd}{\pd t}\begin{pmatrix}
         \big\vert \Psi^{(+E)}_{III}\big\rangle \\  
         \big\vert \Psi^{(+E)}_{IV}\big\rangle
    \end{pmatrix} & = \begin{pmatrix}
        \hat H^{(+E)}_{III} && 0 \\
        0 && -\hat H^{(+E)}_{IV} 
    \end{pmatrix}\begin{pmatrix}
         \big\vert \Psi^{(+E)}_{III}\big\rangle \\  
         \big\vert \Psi^{(+E)}_{IV}\big\rangle
    \end{pmatrix},\quad  \hat{H}_{iso}^{(+E)}  = \hat{H}_{III}^{(+E)}\oplus  \hat{H}_{IV}^{(+E)}  \\ i\hbar \frac{\pd}{\pd t}\begin{pmatrix}
         \big\vert \Psi^{(-E)}_{I}\big\rangle \\  
         \big\vert \Psi^{(-E)}_{II}\big\rangle
    \end{pmatrix} & = \begin{pmatrix}
        \hat H^{(-E)}_{iho} && 0 \\
        0 && -\hat H^{(-E)}_{iho} 
    \end{pmatrix}\begin{pmatrix}
         \big\vert \Psi^{(-E)}_{I}\big\rangle \\  
         \big\vert \Psi^{(-E)}_{II}\big\rangle
    \end{pmatrix},\quad \hat{H}_{iso}^{(-E)}  = \hat{H}_{I}^{(-E)}\oplus \hat{H}_{II}^{(-E)}.
    \end{aligned}
\end{equation}
where the Hamiltonians correspond to each region are functions of corresponding position and momentum operators \eqref{posmomiho}.
Since the positive and negative energy regions are related by the transformations \eqref{BKPE},  the position and momentum wave functions in the region $I$ ($II$) and $III$ ($IV$) swap with each other (i.e., Fourier transform in the region $I$ ($II$) becomes inverse Fourier transform in the region $III$ ($IV$) and vice versa). Working out \eqref{di-sumiho} we obtain (in the units of setting $\omega=1$)
\begin{equation}
\begin{aligned}
        Q_{I}^{1/2}\zeta\LF \frac{1}{2}-\frac{i\vert E\vert }{\hbar} \RF\Psi_{(-E)}\LF Q_{I}\RF  + Q_{III}^{1/2}\zeta\LF \frac{1}{2}+\frac{i \vert E\vert }{\hbar} \RF \Psi_{(+E)}\LF Q_{III}\RF & =0\\
           Q_{II}^{1/2}\zeta\LF \frac{1}{2}-\frac{i\vert E\vert}{\hbar} \RF\Psi_{(-E)}\LF Q_{II}\RF + Q_{IV}^{1/2}\zeta\LF \frac{1}{2}+\frac{i\vert E\vert }{\hbar} \RF \Psi_{(+E)}\LF Q_{IV}\RF & =0\,, 
        \label{boundarycondi}
        \end{aligned}
    \end{equation}
    which generates the zeros of the Riemann zeta function $\zeta\LF \frac{1}{2}\pm i\bar T \RF$.  The above relations \eqref{boundarycondi}, though they seem similar to the BK's quantum boundary condition \eqref{boundarycondiBK}, there is a significant difference, which is "geometrical interpretation". In direct-sum quantization, using phase space geometric SSS, we obtain the geometric interpretation, which is a drawback in Berry and Keating's proposal \cite{Berry1999,Aneva:1999fy}. Direct-sum quantization, by splitting the full phase-space Hilbert space geometrically into SSS, would bring a resolution to the issue of quantum chaos in describing the quantum dynamics of IHO \cite{Berrychaos,Berry1999}. In short, what we achieved here is a description of the IHO quantum state (See \eqref{PNEIHO} and \eqref{ERBIHO}) by mathematical bridges (direct-sum) between various sheets of phase space with different arrows of time. 

\section{Quantum ER bridges and unitarity in Schwarschild, de Sitter, and Rindler spacetimes}

\label{sec:ERDQFT}

In this section, we delve deeper into the consequences of the direct-sum quantum theory for quantum field descriptions in curved spacetime, with particular attention to Rindler spacetime, which represents Minkowski spacetime from the perspective of a uniformly accelerated observer \cite{Kumar:2024oxf}. {The foundations of DQFT are extended to Rindler spacetime, and a renewed understanding of the Reeh-Schlieder theorem and entanglement structure in flat spacetime is derived in detail in \cite{Kumar:2024oxf}.}
Remarkably, Einstein and Rosen’s work also touches upon quantum physics in Rindler-like spacetimes, emphasizing the necessity of constructing "mathematical bridges" to unify seemingly disconnected regions \cite{Einstein:1935tc}. 

\subsection{The mathematical bridges in Rindler spacetime}

Just as phase space horizons in the IHO provide a foundation for geometric quantization with SSS, the Rindler horizons (See Fig.~\ref{fig:Rindler}) serve as the natural basis for defining DQFT in Rindler spacetime. A KG operator in Rindler spacetime is split into 4 components 
\begin{equation}
    \hat \phi\Big\vert_{\rm All\,Rindler} = \frac{1}{\sqrt{2}} \LF\hat\phi_L\oplus \hat \phi_R\rangle \RF \oplus \frac{1}{\sqrt{2}}\LF\hat\phi_F\oplus \hat \phi_P \RF
    \label{splitall}
\end{equation}
with respect to a direct-sum Fock space corresponding to a direct-sum vacuum defined by a commuting set of canonical creation and annihilation operators (details can be found in  Ref.~\cite{Kumar:2024oxf})
\begin{equation}
    \Fc_R = \LF \Fc_L\oplus \Fc_R\RF \oplus  \LF \Fc_F\oplus \Fc_P\RF, \quad \vert 0\rangle = \LF \vert 0\rangle_L\oplus \vert 0\rangle_R\RF \oplus  \LF \vert 0\rangle_F\oplus \vert 0\rangle_P\RF
\end{equation}
{In DQFT, we write the KG operator in Minkowski and Rindler spacetimes as 
\begin{equation}
\begin{aligned}
    \hat{\phi} & = \frac{1}{\sqrt{2}} \LF \hat{\phi}_+\oplus \hat{\phi}_-  \RF = \frac{1}{\sqrt{2}}\begin{pmatrix}
        \hat{\phi}_+ & 0 \\
        0 & \hat{\phi}_- 
    \end{pmatrix}\Bigg\vert_{z^2\gtrsim t^2\, {\rm Minkowski}} \\ &  =    \frac{1}{\sqrt{2}}\LF \hat{\phi}_R\oplus \hat{\phi}_L \RF = \frac{1}{\sqrt{2}} \begin{pmatrix}
        \hat{\phi}_{R} & 0 \\
        0 & \hat{\phi}_{L} 
    \end{pmatrix} \,,
    \label{eqLRM}
    \end{aligned}
\end{equation}
where the subscripts $L,\, R$ represent field operators expressed in the Left Rindler and the Right Rindler coordinates, respectively. Since the Future and Past regions of Rindler spacetime (See Fig.~\ref{fig:Rindler}) contain individually parity conjugate regions, we expand the scalar field operator in Minkowski \eqref{min2}, Future, and Past Rindler spacetimes as 
\begin{equation}
\begin{aligned}
    \hat{\phi} & = \frac{1}{\sqrt{2}} \LF \hat{\phi}_+\oplus \hat{\phi}_-  \RF = \frac{1}{\sqrt{2}}\begin{pmatrix}
        \hat{\phi}_+ & 0 \\
        0 & \hat{\phi}_- 
    \end{pmatrix}\Bigg\vert_{t\gtrsim \vert z\vert \, {\rm Minkowski}} \\ &  =   \frac{1}{\sqrt{2}} \LF \hat{\phi}_{F+}\oplus \hat{\phi}_{F-} \RF = \frac{1}{\sqrt{2}}\begin{pmatrix}
        \hat{\phi}_{F+} & 0 \\
        0 & \hat{\phi}_{F-} 
    \end{pmatrix} 
    \label{eqFP}
    \end{aligned}
\end{equation}
and 
\begin{equation}
\begin{aligned}
    \hat{\phi} & = \frac{1}{\sqrt{2}} \LF \hat{\phi}_+\oplus \hat{\phi}_-  \RF = \frac{1}{\sqrt{2}}\begin{pmatrix}
        \hat{\phi}_+ & 0 \\
        0 & \hat{\phi}_- 
    \end{pmatrix}\Bigg\vert_{t\lesssim -\vert z\vert \, {\rm Minkowski}} \\ &  =   \frac{1}{\sqrt{2}} \LF \hat{\phi}_{P+}\oplus \hat{\phi}_{P-} \RF = \frac{1}{\sqrt{2}}\begin{pmatrix}
        \hat{\phi}_{P+} & 0 \\
        0 & \hat{\phi}_{P-} 
    \end{pmatrix}
    \label{eqFP2}
    \end{aligned}
\end{equation}}
If we quantize a field in \eqref{min2}, according to DQFT, the Minkowski vacuum is a direct sum of two, as we split the quantum field into two components by parity and time reversal, similar to the 4-dimensional case \eqref{eq:qfdisum}. 
As a consequence of this construction, the Minkowski vacuum ($\vert 0_M\rangle$) looks like a pair of quantum states in Rindler spacetime that follow from the Bogoliubov transformations. For example, in the context of Left and Right Rindler regions, the Minkowski vacuum can be written as
\begin{equation}
    \vert 0_M\rangle = \begin{pmatrix}
        \vert 0_{M+}\rangle  \\ 
        \vert 0_{M-}\rangle 
    \end{pmatrix} =  \begin{pmatrix}
      \prod_{\textbf{p}} \frac{1}{\sqrt{\vert \alpha_{kp}^R}\vert} \exp\Bigg[-\LF\frac{\beta^R_{kp}}{2\alpha^{ R_{kp}}}\RF \hat{c}_{R\textbf{p}}^\dagger \hat{c}^\dagger_{R(-\textbf{p})}\Bigg] \vert 0_R\rangle \\
     \prod_{\textbf{p}} \frac{1}{\sqrt{\vert \alpha^L_{kp}}\vert} \exp\Bigg[-\LF\frac{\beta^L_{kp}}{2\alpha^{ L_{kp}}}\RF \hat{c}_{L\textbf{p}}^\dagger \hat{c}^\dagger_{L(-\textbf{p})}\Bigg] \vert 0_L\rangle
    \end{pmatrix}
    \label{LFvac}
\end{equation}
Here $\LF \alpha^R,\,\beta^R \RF$ and $\LF \alpha^L,\,\beta^L \RF$ Bogoliubov coefficients, $\LF c_{L\textbf{p}},\, c_{L\textbf{p}}^\dagger\RF$ and $\LF c_{R\textbf{p}},\, c_{R\textbf{p}}^\dagger\RF$ are creation and annihilation operators corresponding to the Right and Left regions of Rindler spacetime. Thus in this framework any maximally entangled state $\vert \psi_{p}\rangle$  (pure state) is split into two pure state components $\vert\psi_{pL}\rangle,\,\vert\psi_{pR}\rangle$ corresponding to the SSS Hilbert spaces of Left and Right Rindler regions ($\Hc_M=\Hc_L\oplus \Hc_R$).
\begin{equation}
    \vert \psi_{p}\rangle = \frac{1}{\sqrt{2}}\begin{pmatrix}
        \vert \psi_{pL}\rangle \\
        \vert \psi_{pR}\rangle
    \end{pmatrix}
    \label{pure}
\end{equation}
This means the density matrix of the pure state is split into direct-sum of two pure-state density matrices 
\begin{equation}
    \rho_{\psi_{p}} = \frac{1}{\sqrt{2}} \LF \rho_{\psi_{pL}}\oplus \rho_{\psi_{pR}}\RF
\end{equation}
The Von Neumann entropies of the Left and Right states $\LF \vert \Psi_{pL}\rangle,\, \vert \Psi_{pR}\rangle \RF$ vanish because the Left and Right are described by geometric SSS Hilbert spaces. Therefore,
\begin{equation}
    S_{L} = -Tr[\rho_{\psi_{pL}} \ln \rho_{\psi_{pL}}] =0,\quad S_{R} = -Tr[\rho_{\psi_{pR}} \ln \rho_{\psi_{pR}}] =0
    \label{VNE}
\end{equation}
The Von Neumann entropy for $\vert \Psi_P\rangle $ vanishes too since 
\begin{equation}
    S = S_L + S_R =0\,.
\end{equation}
This confirms the state $\vert \Psi_P\rangle$ is a globally pure state, whereas $\vert \Psi_{PL}\rangle$ and $\vert \Psi_{PR}\rangle$ are pure states of a local Rindler observer. 
Note that since the quantum theory on the Left and Right are constructed in SSS corresponding to causally separated spacetime regions, an observer on the Left cannot access any information on the right. However, the Left and Right regions are $\Pc\Tc$ conjugates of each other but separated by Rindler horizons. Though the Left observer cannot access the Right region causally, by observing the pure states $\vert \Psi_{PL}\rangle$, the observer can reconstruct the pure states of the Right region. Thus, both observers share complementary pieces of information in the form of pure states. In this respect, the Rindler horizon acts like a "$\Pc\Tc$ mirror". We can further extend this to the entire Rindler space, which includes both Future and Past, along with Left and Right. Following \eqref{splitall}, any maximally entangled pure state now becomes a direct-sum of 4 pure state components whose individual Von Neumann entropy vanishes; thus, we have a unitary description of QFT in Rindler spacetime.

\subsection{The mathematical bridges in quantum black hole}

\label{sec:QBH}

\begin{figure}
	\centering
	\includegraphics[width=0.7\linewidth]{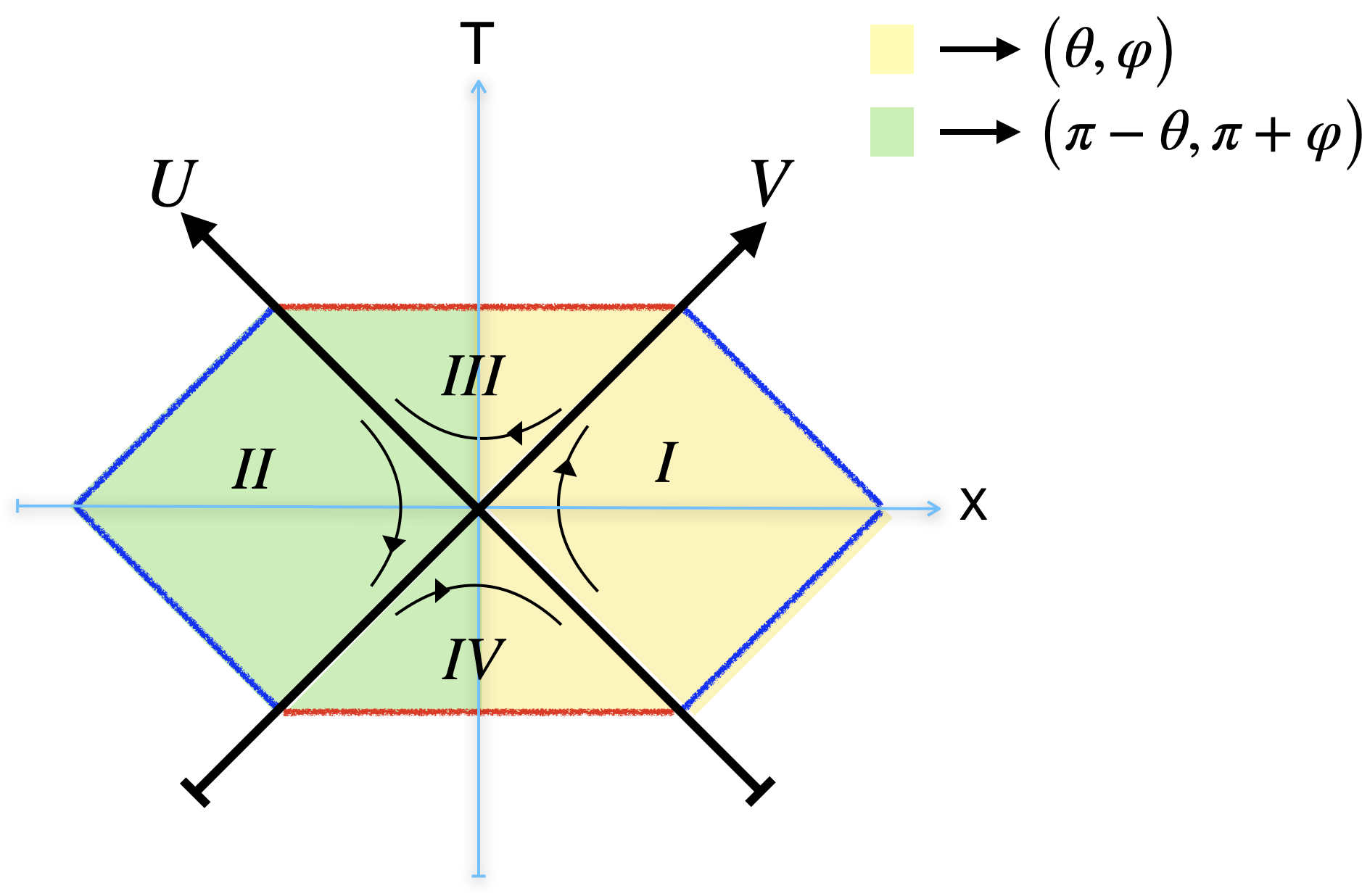}
	\caption{The picture represents the spacetime conformal diagram of quantum SBH according to DQFT. It contains four regions $I,\, II,\, III,\, IV$, which define geometric SSS to describe quantum fields in Schwarzschild spacetime (applying the near-horizon approximation $r\approx 2GM$). In this picture, the regions $I$ ($III$) and $II$ ($IV$) are related by discrete transformation $\LF U,\,V \RF\to \LF -U,\,-V \RF$. {The curved black lines with arrows represent integral curves of killing vector $\pd_t$ \eqref{boostUV} in each region, or in other words, these are the curves of $T^2-X^2 = {\rm constant<0}$ (Region I and II) and $T^2-X^2= {\rm constant}>0$ (Region III and IV). It is trivial to see an analogy between the spacetime of Schwarzschild BH and the phase space of IHO \eqref{phihoeq}. The red horizontal lines are identified with $r=0$. The quantum field components in the exterior are $\hat \Phi_{I}$ and $\hat \Phi_{II}$ at parity conjugate regions with opposite arrows of time $t: -\infty \to \infty $ and $t: \infty \to -\infty $ respectively. Whereas in the interior quantum field (component) $\hat\Phi_{int}$  evolve according to \eqref{intfield}.}   }
	\label{fig:BH}
\end{figure}

We arrive at a novel perspective on quantum fields in the Schwarzschild black hole background, where direct-sum structures provide a coherent mathematical bridge between two disconnected sheets of spacetime. First of all, let us recall the fact that the interior $r<2GM$ and exterior $r>2GM$ are not of the same kind because we cannot treat the time in the same way in the interior and the exterior. Since, in quantum theory, time is a parameter (not an operator), we cannot apply the same quantum theory everywhere. Furthermore, we can notice that the interior and the exterior are related by a discrete operation $U\to -U,\, V\to V$, which takes us from region $I$ to $III$ in Fig.~\ref{fig:BH}, whereas the transformation $U\to U,\, V\to -V$ takes us from region $II$ to $IV$. What this all mean is switching $X\to T,\, T\to X$ or   $X\to -T,\, T\to -X$ respectively. 
All of this indicates that the Hilbert spaces of the interior and exterior of SBH are geometric superselection sectors (SSS), suggesting the KG field operator becomes a direct-sum of two components according to the rules of DQFT. 

For quantizing a scalar field in SBH spacetime, we first perform the expansion of the KG field in spherical harmonics 
\begin{equation}
	\phi \LF U,\, V,\, \theta,\,\varphi \RF = \sum_{\ell,\,m} \frac{\phi_{\ell m} \LF U,\,V \RF}{r}Y^\ell_m\LF \theta,\,\varphi \RF
    \label{fullKGKS}
\end{equation}
where $Y^\ell_m\LF \theta,\,\varphi \RF$ are spherical harmonics. 
Upon substituting \eqref{fullKGKS} in the massless KG field action in SBH spacetime, we can integrate out the spherical harmonics and realize that the effective action for the field $\phi_{\ell m}\LF U,\, V \RF$ can be viewed as massless KG field in 2D in near horizon approximation (for a sufficiently large black hole) (See \cite{Kumar:2023hbj} for more details). 
We can now quantize the field $\Phi=\phi_{\ell m}$(in near-horizon approximation), promoting it to be an operator as 
\begin{equation}
\hat\Phi\Bigg\vert_{r\approx 2GM}  = \hat \Phi_{ext}\oplus \hat \Phi_{int} = 
\frac{1}{\sqrt{2}} \LF \hat{\Phi}_I\oplus \hat{\Phi}_{II} \RF \oplus  \frac{1}{\sqrt{2}} \LF \hat{\Phi}_{III}\oplus \hat{\Phi}_{IV} \RF
\label{eq:bhopsplit}
\end{equation}
where $ext$ and $in$ subscripts indicate the components of the field operator correspond to exterior and interior regions of SBH as defined in \eqref{eq:extint}. The field operator in \eqref{eq:bhopsplit} corresponds to the Fock space with the corresponding geometric SSS as 
\begin{equation}
    \Fc = \LF \Fc_I \oplus \Fc_{II}\RF \oplus \LF \Fc_{III}\oplus \Fc_{IV}\RF
\end{equation}
where {the labels $I,\,II$ represent $\Pc\Tc$ conjugate (defined by \eqref{PTKS}) regions of the exterior $r>2GM$ and the labels $III,\,IV$ indicate $\Pc\Tc$ conjugate regions of the interior $r<2GM$. 
{Explicitly, the geometric SSS to quantize the field in SBH spacetime are defined for the exterior region $r>2GM$ as
\begin{equation}
    \begin{aligned}
   {\rm Region\,I}:& \begin{cases}
        \Omega: \LF \theta,\,\varphi \RF,\quad t: -\infty\to \infty \\
         T = \sqrt{\Big\vert 1-\frac{r}{2GM}\Big\vert } e^{\frac{r}{4GM}} \sinh\!\left( \frac{t}{4M} \right),\quad X= \sqrt{\Big\vert 1-\frac{r}{2GM}\Big\vert } e^{\frac{r}{4GM}} \cosh\!\left( \frac{t}{4M} \right) \\
         T: -\infty \to \infty
    \end{cases}  \\
      {\rm Region\,II}:& \begin{cases}
        \Omega: \LF \pi-\theta,\,\pi+\varphi \RF,\quad t: \infty\to -\infty \\
         T = \sqrt{\Big\vert 1-\frac{r}{2GM}\Big\vert } e^{\frac{r}{4GM}} \sinh\!\left( \frac{t}{4M} \right),\quad X= -\sqrt{\Big\vert 1-\frac{r}{2GM}\Big\vert } e^{\frac{r}{4GM}} \cosh\!\left( \frac{t}{4M} \right) \\ 
         T: \infty\to -\infty 
    \end{cases} 
    \end{aligned}
\end{equation}}
For the interior region $r<2GM$, the geometric SSS are 
\begin{equation}
    \begin{aligned}
   {\rm Region\,III^{\pm}}:& \begin{cases}
       \Omega^+: \LF \theta,\,\varphi \RF,\quad t: -\infty\to 0,\quad \Omega^-: \LF \pi-\theta,\,\pi+\varphi \RF,\quad t: 0\to \infty \\
         X_\pm = -\sqrt{\Big\vert 1-\frac{r}{2GM}\Big\vert } e^{\frac{r}{4GM}} \sinh\!\left( \frac{t}{4M} \right),\quad T_\pm= \sqrt{\Big\vert 1-\frac{r}{2GM}\Big\vert } e^{\frac{r}{4GM}} \cosh\!\left( \frac{t}{4M} \right) 
    \end{cases}  \\
      {\rm Region\,IV^\pm}:& \begin{cases}
        \Omega^-: \LF \pi-\theta,\,\pi+\varphi \RF,\quad t: \infty\to 0,\quad \Omega^+: \LF \theta,\,\varphi \RF,\quad t: 0\to -\infty \\
         X_\pm = -\sqrt{\Big\vert 1-\frac{r}{2GM}\Big\vert } e^{\frac{r}{4GM}} \sinh\!\left( \frac{t}{4M} \right),\quad T_\pm = -\sqrt{\Big\vert 1-\frac{r}{2GM}\Big\vert } e^{\frac{r}{4GM}} \cosh\!\left( \frac{t}{4M} \right) 
    \end{cases} 
    \end{aligned}
    \label{34region}
\end{equation}}
{Note that Regions III and IV are divided into times:  $t: \mp \infty \to 0$ ($III^+,\,IV^-$) and $t: 0 \to \pm \infty$ ($III^-,\,IV^+$). 
For $t: \mp \infty \to 0$, $III^+$ and $IV^-$ are the parity conjugate regions of the interior. Interior $III^+$ region is $X_+>0, T_+>0,\,\Omega^+: \LF \theta,\varphi \RF$ whereas interior $IV^-$ is $X_-<0,\,T_-<0,\,\Omega^-: \LF \pi-\theta, \pi+\varphi \RF$. Therefore, the symmetry \eqref{PTKS} is preserved. Similarly when $t: 0 \to \pm \infty$ the region  ($III^-,\,IV^+$) preserve the same symmetry \eqref{PTKS}. This mean $III^-$ region covers $X_-<0,\,T_->0,\,\Omega^-: \LF \pi-\theta, \pi+\varphi \RF$, where as $IV^+$ covers $X_+>0,\,T_+<0,\,\Omega^+: \LF \theta,\,\varphi \RF$. Therefore, for both $t: \mp \infty \to 0$ ($III^+,\,IV^-$) and $t: 0 \to \pm \infty$ ($III^-,\,IV^+$) the symmetry \eqref{PTKS} is preserved for all times}.

{These are represented in the conformal diagram Fig.~\ref{fig:BH}, which has to be interpreted within the context of our quantum theory (DQFT).  We do not discard symmetries \eqref{PTKS} in our DQFT construction; thus, classical interpretations of parallel Universes, wormholes, and white holes are completely irrelevant in our study.}

The field operators in the geometric SSS are expanded in terms of the corresponding creation and annihilation operators as ({applying the near-horizon approximation $r\approx 2GM$})
\begin{equation}
\begin{aligned}
    \hat\Phi_{I} & = \int \frac{dk}{\sqrt{4\pi \vert k\vert }}\Big[a_{I\,k} e^{- i\vert k\vert T+ ikX}+a_{I\,k}^\dagger e^{ i\vert k\vert T- ikX}   \Big] \\   \hat\Phi_{II} & = \int \frac{dk}{\sqrt{4\pi \vert k\vert }}\Big[ \LF -1 \RF^{\ell} a_{II\,k} e^{i\vert k\vert T- ikX}+ \LF-1\RF^{\ell+m} a_{II\,k}^\dagger e^{- i\vert k\vert T+ ikX}   \Big]\\
     \hat\Phi_{III} & = \int \frac{dk}{\sqrt{4\pi \vert k\vert }}\Big[a_{III\,k} e^{ i\vert k\vert T-ikX}+a_{III\,k}^\dagger e^{ -i\vert k\vert T+ ikX}   \Big] \\   \hat\Phi_{IV} & = \int \frac{dk}{\sqrt{4\pi \vert k\vert }}\Big[ \LF-1\RF^{\ell} a_{IV\,k} e^{-i\vert k\vert T+ ikX}+ \LF-1\RF^{\ell+m} a_{IV\,k}^\dagger e^{ i\vert k\vert T- ikX}   \Big]
    \end{aligned}
\end{equation}
where creation and annihilation operators of each region satisfy canonical commutation relations. 
Regions $I$ and $II$ are related by parity and time reversal $\Pc\Tc$ \eqref{PTKS}, which is why the mode functions are related by the anti-unitary transformation $i\to -i$. The following commutation relations hold as Region I and II parity conjugate (time reversal) regions of the exterior
\begin{equation}
    \Big[ a_{I\,k},\,a_{II\,k^\prime} \Big] = \Big[ a_{I\,k},\,a^\dagger_{II\,k^\prime} \Big]=0.
    \label{I2com}
\end{equation}
The Regions $III$ and $IV$ are also related by $\Pc\Tc$ \eqref{PTKS}, which again explains the respective mode functions change by the anti-unitary transformation $i\to -i$. Similar to 
\begin{equation}
    \Big[ a_{III\,k},\,a_{IV\,k^\prime} \Big] = \Big[ a_{III\,k},\,a^\dagger_{IV\,k^\prime} \Big]=0.
    \label{34com}
\end{equation}
The factors $\LF -1 \RF^{\ell}$ and $\LF -1 \RF^{m}$ are attributed to the properties of spherical harmonics associated with parity and complex conjugation 
\begin{equation}
    Y_{\ell m}\LF \pi-\theta,\,\pi+\varphi \RF = \LF -1 \RF^{\ell} Y_{\ell m}\LF \theta,\,\varphi \RF,\quad Y^\ast_{\ell m}\LF \theta,\,\varphi \RF = \LF -1 \RF^{m}  Y_{\ell m}\LF \theta,\,\varphi \RF\,.
\end{equation}
Note that in regions $III$ and $IV$ we further have an implicit geometric superselection sector with each individually having parity conjugation regions as per Fig.~\ref{fig:BH}. This is very much similar to the Future and Past Kasner regions of Rindler spacetime (See Fig.~\ref{fig:Rindler}). {To take this into account, we split further the operators $a_{III\,k},\,a_{IV\,k}$ to further commuting set of canonical operators as  
\begin{equation}
    \begin{aligned}
     a_{III\,k} & =  a^+_{III\,k}\oplus \LF -1 \RF^{\ell}  a^-_{III\,k} \implies \hat{\Phi}_{III} = \hat{\Phi}_{III^+}\oplus \hat{\Phi}_{III^-}  \\  a_{IV\,k} & =  a^+_{IV\,k}\oplus \LF -1 \RF^{\ell}  a^-_{IV\,k} \implies \hat{\Phi}_{IV} = \hat{\Phi}_{IV^+}\oplus \hat{\Phi}_{IV^-}  
    \end{aligned}
    \label{34split}
\end{equation}
To be precise, the time evolution of quantum field in the interior region $r<2GM$ has to be understood through 
\begin{equation}
    \hat \Phi_{int} = \begin{cases}
    \hat\Phi_{III^+}\oplus \hat\Phi_{IV^-},\quad \vert t\vert: \infty\to 0 \\  \hat\Phi_{III^-}\oplus \hat\Phi_{IV^+},\quad \vert t\vert: 0\to \infty
    \end{cases}
    \label{intfield}
\end{equation}}
This is very much similar to the case of Rindler spacetime (See Eqs.~\eqref{eqFP} and \eqref{eqFP2}). {Recall that in the description of DQFT, we do the combined transformation $\LF T,\,X \RF\to \LF -T,\,-X \RF$ with $\LF \theta,\,\varphi \RF\to \LF \pi-\theta,\,\pi+\varphi \RF$. Thus, according to \eqref{34split}, a quantum field field is realized as direct-sum of a component in Region III at $\LF\theta,\,\varphi\RF$ and another component in Region IV $\LF \theta,\,\varphi \RF\to \LF \pi-\theta,\,\pi+\varphi \RF$ and vice-versa. This is what is depicted in Fig.~\ref{fig:BH}.} 

In Hawking's 1974 paper \cite{Hawking:1975vcx}, the quantum field operator is written as a summation of interior and exterior parts, which means making a superposition of interior and exterior quantum states. Since the concept of time is not the same as we discussed earlier, one cannot write a superposition of quantum states. Our direct-sum operation separates the Hilbert space geometrically into SSS, which avoids any superposition. This follows from the fundamental meaning of time in quantum theory. The second assumption from the Hawking computation is that the interior and exterior components of fields commute $\LT \hat \Phi_{in},\, \hat\Phi_{ext}\RT = 0$ 
based on the intuition that the ingoing state should be independent of the outgoing state. This intuitive argument also led to the initial formulation of the information paradox, which has evolved into many intuitive interpretations in the theories of quantum gravity \cite{Almheiri:2019hni,Almheiri:2020cfm}. 
 However, 't Hooft's calculation of gravitational backreaction from GR and QM implies a non-commutativity of the QFT operators 
 \begin{equation}
 	\LT \hat \Phi_{in},\, \hat\Phi_{ext}\RT = i \hbar \frac{8\pi G}{r_S^2 \LF \ell^2+\ell+1 \RF}
  \label{eq:Infoparadox}
 \end{equation}
 which is extension of \eqref{uvcom} in the context of second-quantization (DQFT) \cite{Kumar:2023hbj,Kumar:2024ahu}. This consideration would modify the initial formulation of information loss. The question of extracting what has formed the BH requires a microscopic (quantum) picture of gravitational collapse, which is an open question still now due to the lack of a concrete way to quantize fields in a dynamical (collapsing) geometry and lack of Planck scale quantum gravity to explain final stages (high curvature regime) of the collapse. The relation \eqref{eq:Infoparadox} is the most fundamental relation that joins gravity and quantum mechanics at the horizon scale, and it is the key to progress in understanding BH physics. Computing the Bogoliubov transformation between the Kruskal vacuum $\vert 0_{K}\rangle = \vert 0_{I}\rangle \oplus \vert 0_{II}\rangle $ and the asymptotic Minkowski vacuum $\vert 0_{M}\rangle = \vert 0_{M+}\rangle \oplus \vert 0_{M-}\rangle $ gives us the Hawking radiation given by the number density\cite{Kumar:2023hbj,Kumar:2024ahu}
 \begin{equation}
     N_{BH} = \frac{1}{e^{2\pi\omega_{BH}/\kappa \hbar}-1}
 \end{equation}
where the surface gravity term $\kappa = \frac{1}{4GM}$ which appears in Bekenstein's Black hole thermodynamics \cite{Bekenstein:1973ur} gives the temperature of Hawking radiation to be $T_{BH}=\frac{\hbar}{8\pi G M}$. 
The result is the same even when calculating Bogoliubov transformation between vacuums of an infinitely long time before and after SBH formation\cite{Mukhanov:2007zz}.

With DQFT, we achieve Hawking radiation in the form of pure states \cite{Kumar:2024ahu} because the density matrix of the maximally entangled pure state is the direct-sum of the pure state components 
\begin{equation}
\rho_P =\rho_{int}\oplus \rho_{ext}
\label{entBH}
\end{equation}
corresponding to the exterior and interior regions of SBH,
that define geometric SSS. 
Analogous to \eqref{VNE}, the Von Neumann entropy corresponding to the density matrices $\rho_{int}$ and $\rho_{ext}$ vanishes, thus $\rho_P$ is the density matrix of a pure state. Recall that the concept of time in the interior is different from the exterior, and we are describing here quantum fields spread across the horizon through geometric SSS, such that unitarity is preserved. 
Thus, any observer who may only access one of the regions in the conformal diagram (Fig.~\ref{fig:BH}) accesses the information in the form of pure states; thus, there is no unitarity loss. Each observer accesses complementary information because fixing a vacuum in any one of the geometric SSS uniquely fixes a vacuum in the rest because of the discrete spacetime transformations that relate to different SSS.  This is called observer complementarity \cite{Susskind:1993if}, which is consistent in the framework of DQFT. {Furthermore, note that the conformal diagram in Fig.\ref{fig:BH}\, does not admit any interpretation involving a white hole or a parallel universe.  All regions shown are physical and correspond to the exterior and interior of a single Schwarzschild black hole (SBH). The symmetry in Eq.\eqref{PTKS} implies that regions I \& II (and III \& IV)  are classically equivalent. From a quantum perspective, they represent different phases in the evolution of quantum fields within the SBH spacetime {at the parity conjugate regions of physical space}.}

\subsubsection{A new understanding of ER=EPR}

{The ER=EPR conjecture, proposed in \cite{Maldacena:2013xja}, posits a deep equivalence between quantum entanglement (Einstein-Podolsky-Rosen correlations) and spacetime connectivity through ER bridges. Traditionally framed within the context of AdS/CFT and thermofield double states, this idea links maximally entangled pairs of quantum systems through non-traversable wormholes connecting distinct regions of spacetime.} {This framework of ER=EPR has been later taken forward to a more complex setup of the gravitational path integral and holography that has resulted in so-called
replica wormholes and entanglement islands \cite{Penington:2019kki,Penington:2019npb,Marolf:2020xie,Jensen:2013ora}. These later constructions fully concede to the idea that unitarity is lost in QFTCS, and argue that some auxiliary states in the BH interior must lead to purification of states after half of the BH is evaporated through Hawking radiation. These auxiliary states are associated with states that belong to  "islands," which are "gravitational regions" in the BH interior entangled with exterior Hawking radiation, which course correct the entropy during the process of evaporation. However, it is important to note that these "islands" emerge in the way saddle points are defined in the Euclidean gravitational path integral, and they cannot be attributed to any physical states. Several investigations \cite{Geng:2020fxl,Goto:2020wnk} questioned the validity of extrapolating Euclidean path integral approaches to the Lorentzian ones. 
 All these investigations, however, overly rely on the tools of AdS/CFT, and the application to asymptotically flat BH spacetimes is not well understood. Furthermore, these formulations adhere to the goal of recovering the so-called Page curve \cite{Page:1993wv}, which again comes from accepting that unitarity is lost, there is traditional entanglement between BH interior and exterior, like open quantum systems, the entanglement entropy grows as BH evaporates, and some new physics (presumably unknown quantum gravity) should bring the entanglement entropy to turn around towards zero after exactly the half of the BH evaporated \cite{Almheiri:2019hni}. The Holography and string theory approaches \cite{Raju:2020smc,Almheiri:2020cfm}, and loop quantum gravity \cite{Ashtekar:2025ptw} are stuck to the idea that some Planck scale physics must save unitarity. We emphasize an alternative notion that the problem of unitarity first emerged through naive quantization of fields in curved spacetime, ignoring the discrete symmetries of the manifold, as we discussed in Sec.~\ref{sec:history}. }

{ In all the investigations that admit unitarity is lost and usual entanglement between BH interior and exterior states, the basic fact that the concept of "time" in the interior and exterior is different is unnoticed at the quantum level. It is well known that the radial coordinate $r$ behaves like time and $t$ becomes spacelike in the BH interior, which makes the interior Schwarzschild spacetime not static but analogous to the Kantowski-Sachs cosmological metric \cite{Doran:2006dq}. All QFTCS construction requires us to define a positive energy state with respect to an arrow of time, since time is not the same in the interior and exterior of BH; one must have a different Hilbert space structure for these two regions. This is exactly what our concept of geometric SSS does.} {Here, we present an alternative and complementary formulation of ER=EPR grounded in DQFT. By studying quantum fields across the Rindler and Schwarzschild horizons, we propose that the entanglement structure implied by ER=EPR arises naturally through the direct-sum Hilbert space formalism, without invoking geometric wormholes.} {To be precise, the outcomes of DQFT \eqref{pure} and \eqref{entBH} indicate a pure state ($\vert \Psi_P\rangle$) is split into direct-sum of two pure state components $\LF \vert \Psi_{wH}\rangle,\, \vert\Psi_{bH}\rangle \RF$ that geometrically evolve  within and beyond the Horizon as
\begin{equation}
    \vert\Psi_P\rangle = \vert \Psi_{wH}\rangle\oplus \vert\Psi_{bH}\rangle \implies \rho_P= \rho_{wH}\oplus \rho_{bH}
    \label{DQFTent}
\end{equation}
in the superselection sector Hilbert spaces. 
$\vert\Psi_P\rangle$ is a pure state here because the Von Neumann entropies of the components $\rho_{wH}$ and $\rho_{bH}$ vanish. 
Thus, any density matrix of a pure state splits into these interior and exterior sectors that preserve the unitarity for every local observer. This construction is applicable also to de Sitter space, but also to spacetimes that are dynamical, in which case geometric SSS must be defined for every constant time spatial hypersurface. 
The mathematical bridges in our case are "direct-sum" operations connecting the regions (for example, left and right wedges of Rindler spacetime or Regions I and II of BHs, see Fig.~\ref{fig:BH}), which are defined via discrete symmetries and inverted harmonic oscillator dynamics. These offer a concrete realization of entanglement through horizon-local quantum correlations.} { To be more specific, in DQFT, any entangled state is split as direct-sum of various components in geometric superselection sectors that are connected by discrete spacetime operations. What we mean by "horizon-local" here is that there is a pure state (component) that keeps evolving as a pure state for any local observer bounded by the horizon. Any pure state (component) beyond the horizon is a mirror state in the corresponding geometric SSS. 
Thus, any local observer (defined by a region with a specific arrow of time) witnesses the horizon-local quantum correlations with unitarity locally reinstated. 
This framework reshapes the ER=EPR correspondence in terms of observable quantum field theoretic structures in real spacetime, offering new insights into the nature of unitarity, observer complementarity, and the emergence of spacetime connectivity. All of this is rooted in the fact that quantum fields in curved spacetime are a collection of IHOs; thus, the usual particle description does not hold near the gravitational horizons.}

{In Table~\ref{tab-ER}, we highlight the conceptual and mathematical parallels between the ER=EPR conjecture and DQFT's realization of it. 
It reveals how spacetime connectivity, typically attributed to wormhole geometry, can alternatively be understood as a consequence of horizon-induced field entanglement and symmetry structure, providing a fresh, horizon-local perspective on quantum gravity phenomena.}

\begin{table}[h!]
\centering
\renewcommand{\arraystretch}{1.3}
\small
\begin{tabularx}{\textwidth}{|X|X|X|}
\hline
\textbf{Aspect} & \textbf{ER=EPR (Maldacena–Susskind)} & \textbf{DQFT's realization of ER=EPR} \\
\hline
\hline
\textbf{Spacetime Setup} & Two-sided asymptotically AdS black holes and Holography & A single Schwarzschild black hole that is asymptotically Minkowski \\
\hline
\textbf{Entanglement Representation} & Tensor product of two entangled systems ($\Hc_L,\,\Hc_R$ representing left the right Rindler Hilbert spaces): $\mathcal{H}_L \otimes \mathcal{H}_R$. Maximal entanglement between two distant black holes & A single quantum field is direct-sum of components in geometric SSS: $\mathcal{H}_L \oplus \mathcal{H}_R$. So any maximally entangled state is direct-sum of components in geometric SSS of a single BH spacetime.\\
\hline
\textbf{Unitarity} & Unitarity is lost for any local observer in the left or right region & Unitarity is reinstated for any local observer in the left or right \\
\hline
\textbf{Geometric Realization} & Einstein–Rosen bridge (non-traversable wormhole) connects two asymptotic boundaries & No geometric wormhole but a quantum field is abridged by geometric superselection sectors across horizons \\
\hline
\textbf{Time Flow} & Forward in both left and right regions (in global AdS time) & Opposite time directions in left and right wedges (e.g., Kruskal or Rindler time) \\
\hline
\textbf{Key Mathematical Tool} & Thermofield double Hamiltonian $H_{\text{TFD}} =  H_R+H_L$ (which is changed from the usual $H_{\text{TFD}} =  H_R-H_L$) & Inverted harmonic oscillator structure near horizons, discrete symmetry identifications \\
\hline
\textbf{Observer Complementarity} & Emerges from dual CFT entanglement and bulk geometry & Built into horizon local QFT: different observers access complementary Hilbert spaces (geometric SSS) \\
\hline
\textbf{Spacetime Connectivity} & Entanglement implies wormhole connectivity (ER=EPR conjecture) & Entanglement encoded in field theory structure across horizon, without wormhole \\
\hline
\end{tabularx}
\caption{Relations between Maldacena and Susskind's ER=EPR conjecture versus a new realization of ER=EPR \cite{Maldacena:2013xja} via DQFT horizon-local entanglement structure. The notation with subscripts $R$ and $L$ denotes right and left Rindler parts of spacetime, which in the Schwarzschild BH case correspond to Region I and Region II of Fig.~\ref{fig:BH}. }
\label{tab-ER}
\end{table}

\subsection{The mathematical bridges in de Sitter spacetime}

{De Sitter spacetime, with its maximally symmetric and horizon structure, provides a rich arena for exploring the foundations of quantum field theory in curved backgrounds. A key insight emerging from our investigation is that, irrespective of the coordinate patch, whether described in the flat FLRW form or in the static chart, a common mathematical structure underlies the quantum theory: a direct-sum formulation of the Hilbert space. This formulation captures the essential parity and time reflection symmetries intrinsic to de Sitter geometry and offers a natural way to define quantum states that span across causally disconnected or antipodally related regions. In the subsections that follow, we illustrate how this DQFT framework manifests concretely in both the flat FLRW and static coordinate representations of de Sitter space, reinforcing the universality of the underlying mathematical bridges.}

\subsubsection{Direct-sum QFT in flat FLRW de Sitter}

Quantizing fields in flat FLRW dS spacetime \eqref{FLRWdS} is widely used in understanding physics related to early Universe cosmology (cosmic inflation in particular). An often taken assumption in the literature is fixing the arrow of time $\tau<0$ before quantization. In DQFT treatment, we preserve the discrete symmetry of spacetime $\tau\to -\tau$ and $\textbf{x}\to -\textbf{x}$ at the quantum level by writing the KG field operator as a direct-sum of the two components which belong to the parity conjugate points of the physical space. For quantizing the KG field, we rescale the field with the scale factor $\phi\to a\phi$ such that the KG action gets to the form of IHO with time-dependent mass \cite{Albrecht:1992kf}
\begin{equation}
    S_{\phi} = \int d\tau d^3x \,\phi \LF -\pd_\tau^2+ \pd^i\pd_i + \frac{2}{\tau^2} \RF \phi 
\end{equation}
Then, in DQFT, we promote the field to an operator as
\begin{equation}
\begin{aligned}
    \hat \phi & = \frac{1}{\sqrt{2}}\LF \hat \phi_+\LF \tau,\,\textbf{x} \RF\oplus \hat \phi_-\LF -\tau,\,-\textbf{x} \RF \RF\\ \hat\phi_{\pm} & = \int \frac{d^3k}{\LF 2\pi \RF^{3/2}\sqrt{2k}} \LT d_{(\pm)\,\textbf{k}}  \phi_{(\pm)k}\LF \pm \tau \RF e^{\pm i\textbf{k}\cdot \textbf{x}} +d^\dagger_{(\pm)\,\textbf{k}}  \phi_{(\pm)k}^\ast\LF \pm\tau \RF e^{\mp i\textbf{k}\cdot \textbf{x}}  \RT  
    \end{aligned}
    \label{dSKG}
\end{equation}
where 
\begin{equation}
    \phi_{(\pm)k} = \frac{A_{(\pm)k}}{\sqrt{2k}} \LF 1\mp \frac{i}{k\tau} \RF e^{\mp ik\tau} +\frac{B_{(\pm)k}}{\sqrt{2k}} \LF 1\pm \frac{i}{k\tau} \RF e^{\pm ik\tau}
\end{equation}
with $A_{(\pm)k},\,B_{(\pm)k}$ being the Bogoliubov coefficients. 
In the above relation, the creation and annihilation operators satisfy $\LT d_{(\pm)\,\textbf{k}},\,d^\dagger_{(\pm)\,\textbf{k}} \RT = 1,\, \LT d_{(\mp)\,\textbf{k}},\,d^\dagger_{(\pm)\,\textbf{k}} \RT=\LT d_{(\mp)\,\textbf{k}},\,d_{(\pm)\,\textbf{k}} \RT = 0 $ preserving causality and locality. These commutation relations imply 
\begin{equation}
\phi_{(\pm)k}\phi_{(\pm)k}^{\ast\prime}- \phi^\prime_{(\pm)k}\phi^{\ast}_{(\pm)k} = \pm i \implies \vert A_{(\pm)k}\vert^2 - \vert B_{(\pm)k}\vert^2 =1. 
\end{equation}
The dS vacuum is 
\begin{equation}
\vert 0\rangle_{dS} = \vert 0_{+}\rangle_{dS} \oplus \vert 0_{-}\rangle_{dS} = \begin{pmatrix} \vert 0_+\rangle_{dS} \\ \vert 0_-\rangle_{dS} \end{pmatrix}
\label{dSvac}
\end{equation}
 which we chose such that we recover the direct-sum Minkowski vacuum \eqref{minvac} in the short-distance limit or sub-horizon $k\gg \vert aH\vert$ (See \eqref{BDdS} below), which we can call the direct-sum Bunch-Davies vacuum. The choice of vacuum depends on the choice of mode functions. Imposing 
 \begin{equation}
     \phi_{(\pm)k} \Bigg\vert_{k\gg \vert aH\vert } \approx \frac{1}{\sqrt{2k}} e^{\mp ik\tau} \implies A_{(\pm)k} =1,\,B_{(\pm)k} =0\,.
     \label{BDdS}
 \end{equation}
Thus, the two components of a single quantum field corresponding to parity-conjugate regions of physical space would then be given by
\begin{equation}
  \hat\phi_\pm =\int \frac{d^3k}{\LF 2\pi \RF^{3/2}\sqrt{2k}}\LT d_{(\pm)\,\textbf{k}}\LF 1\mp\frac{i}{k\tau} \RF e^{\mp ik\tau\pm\textbf{k}\cdot \textbf{x}} + d_{(\pm)\,\textbf{k}}^\dagger \LF 1\pm\frac{i}{k\tau} \RF e^{\pm ik\tau\mp\textbf{k}\cdot \textbf{x}} \RT
\end{equation} 
 This construction joins mathematically the quantum field components $\hat \phi_\pm$ such that we have one expanding Universe with two arrows of time (quantum mechanically). Thus, we create mathematical bridges a l\'a ER and Schr\"{o}dinger's thin rods we described in Sec.~\ref{sec:history}. 

The two-point correlations in de Sitter spacetime are identical at parity-conjugate regions 
\begin{equation}
	\begin{aligned}
		& \frac{1}{a^2}{}_{\rm dS}\langle 0_+\vert \hat{\phi}_{+}\LF \tau,\, \textbf{x} \RF \hat{\phi}_{+}\LF \tau,\, \textbf{x}^\prime \RF\vert 0_+\rangle_{\rm dS} = \\ & \frac{1}{a^2} {}_{dS}\langle 0_-\vert \hat{\phi}_-\LF -\tau,\, -\textbf{x} \RF \hat{\varphi}_{-}\LF -\tau,\, -\textbf{x}^\prime \RF\vert 0_-\rangle_{dS} = \int \frac{dk}{k}\frac{\sin k\xi}{k\xi} \frac{H^2}{4\pi^2}\,, 
	\end{aligned}
	\label{eqcorr}
\end{equation}
where $\xi = \vert \textbf{x}-\textbf{x}^\prime\vert$.

\begin{figure}
    \centering
    \includegraphics[width=0.6\linewidth]{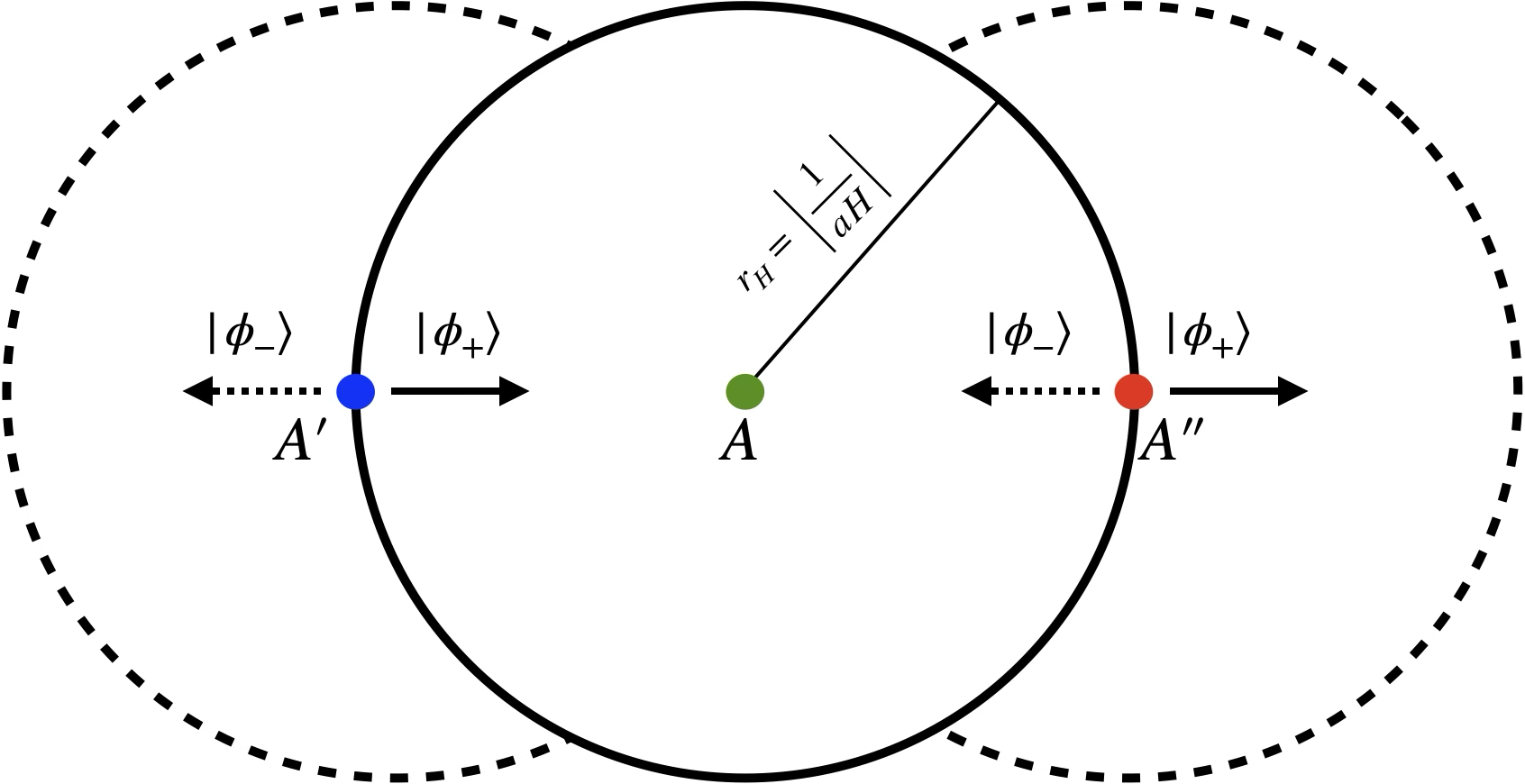}
    \caption{This figure illustrates three comoving (imaginary) observers $A,\,A^\prime,\, A^{\prime\prime}$ with the corresponding comoving horizon radius $r_H= \vert \frac{1}{aH}\vert$ at a moment of dS expansion. We have suppressed here the angular coordinates $\LF \theta,\,\varphi \RF$. Points $A^\prime$ and $A^{\prime\prime}$ are on the horizon of $A$ at antipodal sides of the horizon, that are spacelike separated, i.e., at the angles $\LF \theta,\,\varphi \RF$ and $\LF \pi-\theta,\, \pi+\varphi \RF$ respectively. The dotted circles with the same $r_H$ represent the respective comoving horizons of $A^\prime$ and $A^{\prime\prime}$. In the figure, $\vert \phi_+\rangle ,\,\vert \phi_{-}\rangle $ are defined in \eqref{dSdiaSt}. In the case of exact dS spacetime, both $\vert \phi_{\rm}\rangle $ components would give equal two-point correlations at the parity conjugate points \eqref{eqcorr} due to the $\Pc\Tc$ symmetric dS vacuum \eqref{dSvac}. Whereas in the quasi-dS spacetime, we get unequal correlations \eqref{eq:kmkcoupling} due to the $\Pc\Tc$ asymmetric quasi-dS vacuum \eqref{qdSvac}. }
    \label{fig:desitteru}
\end{figure}

In Fig.~\ref{fig:desitteru}, $\vert \phi_+ \rangle = \hat{d}_{+\,\textbf{k}}^\dagger \vert 0_+\rangle_{dS}$ and $\vert \phi_{-} \rangle = \hat{d}_{(-)\,\textbf{k}}^\dagger \vert 0_-\rangle_{dS}$ form the direct-sum state
\begin{equation}
	\begin{aligned}
		\vert \phi \rangle & = \LF \hat{d}^\dagger_{(+)\,\textbf{k}} \oplus \hat{d}^\dagger_{(-)\,\textbf{k}} \RF \vert 0\rangle_{dS} \\ & = \frac{1}{\sqrt{2}}\begin{pmatrix}
			 \hat{d}^\dagger_{(+)\,\textbf{k}} & 0 \\ 
			 0 &  \hat{d}^\dagger_{(-)\,\textbf{k}} 
		\end{pmatrix}  \begin{pmatrix}
		\vert 0_+\rangle_{dS} \\
		\vert 0_-\rangle_{dS}
	\end{pmatrix} \\ & 
		= \frac{1}{\sqrt{2}} \vert \phi_I\rangle \oplus \frac{1}{\sqrt{2}} \vert \phi_{-}\rangle = \frac{1}{\sqrt{2}}\begin{pmatrix}
			\vert \phi_+\rangle \\ 
			\vert \phi_{-} \rangle
		\end{pmatrix}\,. 
	\end{aligned}
\label{dSdiaSt}
\end{equation} 
 {The concept of the observer complementarity principle \cite{Susskind:1993if,Parikh:2002py} is that all the observers should share complementary information, maintain unitarity within the evolution of states they have access to. We can achieve this in the context of de Sitter space, which is pictorially represented in Fig.~\ref{fig:desitteru}. According to DQFT, every observer in de Sitter space has access to parity conjugate states \eqref{dSdiaSt}. Thus, at any moment of de Sitter expansion, the observer $A$ accesses the states $\vert \phi_+\rangle,\,\vert \phi_-\rangle$ from the antipodal points. All the information beyond the horizon of $A$ gets reflected within the horizon. Similarly, the observer $A^\prime, \, A^{\prime\prime}$ can reconstruct the information beyond their respective horizons by accessing the complementary states within the horizon. }

 {This construction implies a maximally entangled two-particle ($\vert \phi_1\rangle,\,\vert \phi_2\rangle$) (pure) state to be represented by two components corresponding to the parity conjugate regions of physical space give by \cite{Kumar:2023ctp,Gaztanaga:2024whs}
\begin{equation}
    \vert \psi_{12}\rangle = \frac{1}{\sqrt{2}}\begin{pmatrix}
        \vert \psi_{+(12)}\rangle \\ 
        \vert \psi_{-(12)}\rangle 
    \end{pmatrix} = \frac{1}{\sqrt{2}}\begin{pmatrix}
        \sum_{m,n} d^+_{mn} \vert \phi_{+1} \rangle \otimes  \vert \phi_{+2}\rangle \\ 
        \sum_{m,n} d^-_{mn} \vert \phi_{-\,1} \rangle \otimes  \vert \phi_{-\,2}\rangle
    \end{pmatrix} 
    \label{puredS}
\end{equation}
The density matrix of $ \vert \psi_{12}\rangle $ is direct-sum of two pure states in the sectoral Hilbert spaces 
\begin{equation}
    \rho_{12} = \frac{1}{2} \rho_{+(12)}\oplus \frac{1}{2} \rho_{-(12)} 
\end{equation}
and their Von Neumann entropies vanish 
\begin{equation}
    S_{NI} = -Tr\LF \rho_{I(12)}\ln \rho_{I(12)}\RF=0,\quad S_{NII} = -Tr\LF \rho_{II(12)}\ln \rho_{II(12)}\RF=0\,.
\end{equation}
This means we not only have observer complementarity but also have unitarity (pure states evolving into pure states) being maintained for all the states within the horizon. This offers a new perspective on the understanding of QFTCS in an expanding Universe, in contrast to the commonly perceived notion that unitarity is lost in the early Universe \cite{Brandenberger:2021pzy,Hartman:2020khs,Brandenberger:2022pqo,Colas:2024xjy}. 
}
 
\begin{figure}
    \centering    \includegraphics[width=0.8\linewidth]{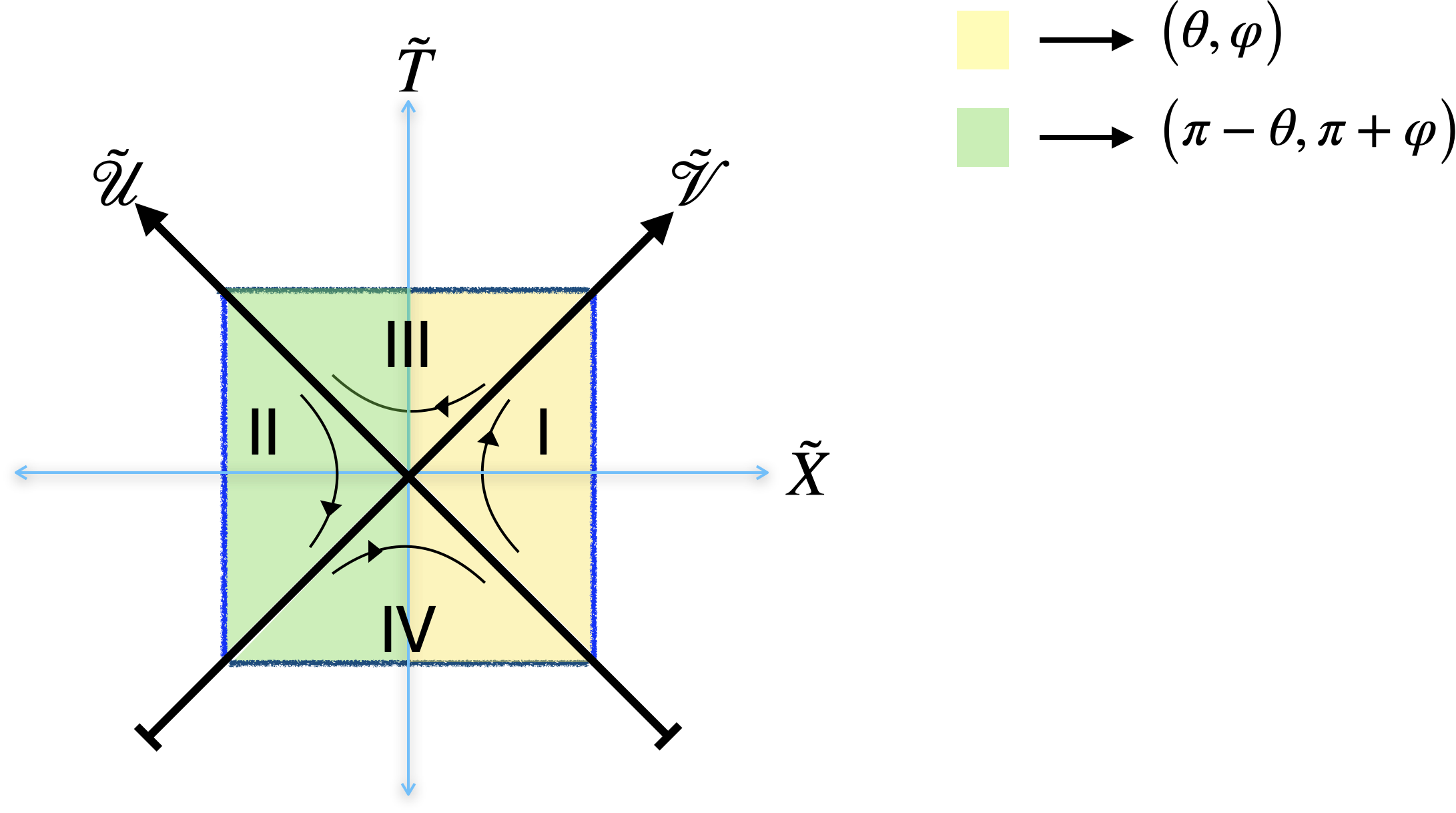}
    \caption{This is a conformal spacetime diagram that represents quantum dS spacetime where a quantum field operator is expressed as a direct-sum of four components corresponding to geometric SSS corresponding to regions I, II, III, IV. The regions I(III) and II(IV) are regions with opposite arrows of time-related by $\LF \tilde \Uc,\,\tilde \Vc \RF\to \LF -\tilde\Uc,\, -\tilde\Vc \RF$. Understanding of this diagram is analogous to the SBH case in Fig.~\ref{fig:BH}. Regions $I$ and $II$ represent the exterior, and $ III$ and $IV$ represent the interior of static dS spacetime (i.e, they are geometric SSS according to DQFT).   }
    \label{fig:deSitter}
\end{figure}

\subsubsection{Direct-sum QFT in static de Sitter}

Together with the symmetry \eqref{tdSsym}, the metric \eqref{statdS} can cover the entire dS spacetime. Similar to SBH and Rindler spacetime, the dS spacetime too has four regions related through discrete coordinate transformations \begin{equation}
    \begin{aligned}
       \Uc  &= -e^{-H \bar{u}}<0,\quad &&\Vc= e^{H \bar{v}}>0\quad &&(\rm Region\,I)\\
        \Uc &= e^{-H \bar{u}}>0,\quad &&\Vc= -e^{H \bar{v}}<0\quad &&(\rm Region\, II) \\
         \Uc &= e^{-H \bar{u}}>0,\quad &&\Vc= e^{H \bar{v}}>0\quad &&(\rm Region\, III) \\
          \Uc &= -e^{-H \bar{u}}<0,\quad &&\Vc= -e^{H \bar{v}}<0\quad &&(\rm Region\, IV)
    \end{aligned}
    \label{UVKruskaldS}
\end{equation}
where 
$\bar u = t-\Tilde{r}_{\ast}$ and $\bar v = t+\Tilde{r}_{\ast}$ with $\Tilde{r}_\ast = \tanh^{-1}\LF H r_s \RF$. DQFT in dS spacetime is analogous (by the construction of geometric SSS) to Rindler and SBH spacetime, and it can be understood through the conformal diagram of quantum dS spacetime depicted in Fig.~\ref{fig:deSitter}. Similar to the SBH case, we achieve unitarity and observer complementarity in dS space due to the geometric construction of quantum theory with SSS Hilbert spaces.

\section{ERBs, direct-sum inflation (DSI) and CMB}
\label{Sec:DSI} 

{This section marks a pivotal point in our exposition, where the theoretical framework developed thus far, including inverted harmonic oscillator dynamics, quantum field theory in spacetimes with horizons, and the direct-sum quantization formalism, culminates in potential observational consequences. Specifically, we propose a novel connection between the mathematical structures underpinning DQFT and anomalies observed in the Cosmic Microwave Background (CMB) \cite{Schwarz:2015cma}, particularly those related to parity asymmetry. Furthermore, we provide a new prediction for primordial gravitational waves for the future observational probes.} 

{The inflationary universe, modeled effectively by quasi-de Sitter spacetime, offers a natural setting where the horizon-induced quantum correlations may leave imprints on primordial fluctuations. In this context, we introduce the idea of Direct-Sum Inflation (DSI): a framework in which quantum states spanning parity conjugated regions, connected via mathematical bridges akin to those in ER=EPR, may manifest observable signatures in the large-scale CMB anisotropies. This opens a concrete avenue for testing the deep theoretical ideas developed throughout this work, providing a possible empirical handle on the quantum structure of spacetime near horizons.}

\subsection{Direct-sum QFT of inflationary quantum fluctuations}

Inflationary background by definition quasi-dS expansion, and it breaks the symmetry \eqref{disdS} or \eqref{tdSsym} i.e., $\LF \tau,\,\textbf{x} \RF\to \LF -\tau,\,-\textbf{x} \RF$ by the time-dependent slow-roll parameters $\LF \epsilon,\,\eta \RF$ \eqref{epsiloneta}. Recall the fact that time is a parameter in quantum theory (in the sense that time reversal is anti-unitary in character), which we extensively discussed in Sec.~\ref{sec:disumQT}. In the context of inflationary quantum fluctuations, we literally promote the gravitational and matter (inflaton) field variables as operators around the dynamically evolving quasi-de Sitter background. This is nothing but linearized quantum gravity, as extensively discussed in \cite{Martin:2004um}, because we do the following operationally (in the units of setting $8\pi G=1$).
 \begin{equation}
        \delta\hat G_{\mu\nu} = \delta \hat T_{\mu\nu}
    \end{equation}
    The curvature perturbation $\zeta = \Psi+\frac{H}{\dot{\bar\phi}}\delta\phi$ is a collective description of metric and inflaton fluctuations. Observationally, we relate the two-point correlations of $\zeta$ with the temperature fluctuations in the CMB. In the framework of DQFT, we express every quantum field as direct-sum of two components, which are geometrically attached to the parity conjugate regions of physical space, which give equal two-point correlations \eqref{eqcorr} in the direct-sum vacuum \eqref{dSvac}. Since the time reversal symmetry \eqref{disdS} is not the symmetry of quasi-dS spacetime, a naive physical expectation would be that the quantum vacuum of quasi-dS spacetime could be such that the two-point correlations could be unequal at parity conjugate regions of physical space. 
    
    In the quasi-dS spacetime, the MS-variable $V_{MS}$ \eqref{ihoinf}, when promoted to an operator in DQFT, becomes a direct-sum of two components 
\begin{equation}
    \hat{V}_{MS} = \frac{1}{\sqrt{2}}\LF \hat V_{MS+}\oplus \hat V_{MS-} \RF,\quad \hat V_{MS\pm} = \int \frac{d^3k}{\LF 2\pi \RF^{3/2}}\Bigg[b_{\textbf{k},\,\pm} V_{k\,\pm}e^{i\textbf{k}\cdot \textbf{x}}+ b^\dagger_{\textbf{k},\,\pm}  V_{k\,\pm}^\ast e^{-i\textbf{k}\cdot \textbf{x}}\Bigg]
    \label{MSDQFT}
\end{equation}
 where $ V_{k\,\pm} = \frac{\sqrt{\mp \pi \tau}}{2} e^{\LF i\nu_s^{\pm}+1\RF} H^{(1)}_{\nu_s^{\pm}}\LF \mp k \tau \RF$ and $\nu_s^\pm \approx \frac{3}{2}\pm \epsilon \pm \eta$. 
\begin{equation}
	\begin{aligned}
	{V}_{k\pm } &  \approx \sqrt{\frac{1}{2k}} e^{\mp ik\tau}\LF 1\mp\frac{i}{k\tau} \RF
	\pm \LF \epsilon+\frac{\eta}{2} \RF \frac{\sqrt{\pi}}{2\sqrt{k}} \sqrt{\mp k\tau} \frac{\pd H^{(1)}_{\nu_s^{\pm}}\LF \mp k\tau\RF}{\pd \nu_s^{\pm}}\Big\vert_{\nu_s^{\pm}=3/2}\equiv v^{\rm dS}_{\pm, k} \LF 1\pm \Delta v\RF
\label{new-vac1BD}
\end{aligned}
\end{equation}
where $H^{(1)}_{\nu_s}\LF z \RF$ is the Hankel functions of the first kind {and
\begin{equation}
    V_{k\pm}^{\rm dS} = \sqrt{\frac{1}{2k}} e^{\mp ik\tau}\LF 1\mp\frac{i}{k\tau} \RF,\quad \Delta v = \LF \epsilon+\frac{\eta}{2} \RF \LT \frac{1}{H_{3/2}^{(1)} \LF  \frac{k}{k_\ast} \RF} \frac{\pd H^{(1)}_{\nu_s}\LF \frac{k}{k_\ast} \RF}{\pd\nu_s} \Bigg\vert_{\nu_s=\frac{3}{2}} \RT
    \label{modeqds}
\end{equation}}
The mode functions of the curvature perturbation can be obtained by classical rescaling (by the factor $\frac{H}{a\dot{\bar\phi}}$ as) 
\begin{equation}
    \zeta_{\pm k} = \LF \frac{H}{a\dot{\bar\phi}}\RF V_{k\pm} = \zeta_{\pm\,k}^{\rm dS}\LF 1\pm \Delta v\RF   \quad ; \quad \zeta_{\pm\,k}^{\rm dS} \equiv \LF\frac{H}{a\dot{\bar\phi}}\RF V_{k\pm}^{\rm dS} 
    \label{eq:vpm}
\end{equation}
 We impose conditions of the vacuum such that we recover the DQFT Bunch-Davies vacuum in the limit $\epsilon\to \eta\to 0$. The DQFT treatment gives two-point correlations for parity conjugate points in physical space, which are unequal due to the asymmetric quantum vacuum imposed by the following discrete transformation  
 \begin{equation}
		\tau\to -\tau \implies t\to -t,\, H\to -H,\, \epsilon\to -\epsilon\,, \eta\to -\eta\,. 
  \label{dsisym}
	\end{equation}
According to DQFT, inflationary quantum fluctuations evolve in $\Pc\Tc$ asymmetric quasi-dS vacuum
\begin{equation}
    \vert 0\rangle_{qdS} =\vert 0_{+}\rangle_{qdS} \oplus \vert 0_{-}\rangle_{qdS} = \begin{pmatrix}
        \vert 0_{+}\rangle_{qdS} \\ \vert 0_{-}\rangle_{qdS} 
    \end{pmatrix}
    \label{qdSvac}
\end{equation}
 leads to a single quantum fluctuation to evolve asymmetrically at parity-conjugate points. Thus, the sign change in \eqref{eq:vpm} indicates a geometric description of fluctuation at parity conjugate points (See the right panel of Fig.~\ref{fig:180}). 
 
 During inflationary expansion, these fluctuations become classical and leave their parity asymmetric imprints as cold and hot structures in the two-dimensional CMB. This is nothing but holography, the imprints of quantum gravity in the bulk on the boundary. This supports 't Hooft original idea of quantum gravity in two dimensions that was propagated to the frameworks of string theory \cite{tHooft:1993dmi,Susskind:1993if}
This is schematically depicted in Fig.~\ref{fig:180}, and the actual data can be visualized in the bottom panel of Fig.~\ref{fig:ERB}.
 Computing the two-point correlations of MS-variable $\hat{V}_{MS}$, we obtain
 \begin{equation}
		\begin{aligned}
			\langle 0_{\rm qdS} \vert \hat{V}_{MS}  \hat{V}_{MS} \vert 0_{\rm qdS} \rangle = &  \,	\frac{1}{2}\langle 0_{\rm qdS{+}} \vert \hat{V}_{MS+}\LF \tau,\,\textbf{x} \RF  \hat{V}_{MS+}\LF \tau,\,\textbf{y} \RF \vert 0_{\rm qdS{+}} \rangle \\ & + \frac{1}{2} \langle 0_{\rm qdS{-}} \vert \hat{V}_{MS-}\LF -\tau,\,-\textbf{x} \RF  \hat{V}_{MS-}\LF -\tau,\,-\textbf{y} \RF \vert 0_{\rm qdS{-}} \rangle\\ 
			=&\, \int \frac{dk}{k}\frac{k^3}{2\pi^2} P_{V} \frac{\sin{kL}}{kL}
		\end{aligned}
  \label{twoVms}
	\end{equation}
	where $L=\vert \textbf{x}-\textbf{y}\vert$ and $P_{V} = \frac{1}{2}\LF\vert V_{k,\,MS+}  \vert^2+\vert V_{k,\,MS-}  \vert^2 \RF$. This results in a power spectrum of curvature perturbation corresponding to the parity conjugate regions of the CMB sky as \cite{Kumar:2022zff,Gaztanaga:2024vtr,Gaztanaga:2024whs} (See Sec.~\ref{Sec:DSIobservational} for more details)
 \begin{equation}
    \Pc_{\zeta\pm}(k)  = \int d^3r e^{-i\textbf{k}.\textbf{L}}	{}_{\rm qdS}\langle 0_{\pm}\vert  \hat \zeta_{\pm}\LF \pm\tau,\,\pm \textbf{x} \RF  \hat \zeta_{\pm}\LF \pm \tau,\, \pm\textbf{x}^\prime \RF\vert 0_\pm\rangle_{\rm qdS} \approx P_\zeta\LF 1\pm \Delta \Pc_v\RF
    \label{eq:pwdsi}
\end{equation} 
where $\Pc_\zeta$ is the near scale invariant (SI) part of the power spectrum $P_\zeta$:
\begin{equation}
		P_\zeta  = \frac{H^2}{8\pi \epsilon} \approx A_s \LF \frac{k}{k_\ast} \RF^{n_s-1}\,. 
		\label{eq:HZpowe}
	\end{equation}
    where $A_s = 2.2\times 10^{-9}$ is called the primordial power spectrum amplitude, and the scalar spectral index $n_s = 1-2\epsilon-\eta= 0.9634\pm 0.0048$ (Planck TT+TE+EE) at $k_\ast = 0.05 \,\text{Mpc}^{-1}$ from the Planck data \cite{Planck:2019evm} and $k_s = \frac{1}{\tau-\tau_{LS}}= 7\times 10^{-5}\,\text{Mpc}^{-1}$ is related to the distance from the CMB surface of last scattering,
and $ j_{\ell}\LF z \RF$ are Bessel functions.
and 
    \begin{equation}
\Delta \Pc_v= \LF 1-n_s \RF\operatorname{Re}\LT \frac{2}{H_{3/2}^{(1)} \LF  \frac{k}{k_\ast} \RF} \frac{\pd H^{(1)}_{\nu_s}\LF \frac{k}{k_\ast} \RF}{\pd\nu_s} \Bigg\vert_{\nu_s=\frac{3}{2}} \RT
		\label{eq:delpPR}
	\end{equation}	
with $H^{(1)}_{\nu_s}\LF z \RF$ is the Hankel functions of the first kind.  
The standard inflation (SI) prediction \cite{Mukhanov:2005sc} corresponds to $
\Delta \Pc_v =0$, while the DSI one is given by Equation~(\ref{eq:pwdsi}). 
    Note that the power spectrum is the Fourier transform of the two-point function of curvature perturbation as shown in \eqref{twoVms} and \eqref{eq:pwdsi}, which contains two unequal contributions due to $\Pc\Tc$ symmetry of dS \eqref{FLRWdS} is broken during inflation. In other words our direct-sum vacuum $\vert 0\rangle_{qdS} = \vert 0_+\rangle_{qdS} \oplus \vert 0_-\rangle_{qdS}$ is $\Pc\Tc$ asymmetric. {We can express \eqref{eq:HZpowe} as the two-point correlation of momentum modes as
    \begin{equation}
        {}_{\rm qdS}\langle 0_\pm 
        \vert \hat\zeta_{\mathbf{k}\pm} \, \hat\zeta_{\mathbf{k}'\pm} \vert  0_{\pm} \rangle_{\rm qdS} 
= (2\pi)^3 \delta^{(3)}(\mathbf{k} + \mathbf{k}') \, \frac{2\pi^2}{k^3} \, \mathcal{P}_{\zeta}\LF 1\pm\Delta\Pc_v \RF\,.
\label{eq:kmkcoupling}
    \end{equation}
 During inflation, quantum fluctuations become squeezed as they are stretched to superhorizon scales, which is analogous to IHO physics, as noted in \cite{Albrecht:1992kf}. This means, the modes when they are deep inside the horizon behave like they are ingoing "Minkowski" states (see \eqref{BDdS}) and when they evolve, we can understand \eqref{eq:kmkcoupling} as pairs of outgoing entangled states with momentum $\textbf{k}$ and $-\textbf{k}$. In DSI, the two-mode squeezed state can be obtained as 
 \begin{equation}
     \vert \Psi_{sq}\rangle =  \begin{pmatrix}
         \prod_{\mathbf{k}} \frac{1}{\sqrt{\vert \alpha^{+}_k\vert }}
\exp\!\Big(- \frac{\beta_k^{+}}{2\alpha_k^{+}} \, \hat{b}^\dagger_{\mathbf{k},\,+} \hat{b}_{-\textbf{k},\,+}^\dagger\Big)\vert 0_+\rangle_{\rm qdS}  \\ 
\prod_{\mathbf{k}} \frac{1}{\sqrt{\vert \alpha^{-}_k\vert} }
\exp\!\Big(- \frac{\beta_k^{-}}{2\alpha_k^{-}} \, \hat{b}^\dagger_{\mathbf{k},\,-} \hat{b}_{-\textbf{k},\,-}^\dagger\Big)\vert 0_-\rangle_{\rm qdS} 
     \end{pmatrix} = \begin{pmatrix}
         \vert \Psi_{sq}^+\rangle \\ 
       \vert \Psi_{sq}^-\rangle  
     \end{pmatrix}
     \label{eq:sqstate}
 \end{equation}
 where $\vert \alpha_k^{\pm}\vert^2-\vert \beta_k^{\pm}\vert^2 =1$ are the Bogoliubov coefficients constrained by the canonical commutation relations of the MS variables $\hat V_{MS\pm}$. The derivation of  \eqref{eq:sqstate} is fairly straightforward, applying the steps in \cite{Albrecht:1992kf,Martin:2019wta} to the context of DSI. The important observation here is that the similarity between \eqref{eq:sqstate} and \eqref{LFvac} or \eqref{pure}. According to \eqref{eq:sqstate}, a pair of entangled fluctuations (correlated with opposite momenta) is generated as two components in the geometric SSS of quasi-dS vacua \eqref{qdSvac}. {This can be viewed as a new understanding of ER=EPR (See Table~\ref{tab-ER}) in the context of CMB antipodal correlations.}
 These two components of \eqref{eq:sqstate} get stretched to superhorizon scales and produce an imprint in the CMB in the form of parity asymmetry, which we shall discuss in the next sections.}  {The parity asymmetry here is entirely different from the theories that introduce parity-violating terms (for example, inflaton coupling to Weyl tensor term as $f(\phi)\epsilon^{\mu\nu\rho\sigma} W_{\tilde \mu\tilde\nu\rho\sigma} W^{\tilde\mu\tilde\nu\tilde\rho\tilde\sigma}$), which do not effect at all the scalar power spectrum, the scalar two-point correlations, by construction as it is widely shown in many works (See \cite{Baumann:2015xxa,Bartolo:2017szm,Creque-Sarbinowski:2023wmb,Komatsu:2022nvu} and references therin). This is because, due to the completely anti-symmetric character of $\epsilon_{\mu\nu\alpha\beta}$ tensor, the two-point scalar correlations would completely cancel out.  } 
    
    Applying the same DQFT quantization to the inflationary tensor modes \eqref{hijexp} yields \cite{Kumar:2022zff}
    \begin{equation}
    \Pc_{h\pm}  = \int d^3r e^{-i\textbf{k}.\textbf{r}}	{}_{\rm qdS}\langle 0_{\pm}\vert  \hat h_{ij\pm}\LF \pm\tau,\,\pm \textbf{x} \RF  \hat h_{ij\pm}\LF \pm \tau,\, \pm\textbf{x}^\prime \RF\vert 0_\pm\rangle_{\rm qdS} \approx P_h\LF 1\pm \Delta \Pc_u\RF
    \label{eq:pTdsi}
\end{equation} 
where 
 \begin{equation}
P_h = A_t\LF \frac{k}{k_\ast} \RF^{n_t},\, r=-8n_t\,,\quad \Delta \Pc_u= \LF  \frac{r}{8}\RF\operatorname{Re}\LT \frac{2}{H_{3/2}^{(1)} \LF  \frac{k}{k_\ast} \RF} \frac{\pd H^{(1)}_{\nu_s}\LF \frac{k}{k_\ast} \RF}{\pd\nu_s} \Bigg\vert_{\nu_s=\frac{3}{2}} \RT
		\label{eq:delpPT}
	\end{equation}	
    where $r =\frac{A_t}{A_s}<0.032$ is the ratio of tensor to scalar power spectrum 
    which is bounded 
    from above with the recent BICEP2/Planck/Keck Array B-mode polarisation data \cite{BICEP:2021xfz,Tristram:2021tvh}. {The tensor power spectrum of DSI \eqref{eq:pTdsi} predicts here an asymmetric amplitude for two-point correlations of tensor modes at the parity conjugate points of physical space. Again, this is distinct from the features of tensor power spectra from parity-violating theories, which predict additional polarizations, chirality, and cosmological birefringence features of primordial gravitational waves \cite{Lue:1998mq,Gluscevic:2010vv,Contaldi:2003zv,Komatsu:2022nvu}. }

{The parity asymmetry in the DSI implies suppression of power in the angular power spectra (both in the scalar and tensor sectors) of even multipoles and an enhancement in the odd multipoles. Therefore, our parity asymmetry is a new, distinct feature that we will study in the next sections. }

\subsection{Parity asymmetry versus even-odd asymmetry}

Probes of CMB such as COBE, WMAP, and Planck have measured the angular correlations of temperature fluctuations  $\Tc(\hat{n}) = \frac{\Delta T(\hat n)}{T_0}$, which can formally be written as the sum of its symmetric (even parity)
	$S(\hat{n})$  and its antisymmetric (odd parity)
	$A(\hat{n})$ components that can be expanded in spherical harmonics as:
	\begin{equation}
 \begin{aligned}
		\Tc(\hat{n}) & = \sum_{\ell,\,m} a_{\ell m} Y_{\ell m} \\ & =S(\hat{n}) + A(\hat{n}) \\ & = \sum_{\ell,\,m} \LF a^S_{\ell m}+ a^A_{\ell m} \RF  Y_{\ell m}\LF \hat{n} \RF  
	  \end{aligned}
	\end{equation}
	where
	\begin{equation}
		\begin{aligned}
			 S(\hat{n}) \equiv   \frac{1}{2} \left[  \Tc(\hat{n}) +  \Tc(-\hat{n})   \right] = S(-\hat{n}), \quad 
			A(\hat{n}) \equiv   \frac{1}{2} \left[   \Tc(\hat{n}) -  \Tc(-\hat{n})  \right]   = -A(-\hat{n})
		\end{aligned}      
		\label{eq:parity}
	\end{equation}
	where $(-\hat{n})$ is parity $\Pc$ conjugate of $(\hat{n})$.   
{The spherical harmonics $Y_{\ell m}$ satisfy:
\begin{equation}
Y_{\ell m}(-\hat{n}) = Y_{\ell m}\LF \pi-\theta,\,\pi+\varphi \RF = \LF -1 \RF^\ell Y_{\ell m}\LF \theta,\,\varphi \RF = \LF -1 \RF^\ell  Y_{\ell m}(\hat{n}). 
\label{eq:Tparity}
\end{equation}
which translates into $a^S_{\ell m}= a_{\ell m}\big\vert_{\ell =even}$ and $a^A_{\ell m}= a_{\ell m}\big\vert_{\ell = odd}$.}    
    Note that  $\Tc(\hat{n})$,  $S(\hat{n})$, and $A(\hat{n})$ are modeled as stochastic random fields and are therefore per se neither a scalar nor a pseudo-scalar under $\Pc$ transformations or isotropic under rotations $R$. By construction, given in \ref{eq:parity}, the decomposition into $S(\hat{n})$ and $A(\hat{n})$  of the particular realization  $\Tc(\hat{n})$ of the random field has a defined parity, but they are not scalar or pseudo scalars. $C_\ell$ is called the angular TT-power spectrum, whose even-odd contributions are given by
 \begin{equation}
		C_{\ell} = \frac{1}{2\ell+1} \sum_m \vert a_{\ell m}\vert^2,\quad
  \quad \text{where}  \quad a_{\ell m}(\Tc=A+S)=   
\left\{ \begin{array}{ll} 
 a_{\ell m}(S) = a_{\ell m}   &  {\text{for}} \quad \ell={\rm even} \\
 a_{\ell m}(A) = a_{\ell m}   & {\text{for}} \quad \ell={\rm odd} \\
\end{array}, \right.
		\label{eq:cl0}
\end{equation}
 or in other words, the $S$ and $A$ maps correspond to the even and odd multipoles $\ell$ of the total map $\Tc$:
  \begin{equation}
C_{\ell} = C_{\ell}^A + C_{\ell}^S \quad , \quad   
C_{\ell}^S =		C_{\ell={\rm even}} = \frac{1}{2\ell+1} \sum_m \vert a_{\ell m}\vert^2,\quad C_{\ell}^A =	
C_{\ell={\rm odd}} = \frac{1}{2\ell+1} \sum_m \vert a_{\ell m}\vert^2\,. 
		\label{eq:cl}
	\end{equation}
{According to \eqref{eq:vpm}, the geometric imprint of parity asymmetric fluctuation in the temperature results in 
\begin{equation}
     \Tc^{DSI}\LF \hat n \RF =  \Tc^{SI}(\hat n) \LF 1+\delta\Ts (\hat n) \RF,\quad \delta\Ts\LF \hat n \RF = -\delta \Ts\LF -\hat n \RF\,
     \label{tempdecomp}
\end{equation}
where 
\begin{equation}
   \delta\Ts\LF \hat n \RF = \sum_{\ell,\,m} \LF -1 \RF^{\ell+1} \delta a_{\ell m} Y_{\ell m} \LF \theta,\,\varphi \RF 
\end{equation}
which positively contributes to odd $\ell$ and negatively contributes to the even $\ell$. Therefore, application of DQFT to single-field inflationary scalar fluctuations (Direct-sum Inflation (DSI)) gives \cite{Gaztanaga:2024vtr,Gaztanaga:2024whs}
\begin{equation}
    C_\ell^{\rm DSI} = C_\ell^{\rm SI} \left[1 + (-1)^{\ell+1} \Delta \mathcal{C}_\ell \right]
    \label{eq:Cl_DQFT}
\end{equation}
where the fractional asymmetric modulation $\Delta \mathcal{C}_\ell$ is
\begin{equation}
    \Delta \mathcal{C}_\ell = \frac{1}{C_\ell^{\rm SI}} \int_0^{k_c} \frac{dk}{k} A_s\left( \frac{k}{k_s} \right)^{n_s - 1} j_\ell^2\left( \frac{k}{k_s} \right) \Delta \mathcal{P}_v(k).
    \label{eq:RDcl}
\end{equation}
where $k_c = 0.02 k_\ast$ is the cut-scale that corresponds to the largest angular scales in the CMB  $\theta>6^\circ$ or $\ell<30$, $k<k_c$ are the first modes that exit the horizon during inflation. The DSI quantum fluctuations are non-Markovian in nature; therefore, the effect of these first modes on the small scales requires further investigation (See Section 5.4 of \cite{Gaztanaga:2024vtr} for more details). 
The standard SI angular power spectrum is:
\begin{equation}
    C_\ell^{\rm SI} = \int_0^{\infty} \frac{dk}{k} A_s\left( \frac{k}{k_\ast} \right)^{n_s - 1} j_\ell^2\left( \frac{k}{k_s} \right)
    \label{eq:SICl}
\end{equation} 
Similar calculation for tensor power spectrum \eqref{eq:pTdsi}  results in the parity asymmetric angular power spectrum for the B-modes as
\begin{equation}
    C_{\ell,\,BB}^{\rm DSI} =C_{\ell,\,BB}^{\rm SI} \left[1 + (-1)^{\ell+1} \Delta \mathcal{C}_{\ell,\,BB} \right]
    \label{clBB}
\end{equation}
where 
\begin{equation}
    C_{\ell,\,BB}^{\rm SI} = \int_0^{\infty} \frac{dk}{k} A_t\left( \frac{k}{k_\ast} \right)^{n_t} T_{\ell,\,BB}^2\left( \frac{k}{k_s} \right)
    \label{eq:SIClB}
\end{equation} 
and
\begin{equation}
    \Delta \mathcal{C}_{\ell,\,BB} = \frac{1}{C_{\ell,\,BB}^{\rm SI}} \int_0^{k_c} \frac{dk}{k} A_t\left( \frac{k}{k_s} \right)^{n_t} T_{\ell,\,BB}^2\left( \frac{k}{k_s} \right) \Delta \mathcal{P}_u(k).
    \label{eq:RDclBB}
\end{equation}
where $T_{\ell,\,BB}$ is the transfer function associated with $B$ modes \cite{Zaldarriaga:1996xe}. In the next subsection, we discuss the observational evidence for the oscillation between even-odd angular power spectra \eqref{eq:Cl_DQFT}. It is worth noting that DSI is the first theoretical model that gives this effect and explains the CMB data better. In the past, the majority of theoretical and observational studies interpreted low-$\ell$ data as power suppression at low multipoles \cite{Contaldi:2003zv}, but these studies ignored the enhancement of power in the odd multipoles as we have shown explicitly in \cite{Gaztanaga:2024vtr} using the latest Planck 2018 data. Several Planck-scale quantum gravity frameworks \cite{Ashtekar:2021izi,Kitazawa:2014mca,Cicoli:2013oba} and also phenomenological models \cite{Sinha:2005mn,Jain:2008dw,Gonzalez:2019icn,Lello:2013mfa,Cicoli:2014bja} addressed this "power suppression", though it is not what the CMB data is indicating to us, as we will see further.    }

\subsection{Observational evidence for DSI in the parity conjugate worlds of CMB}

\label{Sec:DSIobservational}

In this section, we present the first observational test of DQFT in the context of primordial cosmology, which is responsible for the temperature fluctuations in the CMB. {
To assess the significance of the observed parity asymmetry in the CMB, we compare $10^6$ simulated realizations of the data under two models: the Standard Inflation (SI) model and the Direct Sum Inflation (DSI) model. We evaluate the posterior probability $p(M|D)$ of each model $M$ given the data $D$.
This approach contrasts with the standard practice in the CMB community, which typically estimates the likelihood $p(D|M)$—the probability of the data given a specific model—to assess the significance of low multipole anomalies. However, this likelihood-based method presupposes the model and can lead to inflated uncertainties, especially because the $\Lambda$CDM model with SI generally predicts more large-scale power than is observed. This mismatch increases the sampling variance, thereby reducing the apparent significance of observed parity anomalies. For example, the low measured quadrupole $C_2$ has $p(D|M)=2.62\%$ while the posterior value is 29 times smaller: $p(M|D)=0.09\%$, as shown in Table \ref{tab1:annomalies}.
}

Here, following standard CMB analysis, we assume statistical rotational ($R$) isotropy to focus on testing statistical parity ($\Pc$). Under statistical isotropy, the two-point function and power spectrum are defined as:
 \begin{equation}
		w[\theta] \equiv < \Tc(\hat{n}_1) \Tc(\hat{n}_2) > = \sum_{\ell=2}^{\ell_{max}} \frac{2\ell+1}{4\pi} C_\ell \, P_\ell[\cos{\theta}]
		\label{eq:C_theta}
	\end{equation}
where $\theta = \vert \hat n_1-\hat n_2\vert$. 
\begin{figure}
    \centering
    \includegraphics[width=0.6\linewidth]{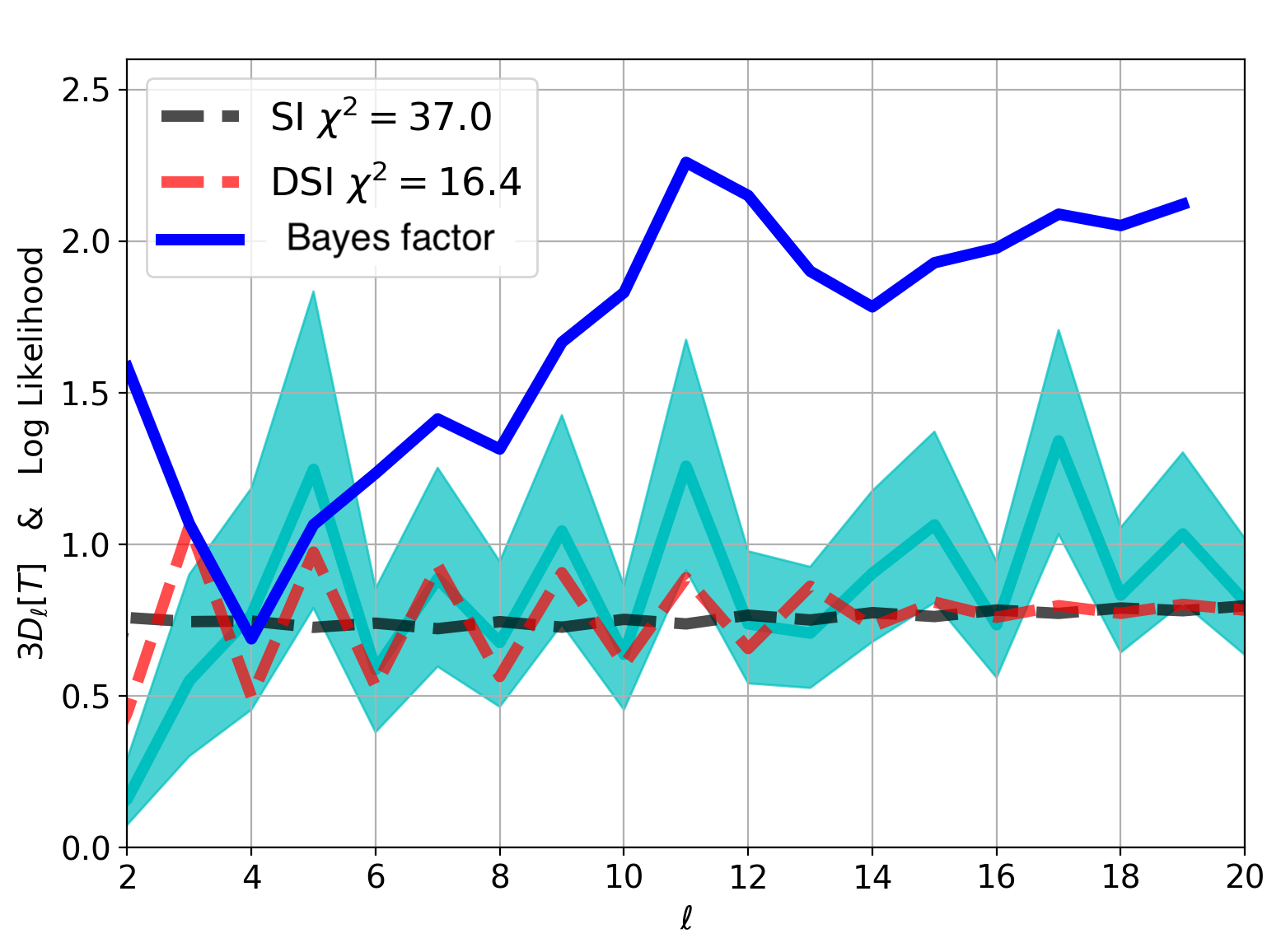}
    \caption{{Mean measured power spectrum $C_\ell$ (shown as $D_\ell \equiv \ell(\ell+1)\, C_\ell / 2\pi$, normalized such that the map has unit pixel variance) from the Planck 2018  CMB temperature map (cyan line). The shaded cyan band shows the 68\% spread from $10^6$ realizations of the mean measured $C_\ell$.
    The black dashed line corresponds to the Standard Inflation (SI) model, i.e., the best-fit scale-invariant $\Lambda$CDM model primarily constrained by the high-$\ell$ range ($30 \leq \ell \leq 2500$). The red dashed line shows the prediction from the Direct-Sum Inflation (DSI) model defined in \ref{eq:Cl_DQFT}, with no additional free parameters.
    The blue line at the bottom shows the Bayes factor, logarithm of the ratio of cumulative posterior likelihoods, $\log \left[ P_{\mathrm{DSI}}(<\ell) / P_{\mathrm{SI}}(<\ell) \right]$, as a function of $\ell$, demonstrating that the DSI model is up to 150 times more likely than SI on large angular scales ($\ell < 20$).}}
    \label{fig:cl}
\end{figure}
{Figure \ref{fig:cl} shows (in cyan) the normalized observed CMB power spectrum for the first multipoles $\ell <20$. The data show a clear even-odd asymmetry and a low quadrupole $C_2$. Both deviate significantly from the scale-invariant prediction of standard inflation (dashed black line).} Decades of analysis of CMB data \cite{Schwarz:2015cma,Muir:2018hjv,Planck:2019evm} with the following quantity 
\begin{equation}
	R_{TT} = \frac{\sum_{\ell =2}^{\ell_{max}} \ell\LF \ell+1 \RF C_{\ell=even} }{\sum_{\ell =3}^{\ell_{max}}\ell(\ell+1)C_{\ell=odd}}\approx 0.79
	\label{RTT}
\end{equation}
 indicates that there is more power (~20\%) in the odd-multipoles compared to even ones for $\ell_{max}\lesssim 20-30$ corresponding to $\theta\gtrsim 6-9^\circ$ (In Fig.~\ref{fig:cl} we show the results for $\ell_{max}=20$, the result does not vary much for  $\ell_{max}=30$ as it is shown in \cite{Gaztanaga:2024vtr}). {This means, on the very large scales, we exactly witness the temperature asymmetries at the antipodal regions of the CMB sky (See the last panel of Fig.~\ref{fig:ERB} and also Fig.~1 of \cite{Gaztanaga:2024vtr}). }

\begin{figure}
    \centering
    \includegraphics[width=0.45\linewidth]{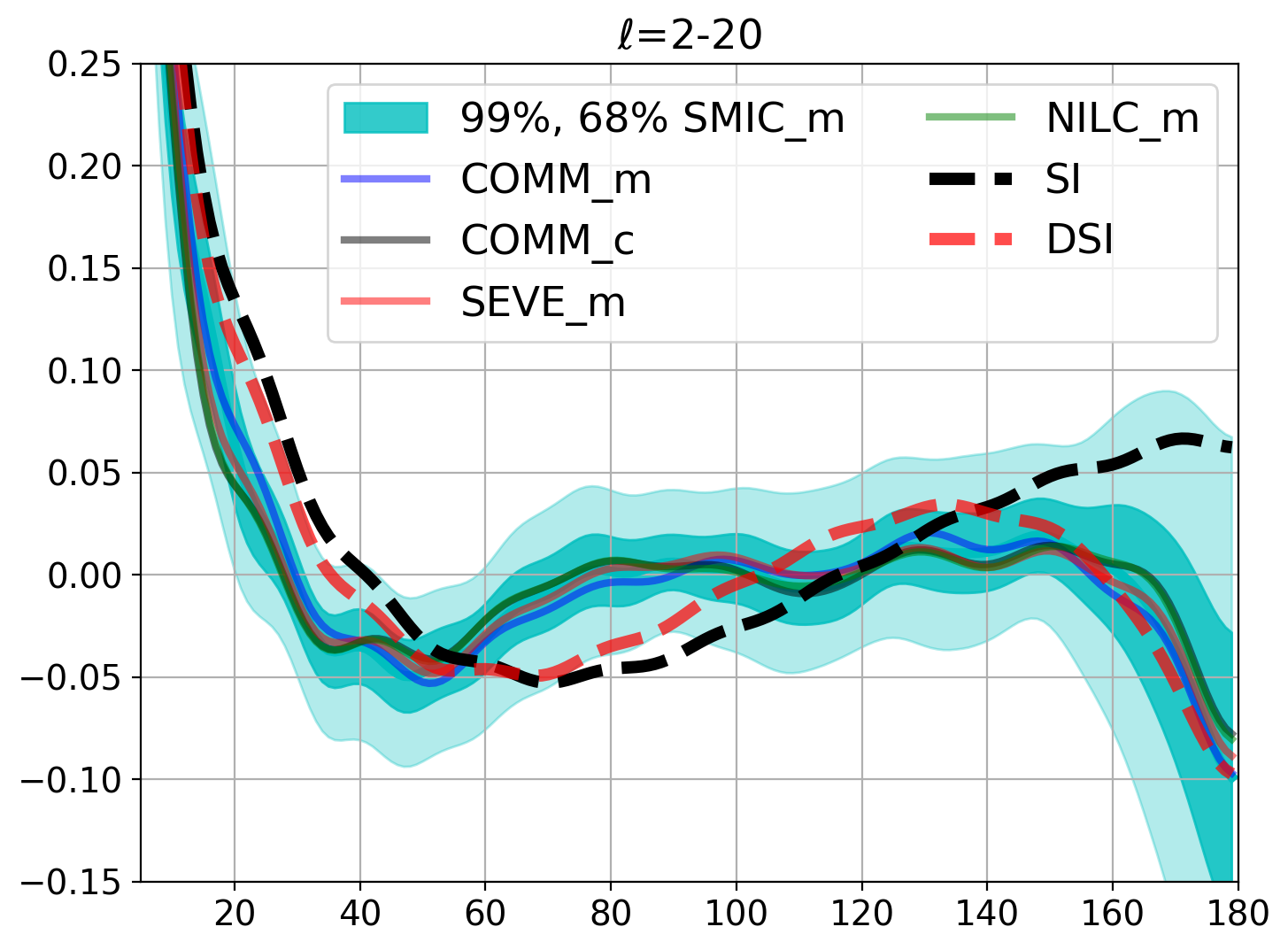}\quad     \includegraphics[width=0.45\linewidth]{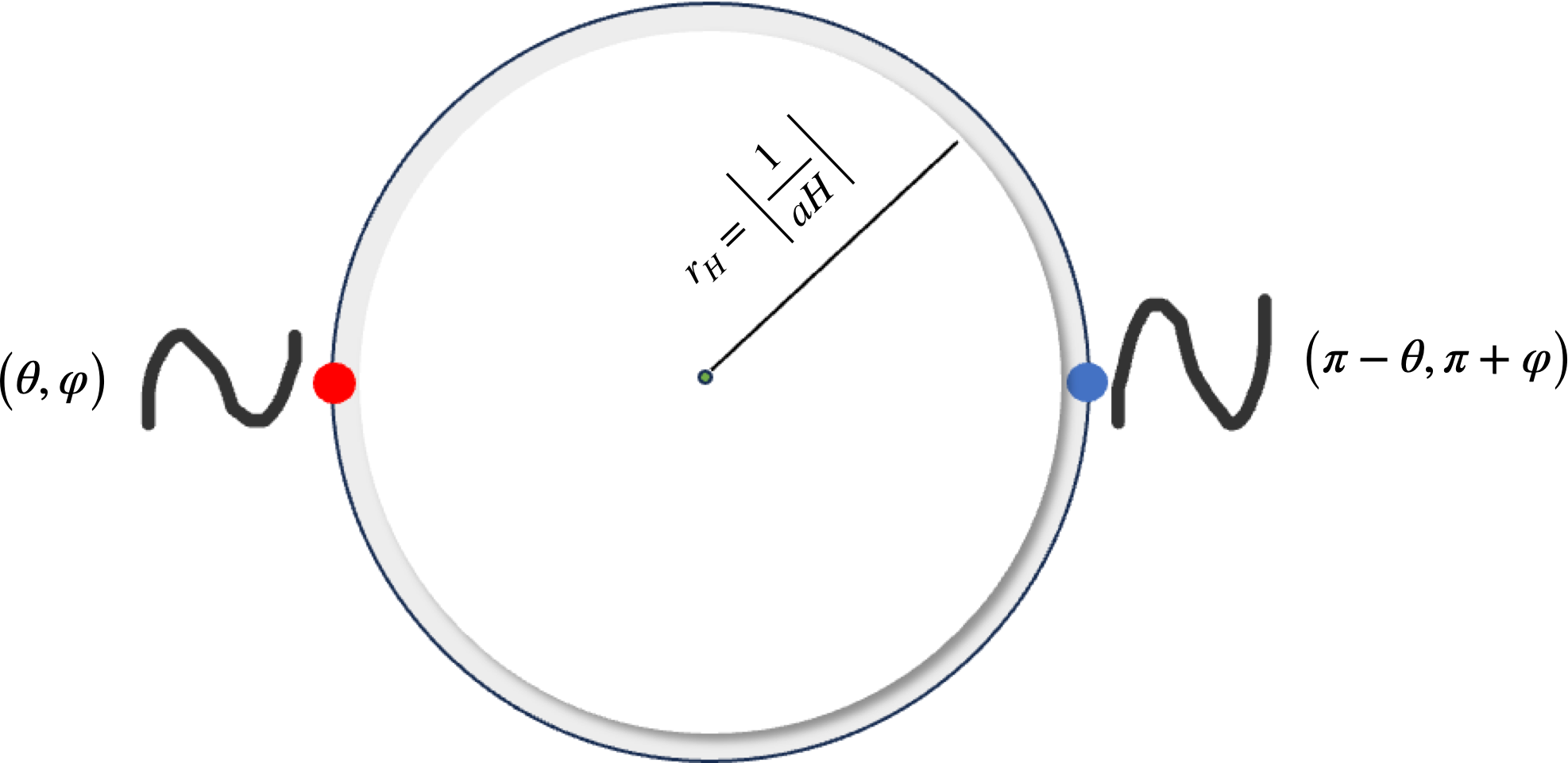}
    \caption{The left panel is for $w(\theta)$ which highlights the temperature correlations at large angular scales $\theta>9^\circ$, the CMB data lines are coloured lines with with 68\% and 99\% errors in the  SMIC$_m$ data \cite{Planck:2019evm}. It is very clear that the direct-sum inflationary (DSI) quantum fluctuations fit better (red-dashed line) than the standard inflation (SI), which is a black dashed line, in particular, the anticorrelation at $\theta = 180^\circ$. The right panel presents the physical (schematic) picture of quantum fluctuations in DSI evolving asymmetrically at parity conjugate points in physical space and leaving their imprints in the CMB when they leave the horizon.   }
    \label{fig:180}
\end{figure}

 Assuming ergodicity, we can instead search for statistical (an)isotropy or statistical parity (a)symmetry in specific statistical quantities measured from the CMB maps. For instance, we can define the { directional two-point function} as:

\begin{equation}
w[\hat{\theta}] \equiv \langle \Tc(\hat{n}_1) \Tc(\hat{n}_2) \rangle
\end{equation}
where the expectation value $\langle \cdots \rangle$ is taken over all pairs of directions. We can then examine if $w[\hat{\theta}]$ is statistically (an)isotropic, meaning whether it remains invariant (or not) within sampling errors under rotations $R$ of $\hat{\theta}$. Furthermore, we can investigate if $w[\hat{\theta}]$ is parity $\Pc-$ symmetric or antisymmetric—that is, whether $w[0] = \pm w[\pi]$, within sampling errors. {From the latest CMB data, we can deduce that $w[0]>0,\,w[\pi]<0$ and $w[0] \neq - w[\pi]$ (See a left panel of Fig.~\ref{fig:180})}. Because $\Pc$ (parity) and $R$ (rotation) are independent symmetries—no combination of $R$ transformations can reproduce $\Pc$—statistical parity symmetry is entirely distinct from statistical isotropy, despite recurring claims to the contrary in the literature (see, e.g., \cite{COMPACT:2024cud} and references therein).
 
{This particular parity asymmetry result is significant to about  3$\sigma$ standard deviations (p-value of $0.7\%$ in Table\ref{tab1:annomalies}). Further evidence for parity asymmetry came from the low quadrupole $C_2$ (which corresponds to a suppressed symmetric component as shown in Table \ref{tab1:annomalies} and Fig.\ref{fig:cl}) and a negative correlation at antipodal separations $w[\pi]$ (also shown in Table \ref{tab1:annomalies} and Fig.\ref{fig:180}).}
Note that the scale invariance characteristic of CMB is only statistically accurate for $\theta<1^\circ$ ($\ell>800$) close to the so-called pivot scale $k_\ast\approx 0.05 {\rm Mpc}^{-1}$.  The even-odd power asymmetry is related to parity, which is a discrete transformation, not anisotropy. Unfortunately, it is interpreted as anisotropy in the literature, which resulted in wrong deductions such as hemispherical or dipolar anisotropy or violation of the cosmological principle\cite{Jones:2023ncn,Smith:2024map}. A severe drawback of these deductions is the lack of sharp definitions of statistical anisotropy and the mistaking of parity with anisotropy. It was shown in \cite{Gaztanaga:2024vtr} that the Universe is statistically homogeneous and isotropic but asymmetric by parity.

\begin{table}

		\caption{Parity posterior probability  $p[M|D]$ ($\times10^2$, i.e. in $\%$) of a model  given the data (based on $10^6$ sky realizations of the data). Each line corresponds to a different CMB parity indicator. We compared two different models: Standard  Inflation quantum fluctuations with LCDM (SI) and direct-sum inflationary quantum fluctuations (DSI). 
			Data is estimated from the Planck 2018 masked SMICA component separation map. Very similar results are found for the other maps.
			In the last column, 'ratio' refers to the ratio of posterior probabilities (i.e., Bayes factor) between the DSI and SI predictions. In the second column, we also show for reference the non-posterior likelihood $p[D|M]$.}
		\begin{center}
			\label{tab1:annomalies}
			\begin{tabular}{l | c | c  c | c}
				Parity  & SI  & SI & DSI  & ratio  \\  indicator & \ $p[D|M]$ & \ $p[M|D]$ \ & \ $p[M|D]$ \ &  \  DSI/SI \\ \hline  \hline 
				$C_2$ & 2.62\ \% & 0.09\ \% & 3.3\ \% & 37 \\         
				$R^{TT}$ & 1.0\ \%  & 0.7\ \% & 39.5\ \% & 56  \\ 
				 $w[\pi]$ & 3.89\ \%  & 1.12\ \% & 45.3\ \% & 40 \\            
				$C_2 ,R^{TT}$  & 0.12\ \% &  0.003\ \% &  1.96\ \% & 653 \\ 
				$w[\pi] ,R^{TT}$  & &  0.45\ \% &  34.6\ \% & 77 \\
				$w[\pi] , C_2$  &  &  0.016\ \% &   2.65\ \% & 166 \\ \hline
				\hline
			\end{tabular}
		\end{center}
	\end{table}

The statement that CMB is scale-invariant is associated with the observational fit of
\begin{equation}
	C_{\ell} = \frac{2}{9\pi} \int_0^{k_c} \frac{dk}{k} j_{\ell}^2\LF \frac{k}{k_s} \RF
	\Pc_\Rc(k),\quad \Pc_\Rc  = A_s\LF \frac{k}{k_\ast} \RF^{n_s-1}
	\label{power-law}
\end{equation}
convoluted with $\Lambda$CDM model for small angular scales $\ell\gtrsim 200$. 

 Often, many cosmologists dismiss the importance of understanding large-scale features of the CMB with a statement that the data fall within the cosmic variance 
\begin{equation}
	\Delta C_\ell = \frac{C_\ell}{\sqrt{(2\ell+1)f_{sky}}} 
	\label{cosmvar}
\end{equation}
of the standard cosmological model with (near) scale-invariance \eqref{power-law}. This sampling variance errors results directly from assuming gaussian statistics in the $C_{\ell} $ definition of Eq.\ref{eq:cl0}.
In \eqref{cosmvar} $f_{sky}$ is the portion of the CMB sky considered in the analysis; usually, one masks the signals from our own galaxy to avoid data contamination from local sources.  This dismissal actually means the incompatibility of \eqref{power-law} with the data, and it is necessary to search for a theory that gives low-cosmic variance and, as such, fits the data better. The $R_{TT}$ in \eqref{RTT} indicates CMB angular power spectra oscillate between even-odd $\ell$ with decreasing amplitude. The literature of theoretical (and phenomenological models) often ignored half of the multipoles (i.e., (odd)-$\ell$) and interpreted data as indicating power suppression at low multipoles\cite{Contaldi:2003zv,Sinha:2005mn}. This misinterpretation has led to numerous works of building speculative models of inflation in the last decades. In a nutshell, both the theoretical and observational studies have corroborated with mutual wrong interpretations over the last two decades and left the CMB anomalies as an unresolved mystery. 

{To assess the significance of the low-multipole anomalies, we need to evaluate the posterior probability $p(M|D)$ of each model $M$ given the data $D$. This Bayesian approach contrasts with the standard practice in the CMB community, which evaluates $p(D|M)$—the likelihood of the data under the SI model. 
The latter approach tends to inflate uncertainties due to the excess power predicted by $\Lambda$CDM at large scales, which increases sampling variance and thereby reduces the apparent significance of observed parity asymmetry.}
	
{Table~\ref{tab1:annomalies} shows that the DSI power spectrum in Eq.~\eqref{eq:Cl_DQFT} is 650 times more probable to fit the data than the standard scale-invariant (SI) inflation model. The posterior probability for the quadrupole $C_2$ alone increases by a factor of 37 — from 0.09\% to 3.3\%. This represents a highly significant improvement. However, the DSI prediction for $C_2$ is still somewhat low (3.3\%), which leaves room for additional suppression of the largest-scale modes, as advocated in \cite{Gaztanaga:2025cun}}. Furthermore, the direct-sum mathematical bridges (ER bridges) between quantum field components at parity conjugate points explain 20\% excess of power in the odd multipoles \eqref{RTT}. 
Similarly, even-odd power asymmetry derived for inflationary graviton fluctuations \eqref{eq:SIClB}, which serves as a test for DSI with future primordial gravitational wave probes \cite{CMB-S4:2016ple,Campeti:2019ylm,LiteBIRD:2022cnt}. Towards the small angular scales in the CMB, parity asymmetry becomes insignificant because high-frequency modes are less affected during inflationary expansion compared to low-frequency modes.
Finally, we depict our observationally consistent new understanding of quantum fields in curved spacetime in analogy with ER bridges in Fig.~\ref{fig:ERB}. Note that the parity asymmetry we found in the CMB is different from the other recent parity-related investigations  \cite{Koh:2023qjq}, which are about small-scale effects due to specific modifications of gravity involving beyond SM degrees of freedom. Parity asymmetry in our context is much more generic due to the combined action of gravity and quantum mechanics. It is attributed to large (angular) scales and found to be insignificant at small (angular) scales in the CMB \cite{Gaztanaga:2024vtr}. 

\section{Conclusions}

\label{Sec:conc}

The reconciliation of general relativity (GR) with quantum mechanics (QM) remains one of the most pressing goals in theoretical physics. A core tension arises from their distinct temporal structures: while GR describes spacetime dynamics via second-order differential equations, quantum theory, especially in the presence of gravitational horizons, relies on first-order temporal evolution, which is in the sense that quantum fields are usually defined through the assumption of positive energy state according to Schr\"{o}dinger equation and the arrow of time. This clash is nowhere more evident than in settings involving event horizons, where issues such as unitarity loss, information paradoxes, and observer complementarity remain unresolved \cite{Almheiri:2019hni,Giddings:2022jda}. Understanding gravitational horizons and the
associated quantum effects play a key role in deciphering the origin and evolution of the Universe from the Big Bang to the present accelerated expansion \cite{Gibbons:1977mu,Ashtekar:2021dab,Gaztanaga:2022fhp,Gaztanaga:2025cun}.

We have examined the historically converging insights of Einstein and Rosen, Schr\"{o}dinger, and 't Hooft, and shown how they point toward a common foundational vision: the quantum description of a single physical universe with consistent time evolution. A crucial element in this investigation is the role of the inverted harmonic oscillators (IHOs), whose deep relevance to quantum theory was highlighted in the seminal work by Berry and Keating~\cite{Berry1999}. We showed that IHOs naturally arise in the quantum field-theoretic description of spacetimes with horizons, indicating that they are not merely mathematical artifacts but fundamental to understanding quantum fields in curved backgrounds.

Nearly a century after Einstein and Rosen's 1935 paper, key ideas about quantum gravity still revolve around their central insight: the necessity of nontrivial spacetime connectivity to preserve unitarity and causal consistency. The ER=EPR conjecture proposed by Maldacena and Susskind~\cite{Maldacena:2013xja} is a modern rephrasing of this idea in the language of quantum entanglement. Yet, unlike many contemporary approaches that invoke multiple parallel universes or many-worlds interpretations, the original perspectives of ER, Schr\"{o}dinger, and 't Hooft remain grounded in a single, unitary physical spacetime.

Adhering to this principle necessitates a reformulation of quantum theory, one that aligns with the geometrical and causal structure of general relativity (GR). Many of the most profound questions in theoretical physics, especially those at the Planck scale and beyond, remain unanswered because they arise precisely at the interface between quantum field theory and curved spacetime dynamics
\footnote{This remains true even when considering quantum gravity in specific curved backgrounds such as de Sitter space~\cite{Witten:2001kn}.}.

While numerous approaches to quantum gravity aim to extend the framework into the ultraviolet, we adopt a complementary strategy: addressing foundational gaps in our understanding of quantum fields in curved spacetime. This path is not only fruitful but essential. As 't Hooft reminds us in a recent article~\cite{tHooft:2023uez}, ``guessing does not often provide for the correct answers, and the best procedure for improving our understanding consists of systematic studies of imperfections that can easily have been overlooked.''%
\footnote{See also recent remarks by Peebles~\cite{Peebles:2024txt}, emphasizing the foundational role of quantum fields in curved spacetime for cosmology and astrophysics.}

Our investigation, grounded in direct-sum quantum theory and its application to IHOs, aims precisely at these foundational issues, offering new directions for reconciling quantum mechanics with gravity in a physically meaningful and observationally relevant way.

In this work, we explored a pathway to resolving these foundational tensions by proposing a novel framework: \emph{direct-sum quantum field theory} (DQFT). This construction stems from the longstanding intuition, going back to Einstein and Rosen in 1935 \cite{Einstein:1935tc}, Schrödinger in 1956 \cite{Schrodinger1956}, and ’t Hooft in recent decades \cite{tHooft:2016qoo}, that a consistent quantum gravitational description of nature must involve two sheets of spacetime, or equivalently, two time directions connected through a quantum bridge.

The key insight of this paper is that gravitational horizons (in black holes, de Sitter, and Rindler spacetimes) induce a natural partitioning of Hilbert space into geometric superselection sectors (SSS), each defined by discrete spacetime symmetries such as parity ($\Pc$) and time reversal ($\Tc$). These sectors form a \emph{direct-sum structure}, within which quantum fields remain globally unitary, even in spacetimes where standard QFT predicts loss of information or thermality.

A unifying element in this picture is the \emph{inverted harmonic oscillator} (IHO), whose quantum properties were reviewed in Section~\ref{sec:IHO}, and which recurs across all horizon-based physics as a universal effective degree of freedom. {In Sec .~\ref {sec:history}} we showed that IHOs naturally encode the dynamics of quantum fields near horizons{, and for a successful description of quantum fields in curved spacetime, we must necessarily understand our physical world with two arrows of time. In Sec.~\ref{sec:disumQT}, we built the direct-sum quantum field theory that gives a new understanding of spacetime based on discrete transformations such as $\Pc\Tc$. This particular quantization structure with two arrows of time at the parity conjugate regions gives a new geometrical understanding of IHO phase space envisioned by Berry and Keating \cite{Berry1999}. Our construction gives an understanding of quantum fields in curved spacetime with direct-sum mathematical bridges that connect the sheets of spacetime related by discrete space-time transformations. These mathematical bridges not only retain the vision of ER, but also restore the unitarity in curved spacetime. According to this framework, a pure state is a direct-sum (not a superposition) of two pure state components that represent the nature of the state within and behind the horizon (See \eqref{pure}, \eqref{entBH}, \eqref{puredS} and also \eqref{DQFTent}). The entanglement gets spread across the gravitational horizon in the form of pure states dictated by discrete spacetime transformations. This means that an observer not only witnesses pure states evolving into pure states but also can reconstruct the information behind the horizon. The gravitational horizons in our construction act quantum mechanically as "mirrors" so that we not only achieve unitarity but also observer complementarity.  This constitutes a first important step towards a consistent construction of QFTCS, which is essential for building quantum gravity \cite{Witten:2001kn,Witten:2021jzq}.  }

In Section~\ref{sec:ERDQFT}, we reinterpreted the ER=EPR conjecture \cite{Maldacena:2013xja} through the lens of DQFT. Rather than requiring geometric wormholes, we argue that entangled quantum states bridging  $\Pc\Tc$-conjugate regions realize mathematical Einstein-Rosen bridges within a single physical spacetime. This reformulation not only preserves unitarity but naturally accommodates observer complementarity and provides a field-theoretic underpinning to quantum connectivity across horizons. Historically, the concept of a thermofield double state, introduced by Israel in 1976 \cite{Israel:1976ur}, involved constructing a pure quantum state in an enlarged Hilbert space that is a tensor product of two copies of the original space. In this formalism, the second Hilbert space was considered fictitious, introduced merely to purify the thermal density matrix of the original system. In contrast, DQFT gives a physical interpretation to both sectors by identifying them with parity-conjugate regions of real spacetime. This reconceptualization removes the need for fictitious degrees of freedom and instead grounds the entanglement structure in observable correlations across gravitational horizons.

Crucially, in Section~\ref{Sec:DSI}, we applied this framework to cosmology, specifically, to the quantum fluctuations generated during inflation. By formulating inflationary quantum field dynamics within the direct-sum structure {(which is named "direct-sum inflation (DSI)"}, we predicted the emergence of parity asymmetry in the primordial power spectrum, a unique signature of quantum gravitational effects. {We showed that this} prediction {of DSI} aligns with well-documented anomalies in the CMB, notably the large-scale parity asymmetry observed in Planck data \cite{Gaztanaga:2024vtr}. {The statistical significance of our result is 650 times better than the standard theory of inflationary fluctuations (See Table.~\ref{tab1:annomalies}). We also predicted the parity asymmetric primordial tensor power spectrum, which leads to large-scale asymmetries in the B-model polarization data (See \eqref{clBB}). This would be a new prediction to test our DSI formalism with the future detection of primordial gravitational waves \cite{CMB-S4:2016ple,LiteBIRD:2022cnt}. Furthermore, we expect that the DQFT could lead to new signatures in the context of even-odd gravitational wave perturbations (the so-called quasi-normal modes \cite{Cardoso:2019apo,Gogoi:2023fow,Rosato:2025byu}) in BH physics, in particular in the context of dynamical BH horizons \cite{Price:1994pm,Ashtekar:2025wnu}. This is a new direction for our future investigations.   }

This connection between our deep theoretical structure and measurable cosmological data marks a significant step forward. Just as parity violation in beta decay (the Wu experiment of 1957) reshaped particle physics \cite{Wu:1957my}, the detection of parity anomalies in the CMB may hint at a new era in our understanding of quantum gravity.

\begin{figure*}
\centering
\includegraphics[width=0.7\linewidth]{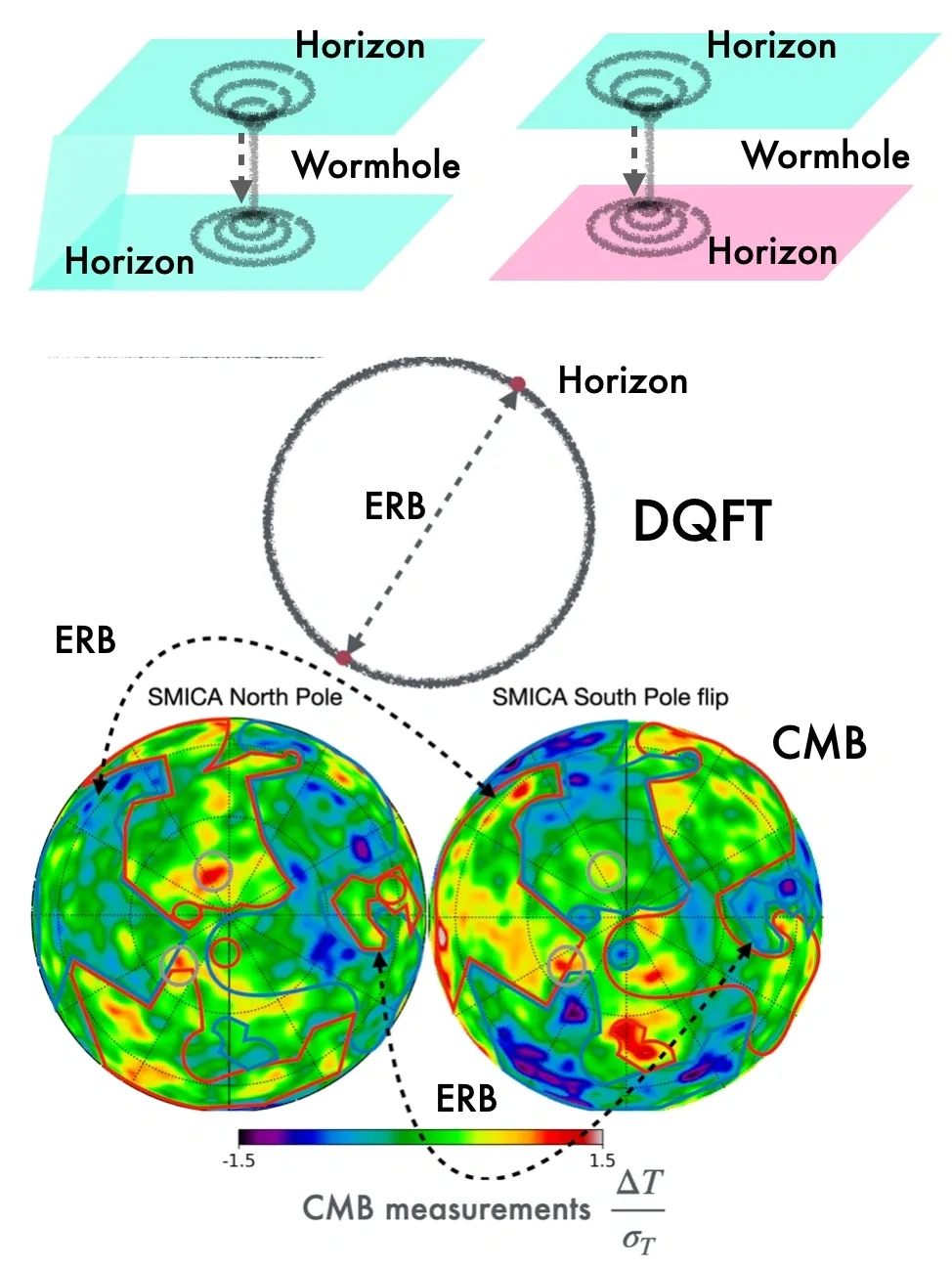}
\caption{
{\bf Einstein-Rosen Bridges (ERB):} "A particle in the physical universe must be described by a mathematical bridge between two sheets of spacetime" (Einstein and Rosen, 1935 \cite{Einstein:1935tc}). 
{\bf Top: Wormhole Interpretation of ERBs-} {(from classical modifications of gravity or introducing exotic matter)} Two configurations of wormholes {by classical considerations of gravity}: (1) connecting distinct sheets of spacetime (top right) or (2) linking space-like separated regions within a single sheet (top left)  {(See \cite{Misner:1957mt, Morris:1988cz, Morris:1988tu,Visser:1995cc,Lobo:2017cay,Rahaman:2013xoa,Rahaman:2015wpx} for more details)}. 
{\bf Middle: Direct-Sum Quantum Field Theory (DQFT) interpretation of ERBs-} 
Quantum theory describes ERBs with two opposing arrows of time, connecting any two antipodal (parity-conjugate) points within the same spacetime and inside the gravitational horizon. Instead of having a single bridge between two separate horizons, there are infinitely many (discrete) bridges within the same horizon.   
{\bf Bottom: Observational Visual evidence for DQFT Interpretation of ERBs-} CMB temperature fluctuations, measured by the Planck 2018 map, reveal a significant antisymmetric mirror pattern at antipodal points. This pattern demonstrates the imprints of a continuous ensemble of ERBs connecting parity-conjugate points.  This supports the DQFT treatment of inflationary quantum fluctuations, which follow the dynamics of a quantum inverted harmonic oscillator and become classical on superhorizon scales. 
These observations align with the principles of quantum gravity and holography \cite{tHooft:1993dmi, Susskind:1993if, Maldacena:1997re} {and bring a new understanding of ER=EPR (See Table.~\ref{tab-ER})}.
}
\label{fig:ERB}
\end{figure*}

\noindent To summarize, the main highlights of our investigation are:
\begin{itemize}
    \item Starting with Einstein and Rosen in 1935 \cite{Einstein:1935tc}, Schr\"{o}dinger in 1956 \cite{Schrodinger1956}, and ’t Hooft in 2016 \cite{tHooft:2016qoo}, all three emphasized the need for a description of quantum fields via two sheets of spacetime, connected through mathematical bridges.
    
    \item Discrete spacetime (a)symmetries play a key role in quantum theory. We proposed a new formulation in which a single quantum state is defined across a direct-sum Hilbert space built from geometric superselection sectors based on the discrete spacetime transformations. This framework enhances our understanding of the Berry-Keating quantization of the IHO \cite{Berry1999}, which is linked to the Riemann zeta function and to quantum chaos.
    
    \item Combining gravity and quantum mechanics requires a new understanding of time. The direct-sum quantum framework provides this and naturally realizes the mathematical bridges envisioned by Einstein and Rosen (see Fig.~\ref{fig:ERB}).
    
    \item Achieving unitarity and observer complementarity in curved spacetimes is a prerequisite for any consistent quantum theory of gravity. DQFT satisfies this by constructing geometric SSS that encode a new understanding of physics across the gravitational horizons.
    
    \item When applied to inflation, this framework predicts parity asymmetry in the primordial spectrum. We showed that this explains the observed CMB parity anomalies and provides observational evidence for underlying quantum gravitational phenomena.
\end{itemize}

\vspace{0.5em}
\noindent Looking forward, the direct-sum quantum structure, rooted in spacetime symmetries and horizon physics, offers a compelling bridge between the quantum and gravitational realms. It re-frames spacetime connectivity not as a geometric accident, but as a consequence of a new type of fundamental quantum entanglement across horizons.

Furthermore, Direct-sum quantum theory also clarifies the conceptual conundrums in quantizing inverted harmonic oscillators (IHOs) and their connection to the Riemann zeta function (See.~\ref{sec:BKDQ}). This development sets the stage for new developments in other related areas such as condensed matter, quantum chemistry, and biophysics, where IHOs play an important role (See \cite{Sundaram:2024ici} and references therein). {Especially, if we can design an experiment that involves time-dependent IHOs, we could create analogous conditions comparable to the context of early Universe inflationary quantum fluctuations. This new step would establish a new platform for developing direct-sum quantum theory.}
In particular, recent developments related to the observation of phase space horizons with BH analogue systems, such as surface gravity water waves \cite{Rozenman:2024ymh}, open new arenas for further exploring our new understanding of ER bridges.

\section*{Acknowledgements}
The authors thank Obinna Umeh, Mathew Hull, Mariam Bouhmadi-L\'opez and Paulo V. Moniz for useful discussions. KSK thanks Gerard 't Hooft for the useful comments. KSK would like to thank the Royal Society for the Newton International Fellowship. EG acknowledges grants from Spain Plan Nacional (PGC2018-102021-B-100) and Maria de
Maeztu (CEX2020-001058-M). This research was funded by Fundaç\~ao para a Ci\^encia e a Tecnologia grant number
UIDB/MAT/00212/2020 and
COST action 23130.

\hspace{0.5cm}
\section*{References}

\bibliographystyle{iopart-num}
\bibliography{ERref}

\providecommand{\newblock}{}
\begin{thebibliography}{100}
\expandafter\ifx\csname url\endcsname\relax
  \def\url#1{{\tt #1}}\fi
\expandafter\ifx\csname urlprefix\endcsname\relax\def\urlprefix{URL }\fi
\providecommand{\eprint}[2][]{\url{#2}}

\bibitem{Einstein:1935tc}
Einstein A and Rosen N 1935 {\em Phys. Rev.\/} {\bf 48} 73--77

\bibitem{Wigner1993}
Wigner E~P 1993 {\em {\"U}ber die Operation der Zeitumkehr in der
  Quantenmechanik\/} (Berlin, Heidelberg: Springer Berlin Heidelberg) pp
  213--226 ISBN 978-3-662-02781-3
  \urlprefix\url{https://doi.org/10.1007/978-3-662-02781-3_15}

\bibitem{Schrodinger1956}
Schr{\"o}dinger E 1956 {\em Expanding Universe\/} (Cambridge University Press)
  ISBN 9780521062213
  \urlprefix\url{https://books.google.co.uk/books?id=UQTsuQEACAAJ}

\bibitem{Hawking:1975vcx}
Hawking S~W 1975 {\em Commun. Math. Phys.\/} {\bf 43} 199--220 [Erratum:
  Commun.Math.Phys. 46, 206 (1976)]

\bibitem{Hawking:1976ra}
Hawking S~W 1976 {\em Phys. Rev. D\/} {\bf 14} 2460--2473

\bibitem{Zeldovich:1971mw}
Zeldovich Y~B and Starobinsky A~A 1971 {\em Zh. Eksp. Teor. Fiz.\/} {\bf 61}
  2161--2175

\bibitem{Starobinsky:1973aij}
Starobinsky A~A 1973 {\em Sov. Phys. JETP\/} {\bf 37} 28--32

\bibitem{Unruh:2017uaw}
Unruh W~G and Wald R~M 2017 {\em Rept. Prog. Phys.\/} {\bf 80} 092002
  (\textit{Preprint} \eprint{1703.02140})

\bibitem{Parikh:2002py}
Parikh M~K, Savonije I and Verlinde E~P 2003 {\em Phys. Rev. D\/} {\bf 67}
  064005 (\textit{Preprint} \eprint{hep-th/0209120})

\bibitem{Giddings:2022jda}
Giddings S~B 2022  (\textit{Preprint} \eprint{2202.08292})

\bibitem{Witten:2001kn}
Witten E 2001 {Quantum gravity in de Sitter space} {\em {Strings 2001:
  International Conference}\/} (\textit{Preprint} \eprint{hep-th/0106109})

\bibitem{Raju:2020smc}
Raju S 2022 {\em Phys. Rept.\/} {\bf 943} 1--80 (\textit{Preprint}
  \eprint{2012.05770})

\bibitem{Raju:2021lwh}
Raju S 2022 {\em Class. Quant. Grav.\/} {\bf 39} 064002 (\textit{Preprint}
  \eprint{2110.05470})

\bibitem{Almheiri:2019hni}
Almheiri A, Mahajan R, Maldacena J and Zhao Y 2020 {\em JHEP\/} {\bf 03} 149
  (\textit{Preprint} \eprint{1908.10996})

\bibitem{Kim:2022pfp}
Kim I~H and Preskill J 2023 {\em JHEP\/} {\bf 02} 233 (\textit{Preprint}
  \eprint{2212.00194})

\bibitem{Ashtekar:2025ptw}
Ashtekar A 2025 {\em Gen. Rel. Grav.\/} {\bf 57} 48 (\textit{Preprint}
  \eprint{2502.04252})

\bibitem{tHooft:2022umh}
't~Hooft G 2022 {How studying black hole theory may help us to quantise
  gravity} {\em {International conference~ on Eternity between Space and Time}:
  {From Consciousness to the Cosmos}\/} (\textit{Preprint} \eprint{2211.10723})

\bibitem{Witten:2021jzq}
Witten E 2022 {\em {Why does quantum field theory in curved spacetime make
  sense? And what happens to the algebra of observables in the thermodynamic
  limit?}\/} (\textit{Preprint} \eprint{2112.11614})

\bibitem{Donoghue:2019ecz}
Donoghue J~F and Menezes G 2019 {\em Phys. Rev. Lett.\/} {\bf 123} 171601
  (\textit{Preprint} \eprint{1908.04170})

\bibitem{tHooft:2018jeq}
't~Hooft G 2018 {\em Front. in Phys.\/} {\bf 6} 81 (\textit{Preprint}
  \eprint{1804.01383})

\bibitem{tHooft:2016qoo}
't~Hooft G 2016 {\em Found. Phys.\/} {\bf 46} 1185--1198 (\textit{Preprint}
  \eprint{1601.03447})

\bibitem{Berry1999}
Berry M~V and Keating J~P 1999 {\em H=xp and the Riemann Zeros\/} (Boston, MA:
  Springer US) pp 355--367 ISBN 978-1-4615-4875-1
  \urlprefix\url{https://doi.org/10.1007/978-1-4615-4875-1_19}

\bibitem{Kumar:2023ctp}
Kumar K~S and Marto J 2025 {\em Symmetry\/} {\bf 17} 29 (\textit{Preprint}
  \eprint{2305.06046})

\bibitem{Kumar:2023hbj}
Kumar K~S and Marto J 2024 {\em PTEP\/} {\bf 2024} 123E01 (\textit{Preprint}
  \eprint{2307.10345})

\bibitem{Kumar:2024oxf}
Kumar K~S and Marto J 2024 {\em Universe\/} {\bf 10} 320 (\textit{Preprint}
  \eprint{2405.20995})

\bibitem{Gaztanaga:2024vtr}
Gazta\~naga E and Kumar K~S 2024 {\em JCAP\/} {\bf 06} 001 (\textit{Preprint}
  \eprint{2401.08288})

\bibitem{Kumar:2022zff}
Kumar K~S and Marto J 2022  (\textit{Preprint} \eprint{2209.03928})

\bibitem{Gaztanaga:2024whs}
Gazta{\~n}aga E and Kumar K~S 2025 {\em Symmetry\/} {\bf 17} 1056
  (\textit{Preprint} \eprint{2403.05587})

\bibitem{Kumar:2024ahu}
Kumar K~S and Marto J 2024 {\em Gen. Rel. Grav.\/} {\bf 56} 143
  (\textit{Preprint} \eprint{2407.18652})

\bibitem{deBoer:2022zka}
de~Boer J {\em et~al.\/} 2022  (\textit{Preprint} \eprint{2207.10618})

\bibitem{Loll:2022ibq}
Loll R, Fabiano G, Frattulillo D and Wagner F 2022 {\em PoS\/} {\bf CORFU2021}
  316 (\textit{Preprint} \eprint{2206.06762})

\bibitem{Coleman:2018mew}
Coleman S 2018 {\em {Lectures of Sidney Coleman on Quantum Field Theory}\/}
  (Hackensack: WSP) ISBN 978-981-4632-53-9, 978-981-4635-50-9

\bibitem{Donoghue:2020mdd}
Donoghue J~F and Menezes G 2020 {\em Prog. Part. Nucl. Phys.\/} {\bf 115}
  103812 (\textit{Preprint} \eprint{2003.09047})

\bibitem{Coleman:1977py}
Coleman S~R 1977 {\em Phys. Rev. D\/} {\bf 15} 2929--2936 [Erratum: Phys.Rev.D
  16, 1248 (1977)]

\bibitem{Subramanyan:2020fmx}
Subramanyan V, Hegde S~S, Vishveshwara S and Bradlyn B 2021 {\em Annals
  Phys.\/} {\bf 435} 168470 (\textit{Preprint} \eprint{2012.09875})

\bibitem{Albrecht:1992kf}
Albrecht A, Ferreira P, Joyce M and Prokopec T 1994 {\em Phys. Rev. D\/} {\bf
  50} 4807--4820 (\textit{Preprint} \eprint{astro-ph/9303001})

\bibitem{Betzios:2020wcv}
Betzios P, Gaddam N and Papadoulaki O 2021 {\em SciPost Phys. Core\/} {\bf 4}
  032 (\textit{Preprint} \eprint{2004.09523})

\bibitem{Ullinger:2022xmv}
Ullinger F, Zimmermann M and Schleich W~P 2022 {\em AVS Quantum Sci.\/} {\bf 4}
  024402

\bibitem{Sundaram:2024ici}
Sundaram S, Burgess C~P and O'Dell D~H~J 2024 {\em New J. Phys.\/} {\bf 26}
  053023 (\textit{Preprint} \eprint{2402.13909})

\bibitem{Schumayer:2011yp}
Schumayer D and Hutchinson D~A~W 2011 {\em Rev. Mod. Phys.\/} {\bf 83} 307--330
  (\textit{Preprint} \eprint{1101.3116})

\bibitem{Sierra:2016rgn}
Sierra G 2019 {\em Symmetry\/} {\bf 11} 494 (\textit{Preprint}
  \eprint{1601.01797})

\bibitem{BALAZS1990123}
Balazs N and Voros A 1990 {\em Annals of Physics\/} {\bf 199} 123--140 ISSN
  0003-4916
  \urlprefix\url{https://www.sciencedirect.com/science/article/pii/0003491690903704}

\bibitem{Sierra:2007du}
Sierra G 2007 {\em Nucl. Phys. B\/} {\bf 776} 327--364 (\textit{Preprint}
  \eprint{math-ph/0702034})

\bibitem{Berrychaos}
Berry M~V 1986 Riemann's zeta function: A model for quantum chaos? {\em Quantum
  Chaos and Statistical Nuclear Physics\/} ed Seligman T~H and Nishioka H
  (Berlin, Heidelberg: Springer Berlin Heidelberg) pp 1--17 ISBN
  978-3-540-47230-8

\bibitem{BARTON1986322}
Barton G 1986 {\em Annals of Physics\/} {\bf 166} 322--363 ISSN 0003-4916
  \urlprefix\url{https://www.sciencedirect.com/science/article/pii/0003491686901429}

\bibitem{Aneva:1999fy}
Aneva B 1999 {\em Phys. Lett. B\/} {\bf 450} 388--396 (\textit{Preprint}
  \eprint{0804.1618})

\bibitem{Almheiri:2020cfm}
Almheiri A, Hartman T, Maldacena J, Shaghoulian E and Tajdini A 2021 {\em Rev.
  Mod. Phys.\/} {\bf 93} 035002 (\textit{Preprint} \eprint{2006.06872})

\bibitem{Maldacena:2013xja}
Maldacena J and Susskind L 2013 {\em Fortsch. Phys.\/} {\bf 61} 781--811
  (\textit{Preprint} \eprint{1306.0533})

\bibitem{Misner:1957mt}
Misner C~W and Wheeler J~A 1957 {\em Annals Phys.\/} {\bf 2} 525--603

\bibitem{Morris:1988tu}
Morris M~S, Thorne K~S and Yurtsever U 1988 {\em Phys. Rev. Lett.\/} {\bf 61}
  1446--1449

\bibitem{Visser:1995cc}
Visser M 1995 {\em {Lorentzian wormholes: From Einstein to Hawking}\/} ISBN
  978-1-56396-653-8

\bibitem{Lobo:2017cay}
Lobo F~S~N (ed) 2017 {\em {Wormholes, Warp Drives and Energy Conditions}\/}
  ({\em Fundamental Theories of Physics\/} vol 189) (Springer) ISBN
  978-3-319-55181-4, 978-3-319-85588-2, 978-3-319-55182-1 (\textit{Preprint}
  \eprint{2103.05610})

\bibitem{Rahaman:2013xoa}
Rahaman F, Kuhfittig P~K~F, Ray S and Islam N 2014 {\em Eur. Phys. J. C\/} {\bf
  74} 2750 (\textit{Preprint} \eprint{1307.1237})

\bibitem{Rahaman:2015wpx}
Rahaman F, Shit G~C, Sen B and Ray S 2016 {\em Astrophys. Space Sci.\/} {\bf
  361} 37 (\textit{Preprint} \eprint{1609.00155})

\bibitem{tHooft:2015pce}
't~Hooft G 2015  (\textit{Preprint} \eprint{1509.01695})

\bibitem{tHooft:2016rrl}
't~Hooft G 2017 {\em Found. Phys.\/} {\bf 47} 1503--1542 (\textit{Preprint}
  \eprint{1612.08640})

\bibitem{Gaztanaga:2025cun}
Gazta{\~n}aga E, Kumar K~S, Pradhan S and Gabler M 2025 {\em Phys. Rev. D\/}
  {\bf 111} 103537 (\textit{Preprint} \eprint{2505.23877})

\bibitem{Hartle:2013tm}
Hartle J~B 2014 {\em {The Quantum Mechanical Arrows of Time}\/} pp 113--128
  (\textit{Preprint} \eprint{1301.2844})

\bibitem{Hartle:1983ai}
Hartle J~B and Hawking S~W 1983 {\em Phys. Rev. D\/} {\bf 28} 2960--2975

\bibitem{Hartle:2007gi}
Hartle J~B, Hawking S~W and Hertog T 2008 {\em Phys. Rev. Lett.\/} {\bf 100}
  201301 (\textit{Preprint} \eprint{0711.4630})

\bibitem{Griffiths:2009dfa}
Griffiths J~B and Podolsky J 2009 {\em {Exact Space-Times in Einstein's General
  Relativity}\/} Cambridge Monographs on Mathematical Physics (Cambridge:
  Cambridge University Press) ISBN 978-1-139-48116-8

\bibitem{Hartman:2017}
Hartman T 2017 {\em {Lecture Notes on Classical de Sitter Space}\/}
  \urlprefix\url{http://www.hartmanhep.net/GR2017/desitter-lectures-v2.pdf}

\bibitem{Bousso:2002fq}
Bousso R 2002 {Adventures in de Sitter space} {\em {Workshop on Conference on
  the Future of Theoretical Physics and Cosmology in Honor of Steven Hawking's
  60th Birthday}\/} pp 539--569 (\textit{Preprint} \eprint{hep-th/0205177})

\bibitem{Hartman:2020khs}
Hartman T, Jiang Y and Shaghoulian E 2020 {\em JHEP\/} {\bf 11} 111
  (\textit{Preprint} \eprint{2008.01022})

\bibitem{Shaghoulian:2021cef}
Shaghoulian E 2022 {\em JHEP\/} {\bf 01} 132 (\textit{Preprint}
  \eprint{2110.13210})

\bibitem{Shaghoulian:2022fop}
Shaghoulian E and Susskind L 2022 {\em JHEP\/} {\bf 08} 198 (\textit{Preprint}
  \eprint{2201.03603})

\bibitem{Balasubramanian:2001rb}
Balasubramanian V, Horava P and Minic D 2001 {\em JHEP\/} {\bf 05} 043
  (\textit{Preprint} \eprint{hep-th/0103171})

\bibitem{Balasubramanian:2021wgd}
Balasubramanian V, Kar A and Ugajin T 2022 {\em Class. Quant. Grav.\/} {\bf 39}
  174001 (\textit{Preprint} \eprint{2104.13383})

\bibitem{Colas:2024xjy}
Colas T, de~Rham C and Kaplanek G 2024 {\em JCAP\/} {\bf 05} 025
  (\textit{Preprint} \eprint{2401.02832})

\bibitem{Brandenberger:2021pzy}
Brandenberger R 2021 {\em LHEP\/} {\bf 2021} 198 (\textit{Preprint}
  \eprint{2102.09641})

\bibitem{Brandenberger:2022pqo}
Brandenberger R and Kamali V 2022 {\em Eur. Phys. J. C\/} {\bf 82} 818
  (\textit{Preprint} \eprint{2203.11548})

\bibitem{Mukhanov:2007zz}
Mukhanov V and Winitzki S 2007 {\em {Introduction to quantum effects in
  gravity}\/} (Cambridge University Press) ISBN 978-0-521-86834-1,
  978-1-139-78594-5

\bibitem{Starobinsky:1980te}
Starobinsky A~A 1980 {\em Phys. Lett. B\/} {\bf 91} 99--102

\bibitem{Baumann:2018muz}
Baumann D 2018 {\em PoS\/} {\bf TASI2017} 009 (\textit{Preprint}
  \eprint{1807.03098})

\bibitem{Mukhanov:2005sc}
Mukhanov V 2005 {\em {Physical Foundations of Cosmology}\/} (Oxford: Cambridge
  University Press) ISBN 978-0-521-56398-7

\bibitem{Birrell:1982ix}
Birrell N~D and Davies P~C~W 1982 {\em {Quantum Fields in Curved Space}\/}
  Cambridge Monographs on Mathematical Physics (Cambridge, UK: Cambridge
  University Press) ISBN 978-0-511-62263-2, 978-0-521-27858-4

\bibitem{Sanchez:1986qn}
Sanchez N~G and Whiting B~F 1987 {\em Nucl. Phys. B\/} {\bf 283} 605--623

\bibitem{tHooft:2024auh}
't~Hooft G 2024  (\textit{Preprint} \eprint{2410.16891})

\bibitem{Israel:1976ur}
Israel W 1976 {\em Phys. Lett. A\/} {\bf 57} 107--110

\bibitem{Zeldovich:1977vgo}
Zel'dovich Y~B and Starobinsky A~A 1977 {\em JETP Lett.\/} {\bf 26} 252

\bibitem{Sasaki:1986hm}
Sasaki M 1986 {\em Prog. Theor. Phys.\/} {\bf 76} 1036

\bibitem{Mukhanov:1988jd}
Mukhanov V~F 1988 {\em Sov. Phys. JETP\/} {\bf 67} 1297--1302

\bibitem{Morris:1988cz}
Morris M~S and Thorne K~S 1988 {\em Am. J. Phys.\/} {\bf 56} 395--412

\bibitem{Cramer:1994qj}
Cramer J~G, Forward R~L, Morris M~S, Visser M, Benford G and Landis G~A 1995
  {\em Phys. Rev. D\/} {\bf 51} 3117--3120 (\textit{Preprint}
  \eprint{astro-ph/9409051})

\bibitem{Witten:2018zxz}
Witten E 2018 {\em Rev. Mod. Phys.\/} {\bf 90} 045003 (\textit{Preprint}
  \eprint{1803.04993})

\bibitem{Buchbinder:2021wzv}
Buchbinder I~L and Shapiro I 2023 {\em {Introduction to Quantum Field Theory
  with Applications to Quantum Gravity}\/} Oxford Graduate Texts (Oxford
  University Press) ISBN 978-0-19-887234-4, 978-0-19-883831-9

\bibitem{Giordano:2023wgx}
Giordano F, Negro S and Tateo R 2023 {\em JHEP\/} {\bf 10} 099
  (\textit{Preprint} \eprint{2307.15025})

\bibitem{Bekenstein:1973ur}
Bekenstein J~D 1973 {\em Phys. Rev. D\/} {\bf 7} 2333--2346

\bibitem{Susskind:1993if}
Susskind L, Thorlacius L and Uglum J 1993 {\em Phys. Rev. D\/} {\bf 48}
  3743--3761 (\textit{Preprint} \eprint{hep-th/9306069})

\bibitem{Penington:2019kki}
Penington G, Shenker S~H, Stanford D and Yang Z 2022 {\em JHEP\/} {\bf 03} 205
  (\textit{Preprint} \eprint{1911.11977})

\bibitem{Penington:2019npb}
Penington G 2020 {\em JHEP\/} {\bf 09} 002 (\textit{Preprint}
  \eprint{1905.08255})

\bibitem{Marolf:2020xie}
Marolf D and Maxfield H 2020 {\em JHEP\/} {\bf 08} 044 (\textit{Preprint}
  \eprint{2002.08950})

\bibitem{Jensen:2013ora}
Jensen K and Karch A 2013 {\em Phys. Rev. Lett.\/} {\bf 111} 211602
  (\textit{Preprint} \eprint{1307.1132})

\bibitem{Geng:2020fxl}
Geng H, Karch A, Perez-Pardavila C, Raju S, Randall L, Riojas M and Shashi S
  2021 {\em SciPost Phys.\/} {\bf 10} 103 (\textit{Preprint}
  \eprint{2012.04671})

\bibitem{Goto:2020wnk}
Goto K, Hartman T and Tajdini A 2021 {\em JHEP\/} {\bf 04} 289
  (\textit{Preprint} \eprint{2011.09043})

\bibitem{Page:1993wv}
Page D~N 1993 {\em Phys. Rev. Lett.\/} {\bf 71} 3743--3746 (\textit{Preprint}
  \eprint{hep-th/9306083})

\bibitem{Doran:2006dq}
Doran R, Lobo F~S~N and Crawford P 2008 {\em Found. Phys.\/} {\bf 38} 160--187
  (\textit{Preprint} \eprint{gr-qc/0609042})

\bibitem{Schwarz:2015cma}
Schwarz D~J, Copi C~J, Huterer D and Starkman G~D 2016 {\em Class. Quant.
  Grav.\/} {\bf 33} 184001 (\textit{Preprint} \eprint{1510.07929})

\bibitem{Martin:2004um}
Martin J 2005 {\em Lect. Notes Phys.\/} {\bf 669} 199--244 (\textit{Preprint}
  \eprint{hep-th/0406011})

\bibitem{tHooft:1993dmi}
't~Hooft G 1993 {\em Conf. Proc. C\/} {\bf 930308} 284--296 (\textit{Preprint}
  \eprint{gr-qc/9310026})

\bibitem{Planck:2019evm}
Akrami Y {\em et~al.\/} (Planck) 2020 {\em Astron. Astrophys.\/} {\bf 641} A7
  (\textit{Preprint} \eprint{1906.02552})

\bibitem{Martin:2019wta}
Martin J 2019 {\em Universe\/} {\bf 5} 92 (\textit{Preprint}
  \eprint{1904.00083})

\bibitem{Baumann:2015xxa}
Baumann D, Lee H and Pimentel G~L 2016 {\em JHEP\/} {\bf 01} 101
  (\textit{Preprint} \eprint{1507.07250})

\bibitem{Bartolo:2017szm}
Bartolo N and Orlando G 2017 {\em JCAP\/} {\bf 07} 034 (\textit{Preprint}
  \eprint{1706.04627})

\bibitem{Creque-Sarbinowski:2023wmb}
Creque-Sarbinowski C, Alexander S, Kamionkowski M and Philcox O 2023 {\em
  JCAP\/} {\bf 11} 029 (\textit{Preprint} \eprint{2303.04815})

\bibitem{Komatsu:2022nvu}
Komatsu E 2022 {\em Nature Rev. Phys.\/} {\bf 4} 452--469 (\textit{Preprint}
  \eprint{2202.13919})

\bibitem{BICEP:2021xfz}
Ade P~A~R {\em et~al.\/} (BICEP, Keck) 2021 {\em Phys. Rev. Lett.\/} {\bf 127}
  151301 (\textit{Preprint} \eprint{2110.00483})

\bibitem{Tristram:2021tvh}
Tristram M {\em et~al.\/} 2022 {\em Phys. Rev. D\/} {\bf 105} 083524
  (\textit{Preprint} \eprint{2112.07961})

\bibitem{Lue:1998mq}
Lue A, Wang L~M and Kamionkowski M 1999 {\em Phys. Rev. Lett.\/} {\bf 83}
  1506--1509 (\textit{Preprint} \eprint{astro-ph/9812088})

\bibitem{Gluscevic:2010vv}
Gluscevic V and Kamionkowski M 2010 {\em Phys. Rev. D\/} {\bf 81} 123529
  (\textit{Preprint} \eprint{1002.1308})

\bibitem{Contaldi:2003zv}
Contaldi C~R, Peloso M, Kofman L and Linde A~D 2003 {\em JCAP\/} {\bf 07} 002
  (\textit{Preprint} \eprint{astro-ph/0303636})

\bibitem{Zaldarriaga:1996xe}
Zaldarriaga M and Seljak U 1997 {\em Phys. Rev. D\/} {\bf 55} 1830--1840
  (\textit{Preprint} \eprint{astro-ph/9609170})

\bibitem{Ashtekar:2021izi}
Ashtekar A, Gupt B and Sreenath V 2021 {\em Front. Astron. Space Sci.\/} {\bf
  8} 76 (\textit{Preprint} \eprint{2103.14568})

\bibitem{Kitazawa:2014mca}
Kitazawa N and Sagnotti A 2015 {\em EPJ Web Conf.\/} {\bf 95} 03031
  (\textit{Preprint} \eprint{1411.6396})

\bibitem{Cicoli:2013oba}
Cicoli M, Downes S and Dutta B 2013 {\em JCAP\/} {\bf 12} 007
  (\textit{Preprint} \eprint{1309.3412})

\bibitem{Sinha:2005mn}
Sinha R and Souradeep T 2006 {\em Phys. Rev. D\/} {\bf 74} 043518
  (\textit{Preprint} \eprint{astro-ph/0511808})

\bibitem{Jain:2008dw}
Jain R~K, Chingangbam P, Gong J~O, Sriramkumar L and Souradeep T 2009 {\em
  JCAP\/} {\bf 01} 009 (\textit{Preprint} \eprint{0809.3915})

\bibitem{Gonzalez:2019icn}
Gonzalez M and Hertzberg M~P 2019 {\em JCAP\/} {\bf 10} 017 (\textit{Preprint}
  \eprint{1904.07249})

\bibitem{Lello:2013mfa}
Lello L, Boyanovsky D and Holman R 2014 {\em Phys. Rev. D\/} {\bf 89} 063533
  (\textit{Preprint} \eprint{1307.4066})

\bibitem{Cicoli:2014bja}
Cicoli M, Downes S, Dutta B, Pedro F~G and Westphal A 2014 {\em JCAP\/} {\bf
  12} 030 (\textit{Preprint} \eprint{1407.1048})

\bibitem{Muir:2018hjv}
Muir J, Adhikari S and Huterer D 2018 {\em Phys. Rev. D\/} {\bf 98} 023521
  (\textit{Preprint} \eprint{1806.02354})

\bibitem{COMPACT:2024cud}
Samandar A {\em et~al.\/} (COMPACT) 2024 {\em JCAP\/} {\bf 11} 020
  (\textit{Preprint} \eprint{2407.09400})

\bibitem{Jones:2023ncn}
Jones J, Copi C~J, Starkman G~D and Akrami Y 2023  (\textit{Preprint}
  \eprint{2310.12859})

\bibitem{Smith:2024map}
Smith A, Copi C~J and Starkman G~D 2024  (\textit{Preprint}
  \eprint{2409.03008})

\bibitem{CMB-S4:2016ple}
Abazajian K~N {\em et~al.\/} (CMB-S4) 2016  (\textit{Preprint}
  \eprint{1610.02743})

\bibitem{Campeti:2019ylm}
Campeti P, Poletti D and Baccigalupi C 2019 {\em JCAP\/} {\bf 09} 055
  (\textit{Preprint} \eprint{1905.08200})

\bibitem{LiteBIRD:2022cnt}
Allys E {\em et~al.\/} (LiteBIRD) 2023 {\em PTEP\/} {\bf 2023} 042F01
  (\textit{Preprint} \eprint{2202.02773})

\bibitem{Koh:2023qjq}
Koh S 2023 {\em New Phys. Sae Mulli\/} {\bf 73} 74--78

\bibitem{Gibbons:1977mu}
Gibbons G~W and Hawking S~W 1977 {\em Phys. Rev. D\/} {\bf 15} 2738--2751

\bibitem{Ashtekar:2021dab}
Ashtekar A, De~Lorenzo T and Schneider M 2021 {\em Adv. Theor. Math. Phys.\/}
  {\bf 25} 1651--1702 (\textit{Preprint} \eprint{2107.08506})

\bibitem{Gaztanaga:2022fhp}
Gaztanaga E 2022 {\em Symmetry\/} {\bf 14} 1849

\bibitem{tHooft:2023uez}
't~Hooft G 2023 {Quantum Foundations as a Guide for Refining Particle Theories}
  {\em {Windows on the Universe}: {30th Anniversary of the Rencontres du
  Vietnam}\/} (\textit{Preprint} \eprint{2312.09396})

\bibitem{Peebles:2024txt}
Peebles P~J~E 2024 {Status of the LambdaCDM theory: supporting evidence and
  anomalies} (\textit{Preprint} \eprint{2405.18307})

\bibitem{Cardoso:2019apo}
Cardoso V, Foit V~F and Kleban M 2019 {\em JCAP\/} {\bf 08} 006
  (\textit{Preprint} \eprint{1902.10164})

\bibitem{Gogoi:2023fow}
Gogoi D~J, {\"O}vg{\"u}n A and Demir D 2023 {\em Phys. Dark Univ.\/} {\bf 42}
  101314 (\textit{Preprint} \eprint{2306.09231})

\bibitem{Rosato:2025byu}
Rosato R~F, Biswas S, Chakraborty S and Pani P 2025 {\em Phys. Rev. D\/} {\bf
  111} 084051 (\textit{Preprint} \eprint{2501.16433})

\bibitem{Price:1994pm}
Price R~H and Pullin J 1994 {\em Phys. Rev. Lett.\/} {\bf 72} 3297--3300
  (\textit{Preprint} \eprint{gr-qc/9402039})

\bibitem{Ashtekar:2025wnu}
Ashtekar A and Krishnan B 2025  (\textit{Preprint} \eprint{2502.11825})

\bibitem{Wu:1957my}
Wu C~S, Ambler E, Hayward R~W, Hoppes D~D and Hudson R~P 1957 {\em Phys.
  Rev.\/} {\bf 105} 1413--1414

\bibitem{Maldacena:1997re}
Maldacena J~M 1998 {\em Adv. Theor. Math. Phys.\/} {\bf 2} 231--252
  (\textit{Preprint} \eprint{hep-th/9711200})

\bibitem{Rozenman:2024ymh}
Rozenman G~G, Ullinger F, Zimmermann M, Efremov M~A, Shemer L, Schleich W~P and
  Arie A 2024 {\em Commun. Phys.\/} {\bf 7} 165

\end{thebibliography}

\end{document}